\input{epsf}

\documentclass[12pt, draftclsnofoot, peerreview, onecolumn]{IEEEtran}

\makeatletter
\def\ps@headings{%
\def\@oddhead{\mbox{}\scriptsize\rightmark \hfil \thepage}%
\def\@evenhead{\scriptsize\thepage \hfil \leftmark\mbox{}}%
\def\@oddfoot{}%
\def\@evenfoot{}}
\makeatother
\usepackage{epsf}
\usepackage{graphicx}
\usepackage[cmex10]{amsmath}
\usepackage{amsthm}
\usepackage{amssymb}
\usepackage{epsfig,latexsym,amsmath,epsf,amssymb,amsfonts}
\usepackage{algorithm, algorithmic}
\usepackage{placeins}
\usepackage{url}
\usepackage{cite}
\usepackage{comment}
\usepackage{subfigure}
\usepackage{lipsum}
\usepackage{physics}
\usepackage{color}
\usepackage{multirow}
\usepackage{pgf,tikz}

\usepackage[top=0.75in, bottom=1in, left=0.625in, right=0.625in]{geometry}
\usepackage{bbm}

\usetikzlibrary{shapes,arrows,automata, calc}

\newtheorem{Definition}{\hskip 0pt Definition}
\newtheorem{Theorem}{\hskip 0pt Theorem}
\newtheorem{Lemma}{\hskip 0pt Lemma}
\newtheorem{Corollary}{\hskip 0pt Corollary}
\newtheorem{Proposition}{\hskip 0pt Proposition}

\newtheorem{Remark}{\hskip 0pt Remark}
\begin{document}
\title{A Hierarchical Game with Strategy Evolution for Mobile Sponsored Content and Service Markets}

\author{
\IEEEauthorblockN{Wenbo Wang,~\IEEEmembership{Member,~IEEE,}
Zehui Xiong,~\IEEEmembership{Student Member,~IEEE,}
Dusit Niyato,~\IEEEmembership{Fellow,~IEEE,}
Ping Wang,~\IEEEmembership{Senior Member,~IEEE} and
Zhu Han,~\IEEEmembership{Fellow,~IEEE}\vspace*{-4mm}}

\thanks{Wenbo Wang, Zehui Xiong, Dusit Niyato and Ping Wang are with the School of Computer Engineering, Nanyang Technological University, Singapore 639798
(email: wbwang@ntu.edu.sg, zxiong002@e.ntu.edu.sg, dniyato@ntu.edu.sg, wangping@ntu.edu.sg).}
\thanks{Zhu Han is with the Department of Electrical and Computer Engineering as well as the Department of Computer Science, University of Houston, TX 77004 USA
(email: zhan2@uh.edu).}\vspace*{-4mm}}

\maketitle
\begin{abstract}
In sponsored content and service markets, the content and service providers are able to subsidize their target mobile users through directly paying the mobile network operator, to lower the price of the data/service access charged by the network operator to the mobile users. The sponsoring mechanism leads to a surge in mobile data and service demand, which in return compensates for the sponsoring cost and benefits the content/service providers. In this paper, we study the interactions among the three parties in the market, namely, the mobile users, the content/service providers and the network operator, as a two-level game with multiple Stackelberg (i.e., leader) players. Our study is featured by the consideration of global network effects owning to consumers' grouping. Since the mobile users may have bounded rationality, we model the service-selection process among them as an evolutionary-population follower sub-game. Meanwhile, we model the pricing-then-sponsoring process between the content/service providers and the network operator as a non-cooperative equilibrium searching problem. By investigating the structure of the proposed game, we reveal a few important properties regarding the equilibrium existence, and propose a distributed, projection-based algorithm for iterative equilibrium searching. Simulation results validate the convergence of the proposed algorithm, and demonstrate how sponsoring helps improve both the providers' profits and the users' experience.
\end{abstract}
\begin{IEEEkeywords}
Sponsored content/service market, global network effects, multi-leader-follower game, evolutionary game, variational inequalities.
\end{IEEEkeywords}

\newpage
\section{Introduction}\label{Sec:Introduction}
\begin{figure}[t]
  \centering
  \begin{tikzpicture} [->,>=stealth',shorten >=1pt, auto,node distance=5cm, inner sep=0pt, bend angle=35, minimum width=4.50mm, minimum height=4.0mm, font=\tiny]
    \node [label={[xshift=-0.4cm, yshift=-2.3cm]{$N$ Over-The-Top Service Providers}}]
      (N1) at (0,0) {\includegraphics[width=0.12\linewidth]{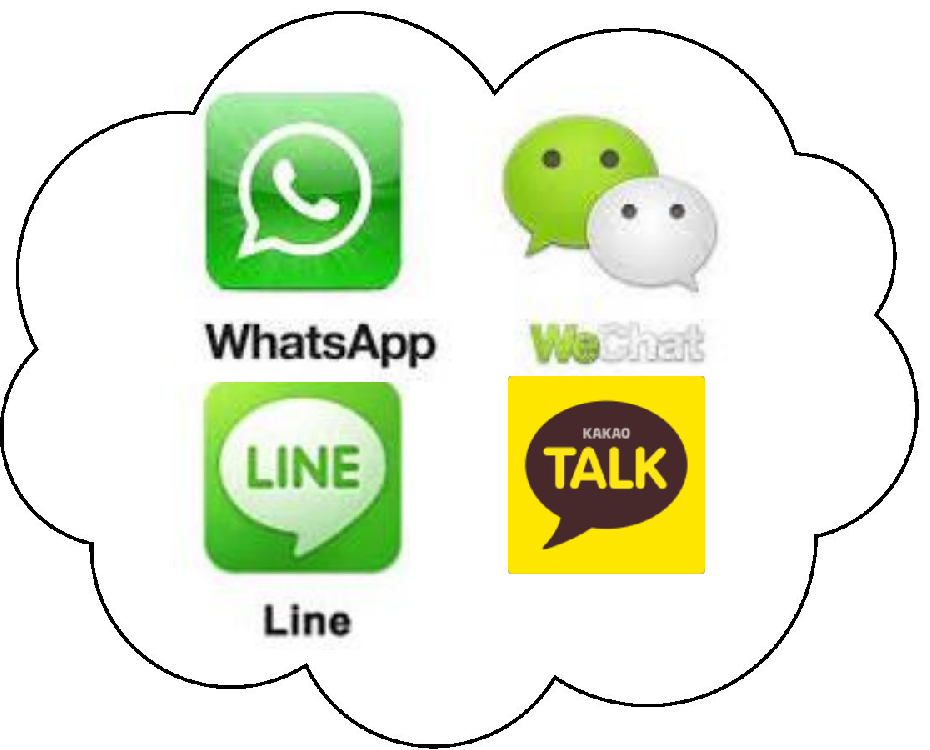}};
    \node [label={[xshift=0.0cm, yshift=0.0cm]{Mobile Network Operator}}]
      (N2) at (4.1,2.0) {\includegraphics[width=0.09\linewidth]{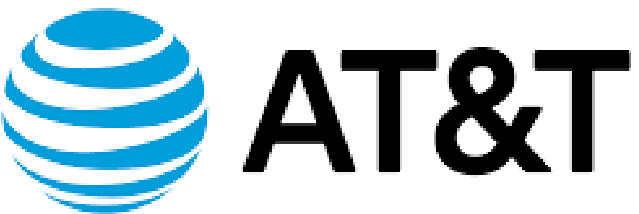}};
    \node [label={[xshift=0.0cm, yshift=-1.7cm]{$M$ Mobile Users}}]
      (N3) at (4.1,-2.0)  {\includegraphics[width=0.09\linewidth]{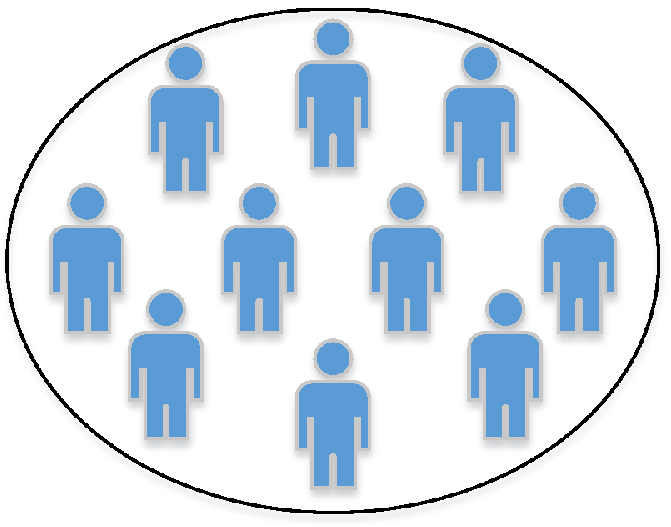}};

    \path [->, line width=1.2pt, every node/.style={sloped,anchor=south,auto=false}]
      (N1) edge [left] node {Distribution Payment} (N2)
      (N1) edge [left] node {Sponsorship Transfer} (N3)
      (N3) edge [below] node {Subscription Payment} (N2);
  \end{tikzpicture}\vspace*{-4mm}
\caption{A schematic example of the sponsored service markets, where all the payments are made to the network operator.}\label{Fig_illustration}
\end{figure}

The recent few years has witnessed a pervasive growth of the mobile data market at about 40\% per year~\cite{index2016global}, thanks to the explosion in the number of mobile applications and daily active Mobile Users (MUs). In response to the market growth, the Internet Service Providers (ISPs) are also upping the ante in their quest for more customers. In 2014, {AT\&T} collaborated with market portal companies like Aquto and launched a billing platform to allow bill transfer from MUs to third-party Content and Service Providers (CSPs). Since then, the sponsored content/service market has developed rapidly with more and more companies identifying the business potential.
For example, Singtel, with the sponsorships provided by the social media networks (e.g., WeChat and Whatapp), has introduced the fixed-rate plans of unlimited Over-The-Top (OTT) service usage in Singapore\footnote{\url{https://sg.news.yahoo.com/wechat-active-discussions-singtel-020012232.html}.}. Verizon, in its competition with other ISPs, offers the FreeBee Data\footnote{\url{https://www.internetservices.verizon.com/mis/freebeeperks/overview/}.}, which allows content providers such as AOL to subside part or all of their data through mobile apps without impacting on users' existing data allotments.
With the ISP-CSP partnerships, this new market structure generates a virtuous cycle~\cite{musacchio2007network}. First, the enhanced data access to the sponsored contents at lower rate encourages a deeper user engagement with the CSPs. This leads to the beneficiary global network effect of the services that the CSPs provide~\cite{1653003}, especially when the engagement stimulates a higher user activity in the social media networks. Second, the CSPs gain more profit from more active user subscriptions, which in return compensates for the sponsorship cost. Third, the ISPs, as content/service distributors, are able to distinguish themselves with featured services in the market competition, and thus obtain more revenue with a larger user population (Figure~\ref{Fig_illustration}). Intuitively, such a market mechanism promises a win-win situation for all the three parties. However, the complexity of analyzing the interactions among the market entities becomes a significant challenge, especially when the market entities seek optimal market strategies non-cooperatively.

In facing such a challenge, we investigate the interactions among the three parties in the sponsored content and service market with the presence of significant global network effects. In this paper, we study a transparent market with a single ISP, i.e., a Mobile Network Operator (MNO), a group of (heterogeneous) CSPs and a large population of MUs. In particular, we focus on the strategy evolution in the population of the MUs and the dynamics of the CSPs' sponsorship provision in the framework of a hierarchical market, where both global network effects and the congestion effect are experienced by the MUs. In the considered hierarchical market, the MNO and the CSPs naturally take the lead to determine the rates of content/service subscriptions and the sponsorship levels. Then, the MUs follow to choose which CSP to subscribe to with bounded rationality. For such a mobile service market, we propose to model the interactions among the market entities as a two-level hierarchical game. The main contributions of this paper include:
\begin{itemize}
 \item [1)] We model the evolution of the sponsored service market as a two-level hierarchical game. On the user (i.e., follower) level, the dynamics of CSP subscriptions by the MUs is formulated as an evolutionary sub-game with the MUs adopting the pairwise proportional imitation protocol. On the provider (i.e., leader) level, the interaction between the MNO and the CSPs is formulated as a  non-cooperative game.
 \item [2)] For the formulated follower sub-game, we investigate the impact of both the congestion and the positive network externalities on the MUs' payoff. We consider the Evolutionary Stable Strategies (ESS) as the solutions of the evolutionary sub-game and investigate the existence and the uniqueness condition for the ESS. Statistically speaking, the proposed evolutionary-game model is especially appropriate for describing the subscription process among a massive population of MUs without full rationality.
 \item [3)] We provide a series of theoretical analyses on the equilibrium properties in the sponsored content and service market. We cast the equilibrium searching problem into a bi-level programming problem and propose a hierarchical strategy updating mechanism. The proposed decentralized equilibrium searching algorithms constitute an incentive-compatible strategy-updating protocol for the three parties in the market, and guarantee the convergence to the Nash Equilibrium (NE).
\end{itemize}

The rest of the paper is organized as follows. Section~\ref{sec_related_work} provides a brief review of the related work. In Section~\ref{sec_model}, we present the model of the sponsored content/service market and propose a hierarchical game-based formulation of the interactions among the three parties therein. In Section~\ref{sec_game_analysis}, we model the user-level sub-game as an evolutionary game and investigate the property of evolutionary dynamics in the CSP-selection process. Further, we model the optimal-pricing mechanism among the providers as a bi-level programming problem, and reveal the condition for the existence of both the NE and the Stackelberg Equilibrium (SE) in the market. In Section~\ref{sec_algorithm}, we propose a distributed projection-based updating scheme of equilibrium searching and provide a practical condition that guarantees its convergence. Section~\ref{Sec:Simulation} presents the simulation results and detailed analysis for the network performance. Section~\ref{Sec:Conclusion} concludes the paper.

\section{Related Work}
\label{sec_related_work}
In the past few years, a line of studies have attempted to provide insight into the dynamics of sponsored content and service markets from different perspectives. In~\cite{7218528}, a three-stage optimization hierarchy was proposed to model the process of selecting content volume, sponsorship levels and unit data price by the MUs, the CSPs and the monopolistic ISP in a pay-as-you-go market. Therein, the CSPs' revenue obtained from ad-clicking was assumed to be proportional to the consumed content volume, and the MUs were allowed to consume contents from different CSPs at the same time. A similar decision hierarchy was considered in~\cite{zhang2014sponsoring}, where the study was limited to a market with one large CSP and one small CSP in terms of their revenue levels. The competitions between the CPSs in two situations, namely, with fixed user number for the CSPs (short-run) and with flexible user number for the CSPs (long-run), were analyzed. The authors of~\cite{zhang2014sponsoring} further expanded their study in~\cite{zhang2015sponsored} to a more general situation with multiple CSPs. In~\cite{zhang2015sponsored}, a two-stage Stackelberg framework was introduced to capture the interactions between the CSPs and the ISP under a known quality of service level required by the MUs. The CSPs decided on whether to sponsor the MUs or not according to the aggregated traffic demand and the ISP's delivery capacity. It is worth noting that many existing studies have adopted similar mathematical models that are able to reflect the decision hierarchy in the markets (e.g., Stackelberg games~\cite{zhang2015sponsored, eldelgawy2015interaction,zhang2016tds} and auctions~\cite{7524559}), mainly owning to the ``price-then-respond'' property of the market structure. However, in the existing studies, the proposed market models are generally subject to non-scalable MU numbers. Meanwhile, the considered payoff/revenue models are usually limited to describing the specific scenarios where the CSPs rely on advertisements for sponsorship compensation~\cite{7218528, zhang2014sponsoring, zhang2015sponsored, eldelgawy2015interaction, zhang2016tds}.

Apart from the interaction among the three parties that displays strong characteristics of hierarchy, the sponsored service market is also characterized by the significant demand-side effects, also known as the \textit{network effects}. Generally, the network effect indicates that the public goods are more valuable to consumers as the number of their consumers increases~\cite{katz1994systems}. If the external benefits with the inclusion of new consumers are the same for all the current consumers, we call it the \textit{global network effect}. Identified as a predominant characteristic of information economies~\cite{brake2016}, network effects plays a significant role in the operations of Internet service markets. The global network effect can be widely observed in social service platforms, e.g., with value-added service providers such as WeChat and C2C e-commerce providers such as Ebay. Due to the user grouping effect, a user's involvement with a social service will result in positive network effects on the other users (e.g., its social friends). For both the MNO and the CSPs, how to attract the next ``marginal'' user has always been among the most important consideration. For the MUs, service sponsoring allows for better differentiation of services, which in return may greatly enhance their own utilities due to the positive network externalities. In the literature, the most relevant network-effect model to that of this paper can be found in~\cite{7835123}, where both the network effect and the congestion effect are jointly modeled by a quadratic payoff function, and a two-stage Stackelberg game is adopted to formulate the interaction between two parties, namely, the MUs and a single wireless provider. Nevertheless, to the best of our knowledge, the study on the impact of network effects on a scalable sponsored content/service market still remains an open issue, which becomes the exact motivation for our research.

\section{Network Model: a Hierarchical Game Interpretation}
\label{sec_model}
\subsection{System model}
\label{sub_sec_model}
We consider a sponsored content/service market as in Figure~\ref{Fig_illustration}, where a large population of $N$ MUs are offered similar but non-compatible mobile social services (e.g., OTT services such as messaging apps) by $M$ competing CSPs. The CSPs rely on the infrastructure (e.g., content delivery servers and backhaul-to-end bandwidth) provided by the monopolistic MNO for service/content delivery. The MNO offers a flat-rate subscription plan to the MUs for unlimited monthly service/data access to each CSP. Each CSP decides on the sponsorship level offered to its customers. Meanwhile, to maximize its profit, the CSP chooses the subsidy level for its customers and a bundle of delivery service to purchase from the MNO. Constrained by the limited budget, an MU independently choose to subscribe to one of the CSPs given the Quality of Experience (QoE) associated with the CSP and the subscription cost after discounting the subsidies.

We note that in practical scenarios, the data price charged by the ISP to the CSPs is usually low\footnote{According to~\cite{odlyzko2014will}, the current charge by ISPs for Content Delivery Network (CDN) services is as low as \$0.01 to \$0.02 per GB, and compared with the whole volume of world Internet traffic, the cost of data volume for the CSPs is very small.}. Therefore, value-added social network service providers are more sensitive to the quality of connectivity (i.e., throughput) since the revenue is made based on relatively little traffic~\cite{odlyzko2014will}. We suppose that the CSPs acquire from the MNO a capped bundle of content-delivery bandwidth and fairly distribute the available bandwidth to their users. Then, we consider that a social network user (i.e., an MU) perceives the QoE of a service based on two factors: the experienced delivery bandwidth, i.e., the internal network effect, and the social gains that it enjoys from the social popularity of the subscribed service, i.e., the global network effect.
Generally, for a pair of the one-time, flat-rate subscription price $p_u$ and the sponsorship level ${\theta_j}\!\in\![0,1]$ offered by CSP $j$, an MU's payoff of choosing CSP $j$ can be defined as:
\begin{equation}
\label{eq_user}
\pi^u_j = {\gamma_{j,1}}u_1(n_j, b_j) + {\gamma_{j,2}}u_2(n_j) - (1-\theta_j)p_u,
\end{equation}
where $n_j$ is the number of MUs subscribing to CSP $j$ and $b_j$ is the CDN bandwidth acquired by CSP $j$ from the MNO. In (\ref{eq_user}), the first term, $\gamma_{j,1}u_1(n_j, b_j)$ is the delivery bandwidth-related QoE function, where $u_1(n_j, b_j)$ is the user's satisfaction level and $\gamma_{j,1}\!>\!0$ is the QoE-sensitivity coefficient. The second term, ${\gamma_{j,2}}u_2(n_j)$ is the extrinsic benefit of an MU owing to the global network effect. $u_2(n_j)$ represents the social benefits due to the global network effect and $\gamma_{j,2}>0$ is the network-effect sensitivity coefficient.

Since the CDN bandwidth $b_j$ is fairly distributed among the $n_j$ subscribers, the QoE function $u_1(n_j, b_j)$ displays the global congestion effect. Namely, for the same amount of CDN bandwidth, an increasing number of subscriptions degenerates each MU's experience of the social network service (e.g., due to longer delay). Then, the user-perceived QoE of the service provided by CSP $j$, $u_1(n_j, b_j)$ is expected to satisfy the conditions of $\frac{\partial{u_1(n_j, b_j)}}{\partial{b_j}}>0$, $\frac{\partial{u_1(n_j, b_j)}}{\partial{n_j}}<0$ and $u_1(n_j=0, b_j)=0$. Meanwhile, we assume that the users' perception of the  social benefit of a service is homogeneous and positively driven by the size of the social network~\cite{katz1994systems}. Compared with the congestion effect $u_1(n_j, b_j)$, the global network effect $u_2(n_j)$ is usually modeled as a monotonically increasing function of $n_j$ with decreasing marginal satisfaction (cf.,~\cite{zhang2014sponsoring,chen2016incentivizing}). Namely, $u_2(n_j)$ is expected to satisfy the conditions of $\frac{\partial{u_2(n_j)}}{\partial{n_j}}>0$, $\frac{\partial^2{u_2(n_j)}}{\partial{n^2_j}}<0$ and $u_2(n_j=0)=0$. An exemplary realization of $u_1(n_j, b_j)$ can be constructed following~\cite{4575128} as:
\begin{equation}
\label{eq_negative_ne}
u_1(n_j, b_j) = \left\{
\begin{array}{ll}
 \log \left(\displaystyle\frac{b_j}{o_j+n_j}\right), & \textrm{if } n_j>0,\\
 0, & \textrm{if } n_j=0,
\end{array}\right.
\end{equation}
where $o_j\!\ge\!1$ is a constant representing the maintenance overhead, and $u_1(n_j)$ is piecewise continuous. Similarly, a typical realization of $u_2(n_j)$ can be constructed as follows:
\begin{equation}
\label{eq_positive_ne}
u_2(n_j)=\log (1+n_j),
\end{equation}

On the providers' side, a CSP $j$ ($1\!\le\! j\!\le\!M$) aims to maximize its profit by choosing the sponsorship level ${\theta_j}\!\in\![0,1]$ as well as negotiating with the MNO about the provision of CDN bandwidth $b_j$ ($b_j\!\ge\!0$). Here, the payment for CDN bandwidth is usually determined by the premium peering agreement between the MNO and each CSP~\cite{Courcoubetis:2016:NPP:2909066.2883610}. The CSP evaluates its profit on the basis of active-user worth, while having to pay the CDN bandwidth price and the user subsidies. Let $u^{\textrm{CSP}}_j(n_j)$ denote the total user worth.
We consider that $u^{\textrm{CSP}}_j(n_j)$ also possesses the property of decreasing marginal return, namely $\frac{\partial{u^{\textrm{CSP}}_j(n_j)}}{\partial{n_j}}>0$, $\frac{\partial^2{u^{\textrm{CSP}}_j(n_j)}}{\partial{n^2_j}}<0$ and $u^{\textrm{CSP}}_j(n_j=0)=0$.
Then, CSP $j$'s utility is:
\begin{equation}
\label{eq_CP}
{\pi^{c}_j} = u^{\textrm{CSP}}_j(n_j) - {p_u}\theta_j {n_j} - {p_c}{b_j},
\end{equation}
where $p_c$ is the contract price for CDN bandwidth $b_j$, and we consider that each CSP has a budget limit on the total spendings as ${p_u}\theta_j {n_j} + {p_c}{b_j}\le\overline{c}_j$. A typical realization of $u^{\textrm{CSP}}_j(n_j)$ can be adopted based on the logarithmic model (cf.~\cite{chen2016incentivizing}):
\begin{equation}
\label{eq_CP_utility}
u^{\textrm{CSP}}_j(n_j) = \sigma_j \log (1 + {n_j}),
\end{equation}
where $\sigma_j$ is the coefficient for the monetary worth of the total active subscriptions for CSP $j$.

Further, we consider that the MNO applies a uniform price $p_u$ for each user's subscription and an indiscriminative price $p_c$ for bandwidth provision to the CSPs. Then, the MNO's revenue is:
\begin{equation}
\label{eq_MNO}
{\pi^{o}} = p_c \sum_{j=1}^{M}b_j + {p_u} \sum_{j=1}^{M}n_j,
\end{equation}
where we assume that the highest prices are limited by the regulator authority as $0\!\le\!p_u\!\le\!\overline{p}_u$ and $0\!\le\! p_c\!\le\!\overline{p}_c$.

\begin{figure}[t]
  \centering
  \begin{tikzpicture} [->,>=stealth',shorten >=1pt, auto,node distance=5cm, inner sep=0pt, bend angle=35, minimum width=1.0mm, minimum height=1.50mm, font=\scriptsize]
      \node [scale=1] (MU1) at (-1.3,0) {\includegraphics[width=0.12\linewidth]{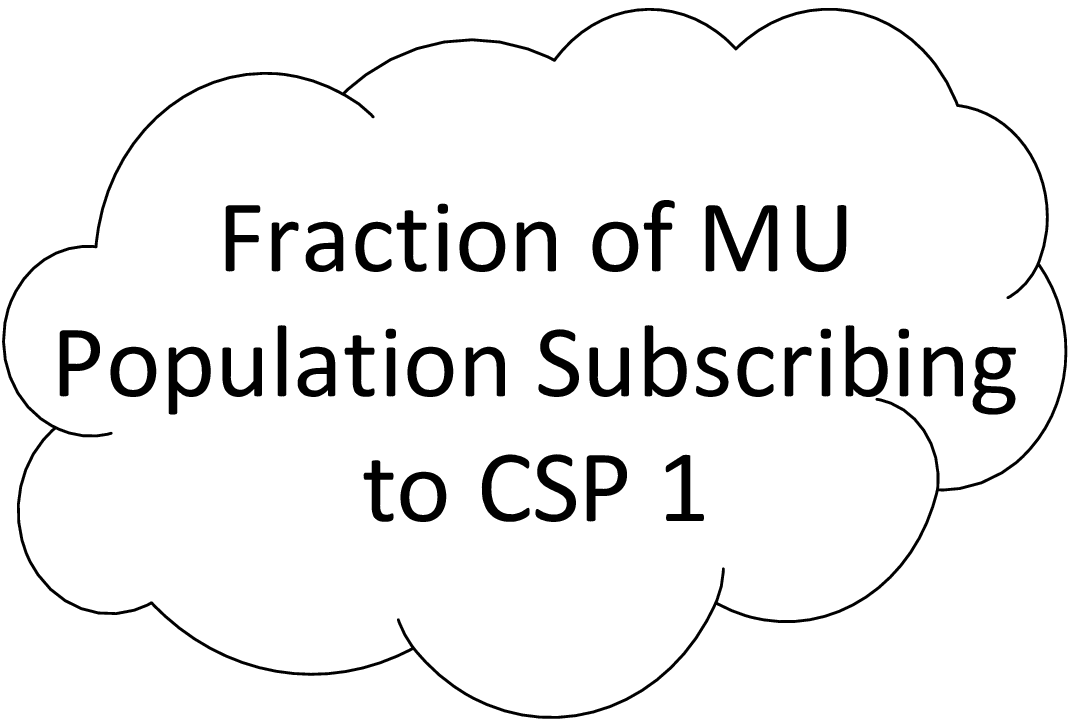}};
      \node [scale=1] (MU2) at (1.6,0) {\includegraphics[width=0.12\linewidth]{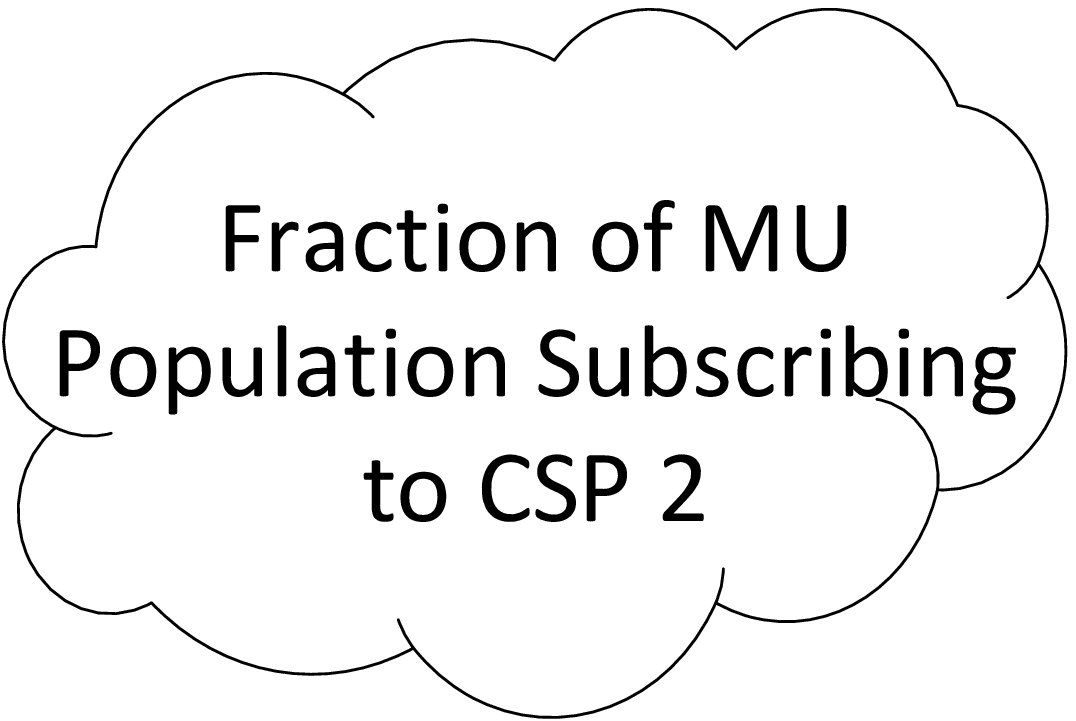}};
      \node [draw=none] (MUdot) at (3.5, 0) {$\ldots$};
      \node [scale=1] (MUM) at (5,0) {\includegraphics[width=0.12\linewidth]{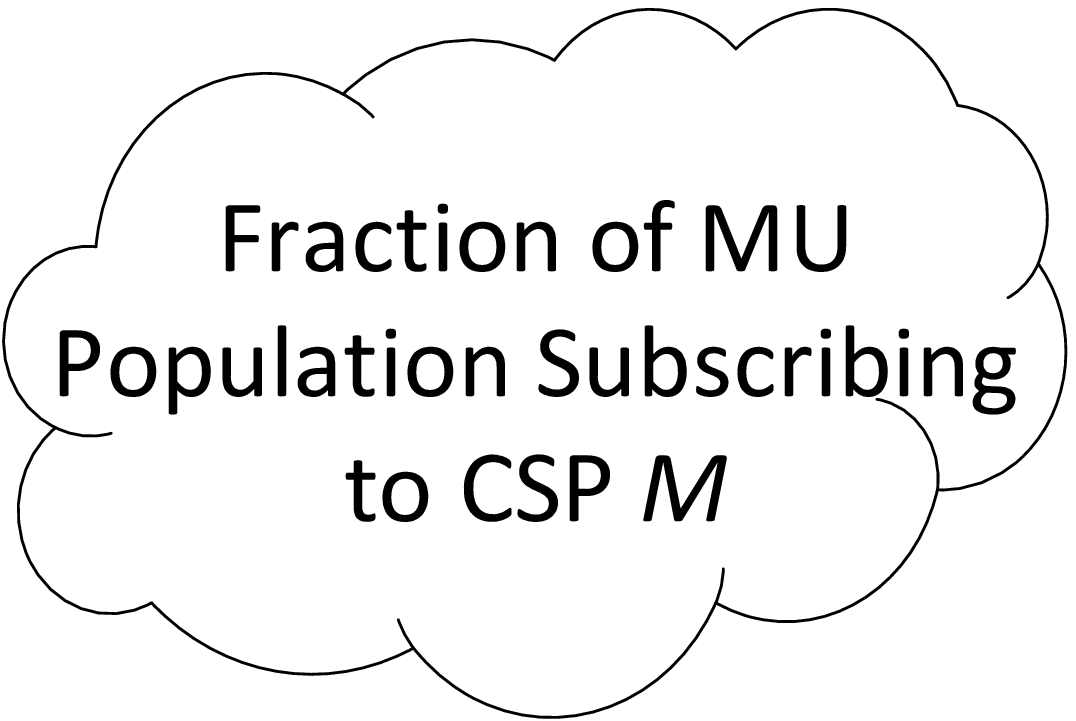}};
      \node [draw, dashed, blue, thick, align=center, minimum height = 2.0cm, text width=9.5cm,
      label={[align=center, blue, draw=none, inner sep=0pt, xshift=-9.3cm, yshift=-0.8cm] east: User-level Evolutionary Sub-game (see Section~\ref{sub_sec_ess})}] (Follower)
      at ([xshift=-1.3cm, yshift=-0.2cm] MUdot.west) {};

      \node [draw, align=center, minimum height=1.6cm, text width = 1cm, scale=0.8] (CSP1) at (3.5, 5) {CSP 1};
      \node [minimum size=2cm, scale=0.8] (CSPdot) at (3.5, 4) {$\vdots$};
      \node [draw, align=center, minimum height=1.6cm, text width = 1cm, scale=0.8] (CSPM) at (3.5, 3) {CSP $M$};

      \node [draw, dashed, thick, minimum height = 3.5cm, text width=3.5cm,
      label={[align=center, draw=none, inner sep=0pt, xshift=-2.2cm]east:Content and \\Service Providers}] (CSPs) at ([xshift=0.3cm, yshift=-0.0cm]CSPdot.east) {};

      \node [draw, align=center, minimum height = 3.5cm, text width=2.2cm] (MNO) at ([xshift=-4.0cm, yshift=-0.0cm] CSPdot.west) {Mobile Network \\Operator};

      \node [draw, dashed, blue, thick, align=center, minimum height = 4.3cm, text width=9.5cm,
      label={[align=center, blue, draw=none, inner sep=0pt, xshift=-9.3cm, yshift=1.8cm] east: Provider-level Non-cooperative Sub-game (see
      Sections~\ref{sub_sec_gne} and~\ref{sec_existence_SE})}] (Leader) at ([xshift=-0.7cm, yshift=0.3cm] CSPdot.west) {};

      \draw [<-, thick] ($(MNO.east) +(0,1.5)$) -- ($(CSP1.west) +(0,0.5)$) node [above,pos=0.25, align=center, xshift=0.9cm, scale=0.8] {Bandwidth Payment: $p_cb_1$};
      \draw [<-, thick] ($(MNO.east) +(0,1.0)$) -- ($(CSP1.west) +(0,0.0)$) node [above,pos=0.25, align=center, xshift=0.9cm, scale=0.8] {Subsidy Transfer: $p_u\theta_1n_1$};
      \draw [->, green, thick] ($(MNO.east) +(0,0.5)$) -- ($(CSP1.west) +(0,-0.5)$) node [black, above,pos=0.25, align=center, xshift=0.9cm, yshift=0.1cm, scale=0.8] {Providng Bandwidth $b_1$};

      \draw [<-, thick] ($(MNO.east) +(0,-0.5)$) -- ($(CSPM.west) +(0,0.5)$) node [above,pos=0.25, align=center, xshift=0.9cm, scale=0.8] {Bandwidth Payment: $p_cb_M$};
      \draw [<-, thick] ($(MNO.east) +(0,-1.0)$) -- ($(CSPM.west) +(0,0.0)$) node [above,pos=0.25, align=center, xshift=0.9cm, scale=0.8] {Subsidy Transfer: $p_u\theta_Mn_M$};
      \draw [->, green, thick] ($(MNO.east) +(0,-1.5)$) -- ($(CSPM.west) +(0,-0.5)$) node [black, above,pos=0.25, align=center, xshift=0.9cm, yshift=0.1cm, scale=0.8] {Providing Bandwidth $b_M$};

      \draw [->, thick] ($(MU1.north) +(0,-0.1)$) -- ($(Leader.south) +(-3.3,0.0)$) node [pos=0.25, align=center, rotate=180, xshift=1.0cm, yshift = 0.2cm,
      style={sloped,anchor=east,auto=false}, scale=0.8] {CSP Selection};
      \draw [->, thick] ($(MU2.north) +(0,-0.1)$) -- ($(Leader.south) +(-0.4,0.0)$) node [pos=0.25, align=center, rotate=180, xshift=1.0cm, yshift = 0.2cm,
      style={sloped,anchor=east,auto=false}, scale=0.8] {CSP Selection};
      \draw [->, thick] ($(MUM.north) +(0,-0.1)$) -- ($(Leader.south) +(3.0,0.0)$) node [pos=0.25, align=center, rotate=180, xshift=1.0cm, yshift = 0.2cm,
      style={sloped,anchor=east,auto=false}, scale=0.8] {CSP Selection};

      \draw [->, thick] ($(Follower.west) +(0,0)$) -- ++(-2mm,0) |- ($(MNO.west) +(0.0,0.0)$) node [pos=0.25, align=center, xshift=2.4cm, yshift = 0.2cm, align=center,
      style={sloped,anchor=east,auto=false}, scale=0.8] {Payment for Subscribing to CSP $j$: $(1-\theta_j)p_u$};
  \end{tikzpicture}\vspace*{-4mm}
\caption{Illustration of the hierarchical game framework for the sponsored service market.}\label{Fig_framework}
\end{figure}

\subsection{Hierarchical Game Formulation}
\label{sub_sec_game_formulation}
Considering the utility functions given in (\ref{eq_user})-(\ref{eq_MNO}), it is natural to cast the market dynamics described in Section \ref{sub_sec_model} into two stages. In the first stage, the MNO and the CSPs negotiate over the prices $p_c$, $p_u$ and the bandwidth $b_j$, and then each CSP $j$ ($1\le\!j\!\le\!M$) chooses the sponsorship level $\theta_j$ in a non-cooperative manner. In the second stage, each MU independently selects one CSP to subscribe to, based on the QoE and the subscription payments associated with the CSPs. Then, we can model the sponsored service market as a two-stage, multi-leader-multi-follower hierarchical game as in Figure~\ref{Fig_framework}. Mathematically, the two-stage hierarchical game can be described in the way of backward induction as follows:

\subsubsection{User-level evolutionary sub-game}
Let $a_0\!=\!(p_u,p_c)$ denote the MNO's pricing strategy and $a_j\!=\!(\theta_j,b_j)$ denote the strategy of CSP $j$ ($1\!\le\!j\!\le\!M$). Given a fixed vector of joint providers' strategy, $\mathbf{a}=[a_0,\ldots,a_M]^{\top}$, the user-level (follower), evolutionary sub-game for CSP selection is defined by a four-tuple: $\mathcal{G}_f\!=\!\langle \mathcal{N}, \mathcal{M}, \mathbf{x}, [\pi^u_j(\mathbf{x},\mathbf{a})]_{j=0}^M \rangle$, where
\begin{itemize}
 \item $\mathcal{N}$ is the single population of active MUs with the cardinality $\vert\mathcal{N}\vert=N$.
 \item $\mathcal{M}\!=\!\{0,1,\ldots, M\}$ is the set of strategies, where $m\!=\!0$ corresponds to the MUs' action of subscribing to no CSP and $m\!\ne\!0$ corresponds to the MUs' action of subscribing to CSP $m$. $\vert\mathcal{M}\vert=1+M$.
 \item $\mathbf{x}\!=\![x_0, x_1,\ldots, x_M]^{\top}$ is the vector of population states, where $x_j$ is the fraction of the MU population choosing CSP $j$ ($n_j\!=\!x_jN$),  and $\mathbf{x}$ is defined in the $M$-simplex $\mathcal{X}\!=\!\{ [x_0, x_1,\ldots, x_M]^{\top}\!\in\!\mathbb{R}^{M+1}\vert\sum_{j=0}^{M}x_j\!=\!1, x_j\!\ge\!0,\forall j\}$.
 \item $[\pi^u_j(\mathbf{x,\mathbf{a}})]_{j=0}^M$ is the vector of MUs' payoffs in the population state $\mathbf{x}$. $\forall j\ne0$, $\pi^u_j(\mathbf{x},\mathbf{a})$ is given by (\ref{eq_user}) with $n_j=x_jN$ and $\pi^u_0(\mathbf{x},\mathbf{a})=0$.
\end{itemize}
Given a joint provider action $\mathbf{a}$, we define the set of negative payoff for the population fractions as $F_f(\mathbf{x},\mathbf{a})\!=\!\left[-\pi^u_j(\mathbf{x},\mathbf{a})\right]_{j\in\mathcal{M}}^{\top}$. Then, from~\cite{Hofbauer20091665} the parametric NE of the user-level sub-game can be defined as follows.
\begin{Definition}[Follower Sub-game NE]
  \label{def_param_follower_NE}
  Given the joint leader action $\mathbf{a}$, an MU population state $\mathbf{x}^*(\mathbf{a})$ is a sub-game NE if for all feasible state $\mathbf{x}$ the following inequality holds
  \begin{equation}
   \label{eq_evolutionary_equilibrium}
   \left(\mathbf{x}-\mathbf{x}^*(\mathbf{a})\right)^{\top}F_f(\mathbf{x}^*(\mathbf{a}),\mathbf{a})\ge0.
  \end{equation}
\end{Definition}

It is worth noting that the NE defined in (\ref{eq_evolutionary_equilibrium}) provides the condition for the equilibrium to be Nash stationary, but not necessarily evolutionary stationary~\cite{Sandholm2009}. Then, it is necessary to check the stability of an NE state for CSP selection. From Definition~\ref{def_param_follower_NE}, suppose that there exists a population state $\mathbf{x}$ trying to invade state $\mathbf{x}^*(\mathbf{a})$ by attracting a small share $\epsilon\!\in\!(0,1)$ of the MUs to switch to state $\mathbf{x}$. Then, an Evolutionary Stable State (ESS) can be defined as follows (cf. (4)-(6) in~\cite{Sandholm2009}):
\begin{Definition}[Follower Sub-game ESS]
  \label{def_param_follower_ESS}
  Given any invading population state $\mathbf{x}$, $\mathbf{x}^*(\mathbf{a})$ is an ESS if there exists a small $\overline{\epsilon}\!\in\!(0,1)$ such that $\forall \epsilon\in(0,\overline{\epsilon})$ the following condition holds
  \begin{equation}
   \label{eq_sub_game_ESS}
   \sum_{j\in\mathcal{M}}x_j^*(\mathbf{a})\pi_j^u\left((1-\epsilon)\mathbf{x}^*(\mathbf{a})+\epsilon\mathbf{x}, \mathbf{a}\right)\ge
   \sum_{j\in\mathcal{M}}x_j(\mathbf{a})\pi_j^u\left((1-\epsilon)\mathbf{x}^*(\mathbf{a})+\epsilon\mathbf{x}, \mathbf{a}\right).
  \end{equation}
  Equivalently, (\ref{eq_evolutionary_equilibrium}) is satisfied by all invading $\mathbf{x}$ and at the equality condition, the following holds
  \begin{equation}
   \label{eq_sub_game_ESS_2}
   \left(\mathbf{x}^*(\mathbf{a})-\mathbf{x}\right)^{\top}F_f(\mathbf{x},\mathbf{a})\le0.
  \end{equation}
\end{Definition}

\subsubsection{Provider-level non-cooperative equilibrium searching problem}
With an MUs' population state $\mathbf{x}$, the non-cooperative equilibrium searching problem among the MNO and the CSPs can also be mathematically described by a three-tuple
$\mathcal{G}_l\!=\!\langle{\mathcal{K}}, \mathcal{A},\pmb\pi\rangle$, where
\begin{itemize}
 \item ${\mathcal{K}}\!=\!\{0,1,\ldots, M\}$ is the set of leaders, where $k\!=\!0$ represents the MNO and $k\!\ne\!0$ represents CSP $k$.
 \item $\mathcal{A}\!=\!\times_{k=0}^{M}\mathcal{A}_k$ is the Cartesian product of the action spaces for player $k\!\in\!{\mathcal{K}}$, $\mathcal{A}_0\!=\!\{(p_u,p_c)\vert 0\!\le\!p_u\!\le\!\overline{{p}_u}, 0\!\le\!p_c\!\le\!\overline{p}_c\}$ and $\mathcal{A}_k\!=\!\{(\theta_k,b_k)\vert \forall k\in\mathcal{K}\backslash\{0\},\theta_k\!\in\![0,1],p_u\theta_kx_kN\!+\!p_cb_k\!-\!\overline{c}_k\le0\}$. Then, $\mathbf{a}\!\in\!\mathcal{A}$.
 \item $\pmb\pi\!=\!\left[\pi^o(\mathbf{a}, \mathbf{x}), \pi^c_1(\mathbf{a}, \mathbf{x}),\ldots, \pi^c_M(\mathbf{a}, \mathbf{x})\right]^{\top}$ is the vector of payoff functions. The MNO's payoff $\pmb\pi_0\!=\!\pi^o(\mathbf{a}, \mathbf{x})$
 is given by (\ref{eq_MNO}), and the CSP $k$'s payoff ($k\!\ne\!0$), $\pmb\pi_k\!=\!\pi^c_k(\mathbf{a})$, is given by (\ref{eq_CP}).
\end{itemize}

The solution to the leaders' non-cooperative equilibrium searching problem depends on the information levels of the MNO of the CSPs. Suppose that the leaders do not anticipate the MUs' responses, then, we obtain the NE of the game based on the best response of the players in the two levels, $\mathbf{x}^*$ and $\mathbf{a}^*$, for which,
\begin{itemize}
 \item $\mathbf{x}^*$ satisfies the condition given in (\ref{eq_evolutionary_equilibrium}).
 \item $\mathbf{a}^*$ is the best response to $\mathbf{x}^*$ and satisfies the following condition:
 \begin{equation}
  \label{eq_best_response}
  \left\{\!\!
  \begin{array}{ll}
   \pi^o(a^*_0, a^*_{-0}, \mathbf{x}^*) \ge \pi^o(a_0, a^*_{-0}, \mathbf{x}^*), &\forall a_0\in\mathcal{A}_0,\\
   \pi^c_j(a^*_j, a^*_{-j}, \mathbf{x}^*) \ge \pi_j^c(a_j, a^*_{-j}, \mathbf{x}^*), &\forall j\in\mathcal{K}\backslash\{0\}, \forall a_j\in\mathcal{A}_j,
  \end{array}\right.
 \end{equation}
 where $a_{-i}$ is the joint adversaries' actions of player $i$ ($\forall i\!\in\!{\mathcal{K}}$). Note that here we slightly abuse the notations of $\pi^u_j$, $\pi^o$ and $\pi^c_j$ in (\ref{eq_user}), (\ref{eq_CP}) and (\ref{eq_MNO}) by giving the corresponding parameters, respectively.
\end{itemize}
On the other hand, if the leaders are able to anticipate the optimistic optimal best response in the follower sub-game, the SE solutions (i.e., the strong SE~\cite{dempe2002foundations}), $\mathbf{x}^*$ and $\mathbf{a}^*$, satisfy the following inequalities:
\begin{equation}
\label{eq_optimal_best_response}
\left\{
\begin{array}{ll}
 \pi^o(a^*_0, a^*_{-0}, \mathbf{x}^*(\mathbf{a}^*)) \ge \pi^o(a_0, a^*_{-0}, \mathbf{x}^*(a_0, a^*_{-0})), &\forall a_0\in\mathcal{A}_0,\\
 \pi^c_j(a^*_j, a^*_{-j}, \mathbf{x}^*(\mathbf{a}^*)) \ge \pi_j^c(a_j, a^*_{-j}, \mathbf{x}^*(a_j, a^*_{-j})), &\forall j\in\mathcal{K}\backslash\{0\}, \forall a_j\in\mathcal{A}_j,
\end{array}\right.
\end{equation}
where $\mathbf{x}^*(\mathbf{a})$ is drawn from the best-response mapping in (\ref{eq_evolutionary_equilibrium}) with respect to $\mathbf{a}$. Further, when the inequalities in (\ref{eq_optimal_best_response}) hold within an open neighbor area of $\mathbf{a}^*$, $U_{\epsilon}(\mathbf{a}^*)=\{\mathbf{a}:\Vert(\mathbf{a}-\mathbf{a}^*)\Vert^2<\epsilon\}$, for all $\mathbf{a}\in\!U_{\epsilon}(\mathbf{a}^*)$, we say that $(\mathbf{x}^*,\mathbf{a}^*(\mathbf{x}^*))$ is a local SE of the hierarchical game~\cite{Leyffer:2010:SMG:1744742.1744748}.

\section{Analysis of the Equilibria in the Hierarchical Game}
\label{sec_game_analysis}

\subsection{Evolutionary Stable Strategies in the User-level Sub-game}
\label{sub_sec_ess}
Given the leaders' strategy $\mathbf{a}$, the MUs' average payoff can be written from (\ref{eq_user}) and (\ref{eq_evolutionary_equilibrium}) as
$\overline{\pi}^u(\mathbf{x},\mathbf{a})=\sum^M_{j=0}x_j\pi_j^u(\mathbf{x},\mathbf{a})$. By the pairwise proportional imitation protocol~\cite{SANDHOLM200181}, the replicator dynamics yields the following system of Ordinary Differential Equations (ODEs), $\forall j\in\mathcal{M}$, which indicates that the subscription growth rate of a CSP is in proportion to the subscription's excess payoff:
\begin{equation}
 \label{eq_replicator_dynamics}
 \begin{array}{ll}
   \displaystyle\dv{x_j}{t}=v_j(\mathbf{x},\mathbf{a})=x_j(\pi_j^u(\mathbf{x},\mathbf{a})-\overline{\pi}^u(\mathbf{x},\mathbf{a}))\\
   =\!x_j\Big(\gamma_{1,j}u_1(x_jN)\!+\!\gamma_{2,j}u_2(x_jN)\!-\!(1\!-\!\theta_j)p_u\!-\!
   \sum\limits_{i=0}^{M}x_i\!\left(\gamma_{1,i}u_1(x_iN)\!+\!\gamma_{2,i}u_2(x_iN)\!-\!(1\!-\!\theta_i)p_u\!\right)\!\Big),
 \end{array}
\end{equation}
where we omit $t$ in $x_j(t)$ for simplicity. It is well-known by the folk theorem in the evolutionary game theory that with the replicator dynamics, state $\mathbf{x}^*$ is the ESS of the user sub-game $\mathcal{G}_f$ if $\mathbf{x}^*$ is a stable rest point of the ODEs~\cite{Sandholm2009}. Therefore, we are interested in identifying the stability property of the sub-game $\mathcal{G}_f$. By (\ref{eq_user}), when the leader strategy $\mathbf{a}$ is fixed, the payoff of MU state $j$, $\pi_j^u(\mathbf{x}, \mathbf{a})$ is only determined by the population state $x_j$. Then, we can verify that $\mathcal{G}_f$ is an evolutionary population potential game~\cite{SANDHOLM200181}:
\begin{Definition}[Potential Evolutionary Game]
\label{def_potential_game}
 Consider a generalized evolutionary game $\mathcal{G}\!=\!\langle \mathcal{N}, \mathcal{M}, \mathbf{x}, F(\mathbf{x})\!=\![F_j(\mathbf{x})]_{j\in\mathcal{M}}\rangle$ with the vector of payoff function $F(\mathbf{x}):\mathbb{R}^n_+\rightarrow\mathbb{R}^n$. $\mathcal{G}$ is a full potential game if there exists a continuously differentiable function $f(\mathbf{x}): \mathbb{R}^n_+\rightarrow\mathbb{R}$ satisfying $\pdv{f(\mathbf{x})}{x_j}=F_j(\mathbf{x})$, $\forall j\in\mathcal{M}$.
\end{Definition}

\begin{Lemma}
 \label{le_potential_game}
 Given any feasible joint leader action $\mathbf{a}$, $\mathcal{G}_f$ is a population potential game.
\end{Lemma}
\begin{proof}
 By \cite{SANDHOLM200181}, $\mathcal{G}_f$ is a potential game if and only if it satisfies the property of full externality symmetry:
 \begin{equation}
  \label{eq_proof_partial_order_two}
  \frac{\partial{\pi^u_j(\mathbf{x},\mathbf{a})}}{\partial{x_i}}=\frac{\partial{\pi^u_i(\mathbf{x},\mathbf{a})}}{\partial{x_j}}, \forall i,j\in\mathcal{M}.
 \end{equation}
 From (\ref{eq_user}), we note that $\pi_j^u(\mathbf{x},\mathbf{a})$ is only determined by the population state $x_j$. Then, we can easily derive
 \begin{equation}
  \label{eq_proof_partial_order_one}
  \frac{\partial{\pi^u_j(\mathbf{x},\mathbf{a})}}{\partial{x_i}}=\frac{\partial{\pi^u_i(\mathbf{x},\mathbf{a})}}{\partial{x_j}}=0, \forall j\ne i,
 \end{equation}
which satisfies (\ref{eq_proof_partial_order_two}). Therefore Lemma \ref{le_potential_game} holds.
\end{proof}

\begin{Lemma}
 \label{le_zero_sum_rate}
 Under the replicator dynamics given in (\ref{eq_replicator_dynamics}), the population state $\mathbf{x}(t)$ always falls into the $M$-simplex $\mathcal{X}$ from any $\mathbf{x}(0)\in\mathcal{X}$ with a sum of changing rate
 \begin{equation}
  \label{eq_zero_rate}
  \sum_{j=0}^{M}\dv{x_j(t)}{t}= \sum_{j=0}^{M}x_j(t)(\pi^u_j(\mathbf{x}(t),\mathbf{a})-\overline\pi^u(\mathbf{x}(t),\mathbf{a}))=0.
 \end{equation}
\end{Lemma}
\begin{proof}
 Let $v_j(\mathbf{x}(t))\!=\!x_j(t)\!\left(\pi_j^u(\mathbf{x}(t),\mathbf{a})\!-\!\overline{\pi}^u(\mathbf{x}(t),\mathbf{a})\right)$. From (\ref{eq_replicator_dynamics}), we can omit $\mathbf{a}$ in $\pi_j^u(\mathbf{x}(t),\mathbf{a})$ and obtain
 \begin{equation}
  \label{eq_proof_sum_rate}
  \begin{array}{ll}
  \displaystyle\sum_{j=0}^{M}\!\dv{x(t)}{t}\!=\!\sum_{j=0}^{M}\!v_j(\mathbf{x}(t))\!=\!\sum_{j=0}^{M}x_j(t)\!\left(\pi_j^u(\mathbf{x}(t))\!-\!\overline{\pi}^u
  (\mathbf{x}(t))\right)
  =\displaystyle\sum_{j=0}^{M}x_j(t)\pi_j^u(\mathbf{x}(t))-\overline{\pi}^u(\mathbf{x}(t))=0.
  \end{array}
 \end{equation}
From (\ref{eq_proof_sum_rate}), we have $\sum_{j=0}^Mx_j(t)=\sum_{j=0}^{M}x_j(0)$. Since for $x_j(t)\!=\!0$ and $x_j(t)\!=\!1$, $\dv{x_j(t)}{t}\!=\!0$, we have $0\!\le\!x_j(t)\!\le\!1$. Therefore, $\mathbf{x}(t)$ always falls into the $M$-simplex.
\end{proof}

For a general replicator dynamics-based system, a rest point of the dynamic function $V=[v_j(\mathbf{x}(t))]_{j=0}^M$ given by
(\ref{eq_replicator_dynamics}) may not necessarily be the ESS of the game~\cite{Sandholm2009}. Therefore, in order to obtain the ESS of game $\mathcal{G}_f$, we need to examine the stability of the rest points of the replicator dynamics. Based on Lemma~\ref{le_potential_game} and Lemma~\ref{le_zero_sum_rate}, we are able to obtain the following property regarding the stability of game $\mathcal{G}_f$'s equilibria:
\begin{Theorem}
 \label{thm_global_convergence}
 Every NE of $\mathcal{G}_f$ is evolutionarily stable in the interior of $\mathcal{X}'=\mathcal{X}\cap\{x_0=0\}$.
\end{Theorem}

\begin{proof}
By Lemma~\ref{le_potential_game}, $\mathcal{G}_f$ is a population potential game. Then, by Definition~\ref{def_potential_game}, there exists a potential function
$f(\mathbf{x})$ such that $\pdv{f(\mathbf{x})}{x_j}\!=\!\pi^u_j(\mathbf{x}), \forall j\!\in\!\mathcal{M}$. Note that $\pi^u_j$ only depends on the local state $x_j$. By Proposition 3.1 in~\cite{SANDHOLM200181}, for a potential population game, the set of NE states $\mathbf{x}^*$ coincides with the solution set of the following local maximization problem:
 \begin{equation}
  \label{eq_proof_global_maximization}
  \mathbf{x}^*=\arg\max_{\mathbf{x}\in\mathcal{X}}\left(f(\mathbf{x})=\sum_{j\in\mathcal{M}}\int_{0}^{x_j}\pi^u_j(z,\mathbf{a}){d}z\right).
 \end{equation}
Then, for each local maximum solution $\mathbf{x}^*$, we can find an $\epsilon>0$ such that $f(\mathbf{x}^*)\!\ge\!f(\mathbf{x})$ for any $\mathbf{x}\in\mathcal{C}=\mathcal{B}_{\epsilon}(\mathbf{x}^*)\cap\mathcal{X}'$, where $\mathcal{B}_{\epsilon}(\mathbf{x}^*)$ is the $\epsilon$-ball centered at $\mathbf{x}^*$. According to Definition 2.6 in~\cite{Weibull1997}, the asymptotically stable state of the ODEs given in (\ref{eq_replicator_dynamics}) is guaranteed to be an ESS. Thereby, we can design a Lyapunov function $L(\mathbf{x}(t))\!=\!f(\mathbf{x}^*)\!-\!f(\mathbf{x}(t))$ such that $L(\mathbf{x}(t))\ge0,\forall \mathbf{x}(t)\in\mathcal{C}\backslash\{\mathbf{x}^*\}$ and $L(\mathbf{x}(t))\!=\!0$ at $\mathbf{x}^*$. Following the Lyapunov Theorem~\cite{Sastry1999}, we have
\begin{equation}
 \label{eq_proof_lyapunov}
 \dv{L(\mathbf{x}(t))}{t}=-\left(\pdv{f(\mathbf{x}(t))}{\mathbf{x}(t)}\right)^{\top}\dv{\mathbf{x}(t)}{t},
\end{equation}
where $\pdv{f(\mathbf{x}(t))}{\mathbf{x}(t)}\!=\![\pi^u_0(\mathbf{x}(t),\mathbf{a}),\ldots, \pi^u_M(\mathbf{x}(t),\mathbf{a})]^{\top}$ and $\dv{\mathbf{x}(t)}{t}\!=\![v_0(\mathbf{x}(t),\mathbf{a}), \ldots, v_M(\mathbf{x}(t),\mathbf{a})]^{\top}$ (see also (\ref{eq_replicator_dynamics})). If we omit $\mathbf{x}(t)$ and $\mathbf{a}$ in $\pi_j(\mathbf{x}(t),\mathbf{a})$ for conciseness, then, expanding (\ref{eq_proof_lyapunov}) leads to:
\begin{eqnarray}
 \label{eq_proof_lyapunov_expand}
 \begin{array}{ll}
 \displaystyle\dv{L(\mathbf{x})}{t}=-\displaystyle\sum_{j\in\mathcal{M}}\pi^u_jv_j=-\displaystyle\sum_{j\in\mathcal{M}}x_j\pi^u_j\left(\pi^u_j-\overline{\pi}^u\right)\\
 \stackrel{(\ref{eq_zero_rate})}{=}-\displaystyle\sum_{j\in\mathcal{M}}x_j\pi^u_j\left(\pi^u_j-\overline{\pi}^u\right)-\overline{\pi}^u\sum_{j\in\mathcal{M}}x_j
 (\pi^u_j-\overline\pi^u)
 \!=\!-\!\displaystyle\sum_{j\in\mathcal{M}}x_j((\pi^u_j)^2\!-\!(\overline{\pi}^u)^2)\!=\!-\!E_{x_j}\left\{(\pi^u_j-\overline{\pi}^u)^2\right\}\le0.
 \end{array}
\end{eqnarray}

By the definition of NE, we know that all the rest points in (\ref{eq_replicator_dynamics}) violating the NE occur on the boundary of $\mathcal{X}'$. This is because the extinct states do not revive, and hence the rest points at the boundary are unstable~\cite{SANDHOLM200181}. By the Karush-Kuhn-Tucker (KKT) Theorem, for a stable point $\mathbf{x}^*$ in the interior of $\mathcal{X}'$, we have the following KKT system $\forall j\in\mathcal{M}\backslash\{0\}$ from the optimization problem in (\ref{eq_proof_global_maximization}):
\begin{equation}
\label{eq_proof_kkt}
 \left\{\begin{array}{ll}
 \pi^u_j(x^*_j,\mathbf{a})=\mu^u-\lambda^u_j, \\
 \lambda^u_jx^*_j=0, \quad \lambda^u_j\ge0,
 \end{array}\right.
\end{equation}
where $\mu^u$ is the Lagrange multiplier for the active constraint $\sum_{j\in\mathcal{M}\backslash\{0\}}x_j\!=\!1$ and $\lambda^u_j$ is the KKT multiplier for the constraint $x_j\!\ge\!0$. Since $\forall j\ne0$, $x_j\!\ne\!0$ in the interior of $\mathcal{X}'$, then we have $\lambda^u_j\!=\!0$ in (\ref{eq_proof_kkt}). Therefore, $\forall j\ne0$, $\pi^u_j(x^*_j,\mathbf{a})=\mu^u$ and the solution of (\ref{eq_proof_kkt}) is equivalent to the solution of $E_{x_j}\left\{(\pi^u_j-\overline{\pi}^u)^2\right\}\!=\!0$ (cf. (\ref{eq_proof_lyapunov_expand})). By applying Proposition 3.1 of \cite{SANDHOLM200181} again, we know that the equality in (\ref{eq_proof_lyapunov_expand}) holds only when $\mathbf{x}\!=\!\mathbf{x}^*$, so $\mathbf{x}^*$ is either locally uniformly asymptotically stable (if $\mathbf{x}^*$ is an isolated solution) or Lyapunov stable in a locally asymptotically attractor area (if $\mathbf{x}^*$ is in a connected set of solutions). Thereby, by Proposition 2.6 in~\cite{Weibull1997}, the proof of Theorem~\ref{thm_global_convergence} is completed.
\end{proof}

Further, since $\forall j\!\in\!\mathcal{M}$, $\pi^u_j(\mathbf{x}, \mathbf{a})$ is upper bounded, by the Weierstrass Theorem~\cite{sundaram1996first}, there exists a global maximum of $f(\mathbf{x})$ on $\mathcal{X}'$. Suppose that this global maximum is achieved at $\hat{\mathbf{x}}^*$. Following the same technique of proving Theorem~\ref{thm_global_convergence}, we can find another Lyapunov function $\hat{L}(\mathbf{x}(t))\!=\!f(\hat{\mathbf{x}}^*)\!-\!f(\mathbf{x}(t))$ such that $\hat{L}(\mathbf{x}(t))\!\ge\!0,\forall \mathbf{x}(t)\!\in\!\mathcal{X}'$. Moreover, $\hat{L}(\mathbf{x}(t))\!=\!0$ only holds at the NE. By employing (\ref{eq_proof_lyapunov_expand}) again, we know that  $\hat{L}(\mathbf{x}(t))$ is strictly decreasing at the non-equilibrium points (excluding the boundary). Then, there is no unstable equilibrium that attracts an evolutionary trajectory on the interior of $\mathcal{X}'$ (see Corollary~\ref{cor_ess}).
\begin{Corollary}
 \label{cor_ess}
 On the interior of $\mathcal{X}'$, the replicator dynamics in (\ref{eq_replicator_dynamics}) always converges to an ESS.
\end{Corollary}

From~\cite{Hofbauer20091665}, we know that an evolutionary game is a stable game when it admits a concave potential function. Namely, all the equilibrium states are globally neutrally stable and form a convex set. Therefore, if the potential function of the follower sub-game $f(\mathbf{x})$ is concave, the globally neutral stability of the equilibrium states is ensured on the entire simplex $\mathcal{X}$. Furthermore, the condition for game $\mathcal{G}_f$'s NE to be unique (i.e., globally evolutionarily stable) can be obtained through the analysis of the maximization problem in (\ref{eq_proof_global_maximization}) based on the strict concavity of the potential function $f(\mathbf{x})$:
\begin{Corollary}
 \label{cor_unique_ess}
 For the replicator dynamics in (\ref{eq_replicator_dynamics}), a unique interior ESS exists if:
 \begin{equation}
  \label{eq_strict_concavity_condition}
  \gamma_{j,1}\dv{u_1(x_jN)}{x_j}+\gamma_{j,2}\dv{u_2(x_jN)}{x_j}<0, \forall j\in\mathcal{M}\backslash\{0\}.
 \end{equation}
\end{Corollary}
\begin{proof}
 According to our discussion in the proof of Theorem~\ref{thm_global_convergence}, the condition of a unique NE in the sub-game $\mathcal{G}_f$ is equivalent to the condition of a unique solution to the problem give in (\ref{eq_proof_global_maximization}) (cf. Proposition 3.1 in~\cite{SANDHOLM200181}). Therefore, we only need to check the condition for $f(\mathbf{x})$ in (\ref{eq_proof_global_maximization}) to be strictly concave. In the interior of $\mathcal{X}'$, we only need to examine $f(\mathbf{x})$ with respect to the non-zero population states $\mathbf{x}'=[x_j]^M_{j=1}$. Since $\pi_j(\mathbf{x}, \mathbf{a})$ is determined only by the local state $x_j$, and $\pdv{f(\mathbf{x})}{x_j}\!=\!\pi^u_j(\mathbf{x}), \forall j\in\mathcal{M}\backslash\{0\}$, it suffices to show that the following diagonal Hessian matrix of $f(\mathbf{x}')$,
 \begin{equation}
  \label{eq_hessian}
  \pdv[2]{f(\mathbf{x}',\mathbf{a})}{{\mathbf{x}'}}=\mathop{\textrm{Diag}}\left(\dv{\pi_1(\mathbf{x},\mathbf{a})}{x_1},\ldots, \dv{\pi_M(\mathbf{x},\mathbf{a})}{x_M}\right),
 \end{equation}
 is negative definite. Then, it is sufficient to have $\forall j\in\mathcal{M}\backslash\{0\}$,
 \begin{equation}
  \label{eq_negative_definite}
  \begin{array}{ll}
   0>\gamma_{j,1}\displaystyle\dv{u_1(x_jN)}{x_j}+\gamma_{j,2}\displaystyle\dv{u_2(x_jN)}{x_j}.
  \end{array}
 \end{equation}
 Then, by Theorem~\ref{thm_global_convergence}, Corollary~\ref{cor_unique_ess} immediately holds.
\end{proof}
\begin{Remark}
  \rm
  With the realization of $u_1(x_jN)$ and $u_2(x_jN)$ in (\ref{eq_negative_ne}) and (\ref{eq_positive_ne}), respectively, from (\ref{eq_strict_concavity_condition}) we obtain
\begin{equation}
  \label{eq_practical_concave}
  0>\displaystyle\dv{\pi_j(\mathbf{x}',\mathbf{a})}{x_j}=\displaystyle\frac{\gamma_{j,2}N}{1+Nx_j}-\displaystyle\frac{\gamma_{j,1}N}{o_j+Nx_j}.
\end{equation}
Namely, when $\gamma_{j,2}(o_j+Nx_j)<\gamma_{j,1}(1+Nx_j)$, the follower sub-game with the utilities given by (\ref{eq_negative_ne}) and (\ref{eq_positive_ne}) for the MUs and the CSPs admits a unique ESS. Intuitively, a service needs to first function properly in order to survive in the intensively competing market. Then, under the assumption of utility realization in (\ref{eq_negative_ne}) and (\ref{eq_positive_ne}), we can assume that the QoE sensitivity coefficient $\gamma_{j,1}$ is sufficiently larger than the network effect sensitivity $\gamma_{j,2}$ due to the higher priority on the proper functioning of the services. In this case\footnote{Similar assumptions can be found in~\cite{REGGIANI201616}, where the congestion effect is modeled as the cost and rational MUs only select the services with positive net payoffs.}, the MUs' strategy evolution is able to converge to the unique ESS from any initial point.
\qed
\end{Remark}

\subsection{Non-hierarchical Equilibrium in the Game}
\label{sub_sec_gne}
By Theorem~\ref{thm_global_convergence}, the equilibrium-searching problem of the replicator dynamics in (\ref{eq_replicator_dynamics}) can be converted into the constrained convex optimization problem described in (\ref{eq_proof_global_maximization}). From Definition~\ref{def_param_follower_NE} and (\ref{eq_best_response}), the non-hierarchical equilibrium, i.e., the NE of the game is obtained when all the players take their actions simultaneously. If we treat the entire MU population as a super player with the joint state $\mathbf{x}$ being its continuous strategy, by Theorem \ref{thm_global_convergence}, the set of NE in the game will be equivalent to the set of NE in the following auxiliary constrained game defined by (\ref{eq_proof_global_maximization_added_constraint})-(\ref{eq_best_response_mno_no_constraint}):
\begin{itemize}
  \item MUs' subscription selection problem:
\begin{eqnarray}
  \label{eq_proof_global_maximization_added_constraint}
   \begin{array}{ll}
  \mathbf{x}^*=\arg\max\limits_{\mathbf{x}}&\left(f(\mathbf{x})=\sum\limits_{j\in\mathcal{M}}\displaystyle\int_{0}^{x_j}\pi^u_j(z,\mathbf{a}){d}z\right),\\
  \qquad\qquad \textrm{s.t.} & \displaystyle\sum_{j\in\mathcal{M}\backslash\{0\}}x_j=1, \forall\!j\!\in\!\mathcal{M}\backslash\{0\}:\; x_j\ge0.\\
   \end{array}
\end{eqnarray}
\item CSP $j$'s strategy selection problem ($0\!\le\!j\!\le\!M$):
\begin{eqnarray}
 \label{eq_best_response_cp_no_constraint}
 \begin{array}{ll}
  a^*_j\!=\!\arg\max\limits_{a_j=(\theta_j, b_j)}&\left(\pi_j^c(\theta_j, b_j)=u^{\textrm{CSP}}_j(x_jN)\!-\!p_u\theta_jx_jN\!-\!p_cb_j\right),\\
  \qquad\qquad\;\; \textrm{s.t.} & g_j(x_j,\theta_j,b_j, p_u, p_c)=p_u\theta_jx_jN+p_cb_j-\overline{c}_j\le0,\\
  & b_j\ge0, \quad0\le\theta_j\le1.
 \end{array}
\end{eqnarray}
\item MNO's price selection problem:
\begin{eqnarray}
 \label{eq_best_response_mno_no_constraint}
 \begin{array}{ll}
  a^*_0\!=\!\arg\max\limits_{a_0=(p_u, p_b)}&\left(\pi^o(p_u,p_b)=p_c\displaystyle\sum_{j=1}^{M}b_j + {p_u}N\displaystyle\sum_{j=1}^{M}x_j\right),\\
  \qquad\qquad\;\; \textrm{s.t.} & 0\le p_u\le\overline{p}_u,\quad 0\le p_c\le\overline{p}_c.
 \end{array}
\end{eqnarray}
\end{itemize}
From the constraints of the auxiliary game, we note that the strategy space of each player is jointly determined by its adversaries' strategies. Therefore, the equilibrium searching problem in the auxiliary game becomes a Generalized Nash Equilibrium (GNE) problem~\cite{facchinei201012}. This naturally leads to the idea of applying the mathematical tool of Generalized Quasi-Variational Inequality (GQVI)~\cite{facchinei201012} to examine the property of the GNE. Before proceeding, we first provide the definition of the GQVI problem as follows:
\begin{Definition}[GQVI~\cite{facchinei201012}]
  \label{def_VI}
 For a given set in a Euclidean space $\mathcal{S}\in\mathbb{R}^n$ and a mapping $F:\mathcal{S}\rightarrow\mathbb{R}^n$, the GQVI problem denoted as $\mathop{\textrm{VI}} (\mathcal{S},F)$, is to find a pair of vector $\mathbf{s}^*\!\in\!\mathcal{S}(\mathbf{s}^*)$ and $\mathbf{y}^*\!\in\!F(\mathbf{s}^*)$ such that:
 \begin{equation}
  \label{eq_vi_definition}
  (\mathbf{s}-\mathbf{s}^*)^{\top} \mathbf{y}^*\ge 0, \forall\mathbf{s}\in\mathcal{S}(\mathbf{s}^*).
 \end{equation}
\end{Definition}

Let $F_l(\mathbf{x},\mathbf{a})\!=\!\left[-\nabla_{a_0}\pi^o(\mathbf{x},\mathbf{a}), \left(-\nabla_{a_j}\pi^c(\mathbf{x}, \mathbf{a})\right)_{1\le j\le{M}}\right]^{\top}$ define the mapping based on the gradients of the objective functions in (\ref{eq_best_response_cp_no_constraint}) and (\ref{eq_best_response_mno_no_constraint}), and $F_f(\mathbf{x},\mathbf{a})\!=\!\left[-\pi^u_j(\mathbf{x},\mathbf{a})\right]_{j\in\mathcal{M}}^{\top}$ define the mapping based on the gradients of the potential function $f(\mathbf{x}, \mathbf{a})$ in (\ref{eq_proof_global_maximization_added_constraint}). Then, by concatenating the two mappings $F_l(\mathbf{x},\mathbf{a})$ and $F_f(\mathbf{x}, \mathbf{a})$, we introduce the following GQVI problem, which aims to find a pair of $(\mathbf{x}^*,\mathbf{a}^*)$ to satisfy the GQVI condition given in Definition~\ref{def_VI}:
\begin{equation}
 \label{eq_auxiliary_gqvi}
 \left[\begin{array}{ll}
       \mathbf{x}-\mathbf{x}^*\\
       \mathbf{a}-\mathbf{a}^*
      \end{array}
\right]^{\top}
 \left[\begin{array}{ll}
      F_f(\mathbf{x}^*,\mathbf{a}^*)\\
      F_l(\mathbf{x}^*,\mathbf{a}^*)
      \end{array}
\right]\ge0, \forall (\mathbf{x},\mathbf{a})\in\mathcal{C}({\mathbf{x}^*,\mathbf{a}^*}),
\end{equation}
where $\mathcal{C}({\mathbf{x},\mathbf{a}})$ is defined by the following system of inequalities extracted from (\ref{eq_proof_global_maximization_added_constraint})-(\ref{eq_best_response_mno_no_constraint}):
\begin{equation}
 \label{eq_feasible_space}
 \left\{\begin{array}{ll}
 \forall\!j\!\in\!\mathcal{M}\backslash\{0\}:\quad g_j(x_j,\theta_j,b_j, p_u, p_c)=p_u\theta_jx_jN+p_cb_j-\overline{c}_j\le0,\\
 \forall\!j\!\in\!\mathcal{M}\backslash\{0\}:\quad x_j\ge0,\quad b_j\ge0, \quad0\le\theta_j\le1,\\
 0\le p_u\le\overline{p}_u,\quad 0\le p_c\le\overline{p}_c, \\
 \sum_{j\in\mathcal{M}}x_j=1.
 \end{array}\right.
\end{equation}

By the well-known solution-existence theorem of the GQVI problem (cf. Proposition 12.3 in~\cite{facchinei201012}), the set of GNE points in the auxiliary game, $({\mathbf{x}^*, \mathbf{a}}^*)$, is equivalent to the set of solutions to the GQVI problem given by (\ref{eq_auxiliary_gqvi}), if the local strategy space is compact and convex for any adversaries' strategy, and the objective function of each player is concave in the local strategy for any adversaries' strategy. Then, to ensure the existence of the GNE in the auxiliary game, it suffices to provide the conditions for (a) the GNE problem defined by (\ref{eq_proof_global_maximization_added_constraint})-(\ref{eq_best_response_mno_no_constraint}) to be equivalent to the GQVI problem given by (\ref{eq_auxiliary_gqvi}), and (b) the solution set of the GQVI problem to be non-empty. With this in mind, we can derive the following condition that guarantees the existence of the NE in the game:
\begin{Theorem}
 \label{thm_existence_NE}
 Suppose that the potential function $f(\mathbf{x},\mathbf{a})$ defined in (\ref{eq_proof_global_maximization}) is concave with respect to $\mathbf{x}$, namely, $\gamma_{j,1}\dv{u_1(x_jN)}{x_j}+\gamma_{j,2}\dv{u_2(x_jN)}{x_j}\le0$, $\forall j\!\in\!\mathcal{M}\backslash\{0\}$. Then, at least one NE exists in the hierarchical game.
\end{Theorem}
\begin{proof}
It is straightforward that in the inequality constraint $g_j(x_j,\theta_j,b_j, p_u, p_c)\le0$ for CSP $j$ in (\ref{eq_feasible_space}), the function $g_j(x_j,\theta_j,b_j, p_u, p_c)$ is linear with respect to $(\theta_j, b_j)$ for any $(x_j, p_u, p_c)$. Then, (\ref{eq_feasible_space}) defines for each type of the players a compact, convex sub-space as $\mathcal{C}_{\mathbf{x}}(\mathbf{a})\!=\!\left\{\mathbf{x}:\sum_{j\!\in\!\mathcal{M}}x_j\!=\!1, x_j\!\ge\!0,\forall j\in\mathcal{M}\right\}$, $\mathcal{C}_{j\ne0}(\mathbf{x},a_0)\!=\!\left\{(\theta_j,b_j):b_j\!\ge\!0,0\!\le\!\theta_j\le1, g_j(\theta_j,b_j; x_j, p_u, p_c)\!\le\!0\right\}$ and $\mathcal{C}_0\!=\!\left\{(p_u,p_c):0\!\le\!p_u\!\le\!\overline{p}_u, 0\le p_c\le\overline{p}_c\right\}$, respectively. Observing the objective functions in (\ref{eq_best_response_cp_no_constraint}) and (\ref{eq_best_response_mno_no_constraint}), we note that $\pi_j^c(\theta_j, b_j)$ and $\pi^o(p_u,p_b)$ are both linear functions of their local strategies. Therefore, by Proposition 12.3 in~\cite{facchinei201012}, the set of the GNE in the auxiliary game defined by (\ref{eq_proof_global_maximization_added_constraint})-(\ref{eq_best_response_mno_no_constraint}) will be equivalent to the set of solutions to the GQVI problem defined by (\ref{eq_auxiliary_gqvi}), as long as the objective function in (\ref{eq_proof_global_maximization_added_constraint}) is also a concave function of the MUs' states $\mathbf{x}$. By applying the same method in the proof of Corollary~\ref{cor_unique_ess}, we can show that the property of concavity for (\ref{eq_proof_global_maximization_added_constraint}) holds with the inequality condition given by Theorem~\ref{thm_existence_NE}. Then, the equivalence between the solution sets of the GNE and the GQVI problems is established.

Since the GNE in the auxiliary game is the same as the NE in the hierarchical game, it now suffices to prove that the solution set of the GQVI problems is non-empty to ensure the existence of the NE in the hierarchical game. By Proposition 12.7 in~\cite{facchinei201012}, the set of solution to the GQVI problem is non-empty as long as the gradient-based mapping $H(\mathbf{x},\mathbf{a})=\left[F^{\top}_f(\mathbf{x}^*,\mathbf{a}^*), F^{\top}_l(\mathbf{x}^*,\mathbf{a}^*)\right]^{\top}$ is continuous with respect to the local strategies of each player. This condition immediately holds since $\pi_j^c(\mathbf{x},\mathbf{a})$ and $\pi^o(\mathbf{x},\mathbf{a})$ are both first-order continuous and $\pi^u_j(\mathbf{x},\mathbf{a})$ is piecewise continuous. Then, the proof of Theorem~\ref{thm_existence_NE} is completed.
\end{proof}

\subsection{Hierarchical Equilibrium in the Game}
\label{sec_existence_SE}
In contrast to the NE based on simultaneous play, the derivation of the SE in (\ref{eq_optimal_best_response}) requires that the ESS of the follower sub-game is drawn from a best reply-based point-to-set mapping from the leaders' strategies. Namely, the providers make their optimal decisions while knowing how the MUs will react to their market strategies. Therefore, SE searching can be expressed in the form of a sequence of bi-level programming problems of each CSP and the MNO, with the optimization problem given in (\ref{eq_proof_global_maximization}) being the lower-level equilibrium constraint for each of them~\cite{dempe2002foundations}. Compared with (\ref{eq_best_response_cp_no_constraint}), the solution to the best-response problem of CSP $j$ ($\forall j\in{\mathcal{K}\backslash\{0\}}$) with the follower equilibrium constraint can be expressed as follows:
\begin{eqnarray}
 \label{eq_best_response_cp}
 \begin{array}{ll}
  a^*_j\!=\!\arg\max\limits_{a_j=(\theta_j, b_j)}&\left(u^{\textrm{CSP}}_j(x_jN)\!-\!p_u\theta_jx_jN\!-\!p_cb_j\right),\\
  \qquad\qquad\;\; \textrm{s.t.} & p_u\theta_jx_jN+p_cb_j-\overline{c}_j\le0,\\
  &
  \begin{array}{ll}
  \mathbf{x}\in\arg\max\limits_{\mathbf{x}}\left(\displaystyle\sum_{j\in\mathcal{M}}\int_{0}^{x_j}\pi^u_j(z,\mathbf{a}){d}z\right),
  \quad\textrm{s.t.}\displaystyle\sum_{j\in\mathcal{M}\backslash\{0\}}x_j=1.
  \end{array}
 \end{array}
\end{eqnarray}
For conciseness, in (\ref{eq_best_response_cp}) the feasibility conditions of the parameters, i.e., $x_j\ge0$, $b_j\ge0$, $0\le\theta_j\le1$, $0\!\le\!p_u\le\overline{p}_u$ and $0\!\le\!p_c\le\overline{p}_c$, are not included in the constraints. Similarly, for the MNO, the best-response problem with the same equilibrium constraint can be expressed by
\begin{eqnarray}
 \label{eq_best_response_mno}
 \begin{array}{ll}
  a^*_0\!=\!\arg\max\limits_{a_0=(p_u, p_b)}&\left(p_c\displaystyle\sum_{j=1}^{M}b_j + {p_u}N\displaystyle\sum_{j=1}^{M}x_j\right),\\
  \qquad\qquad\;\; \textrm{s.t.} &\forall\!j\!\in\!\mathcal{M}\backslash\{0\}:\; p_u\theta_jx_jN+p_cb_j-\overline{c}_j\le0,\\
  & \begin{array}{ll}
  \mathbf{x}\in\arg\max\limits_{\mathbf{x}}\left(\displaystyle\sum_{j\in\mathcal{M}}\int_{0}^{x_j}\pi^u_j(z,\mathbf{a}){d}z\right),
  \quad\textrm{s.t.}\displaystyle\sum_{j\in\mathcal{M}\backslash\{0\}}x_j=1.
  \end{array}
 \end{array}
\end{eqnarray}

When the optimization problems given by (\ref{eq_best_response_cp}) and (\ref{eq_best_response_mno}) are solved jointly, the solution vector $(\mathbf{x}^*,\mathbf{a}^*)$ provides an optimistic, locally optimal solution in the sense of bi-level programming~\cite{dempe2002foundations}. Again, by considering the MU population as a super player with strategy $\mathbf{x}$, SE searching in (\ref{eq_best_response_cp}) and (\ref{eq_best_response_mno}) can be interpreted as a multi-leader-common-follower game. By the proof of Theorem~\ref{thm_global_convergence}, the set of follower sub-game NE, namely, the ESS in the interior of $\mathcal{X}$ is equivalent to the set of solutions to the KKT systems given in (\ref{eq_proof_kkt}). Then, we can replace the lower-level optimization problem in the constraints of (\ref{eq_best_response_cp}) and (\ref{eq_best_response_mno}) by the KKT conditions in (\ref{eq_proof_kkt}) and obtain the equivalent Equilibrium Programming with Equilibrium Constraints (EPEC) problem described jointly by (\ref{eq_best_response_cp_KKT}) and (\ref{eq_best_response_mno_KKT}) as follows:
\begin{eqnarray}
 \label{eq_best_response_cp_KKT}
 \begin{array}{ll}
  (a^*_j, \tilde{\mathbf{x}}^*)\!=\!\arg\max\limits_{(a_j,\tilde{\mathbf{x}})}\left(\pi^c_j(x_j,a_j,a_0)=u^{\textrm{CSP}}_j(x_jN)\!-\!p_u\theta_jx_jN\!-\!p_cb_j\right),\\
  \qquad\qquad\qquad \textrm{s.t.} \quad g_j(x_j,a_j, a_0)=p_u\theta_jx_jN+p_cb_j-\overline{c}_j\le0,\\
  \qquad\qquad\qquad\qquad \forall  i\!\in\!\mathcal{M}:\; h_i(x_i,a_i, a_0) = \pi^u_i(x_i,a_i, a_0)-\mu^u+\lambda^u_i=0, \\
  \qquad\qquad\qquad\qquad \forall i\!\in\!\mathcal{M}:\; \lambda^u_ix_i=0, x_i\ge0, \lambda^u_i\ge0,\\
  \qquad\qquad\qquad\qquad \sum_{i\in\mathcal{M}}x_i=1,
 \end{array}
\end{eqnarray}
$\forall j\!\in\!\mathcal{K}\backslash\{0\}$ and
\begin{eqnarray}
 \label{eq_best_response_mno_KKT}
 \begin{array}{ll}
  (a^*_0, \tilde{\mathbf{x}}^*)\!=\!\arg\max\limits_{(a_0, \tilde{\mathbf{x}})}\left(\pi^o(\mathbf{x},\mathbf{a})=p_c\sum\limits_{j=1}^{M}b_j + {p_u}N \sum\limits_{j=1}^{M}x_j\right),\\
  \qquad\qquad\qquad \textrm{s.t.} \quad \forall j\!\in\!\mathcal{M}\backslash\{0\}:\; , g_j(x_j,a_j, a_0)=p_u\theta_jx_jN+p_cb_j-\overline{c}_j\le0,\\
  \qquad\qquad\qquad\qquad \forall j\!\in\!\mathcal{M}:\; h_j(x_j,a_j, a_0) = \pi^u_j(x_j,a_j,a_0)-\mu^u+\lambda^u_j=0, \\
  \qquad\qquad\qquad\qquad \forall j\!\in\!\mathcal{M}:\; \lambda^u_jx_j=0, x_j\ge0, \lambda^u_j\ge0,\\
  \qquad\qquad\qquad\qquad \sum_{j\in\mathcal{M}}x_j=1,
 \end{array}
\end{eqnarray}
where $\tilde{\mathbf{x}^*}\!=\!(\mathbf{x}^*, \lambda^u_1,\ldots, \lambda^u_M, \mu^u)$ is the combination of the MU population state and the multipliers that constructs the Lagrangian function corresponding to the KKT conditions in (\ref{eq_proof_kkt}) for the follower sub-game.
The EPEC problem defined by (\ref{eq_best_response_cp_KKT}) and (\ref{eq_best_response_mno_KKT}) is featured by the complementary slackness condition in the constraints, which is introduced by the KKT condition of the follower sub-game.
Alongside the non-convexity introduced by the follower-level KKT conditions, it is also well-known that the classical KKT theorem on the necessary optimality conditions is in jeopardy in the context of the above EPEC problem. The reason lies in the presence of the complementary slackness conditions in the constraints, since with them no standard constraint qualification is satisfied at any feasible $(a_j, \tilde{\mathbf{x}})$ in either (\ref{eq_best_response_cp_KKT}) or (\ref{eq_best_response_mno_KKT}) (cf., Theorem 5.11 in \cite{dempe2002foundations}). Thus, an alternative approach is needed for analyzing the SE in the considered game.

Following Scholtes' regularization scheme~\cite{doi:10.1137/S1052623499361233} (cf. its generalization in~\cite{su2004sequential}) for the Mathematical Programs with Equilibrium Constraints (MPEC), the index sets of the active constraints for a feasible strategy $(\mathbf{a},\tilde{\mathbf{x}})$ in the joint problem given by (\ref{eq_best_response_cp_KKT}) and (\ref{eq_best_response_mno_KKT}) can be defined as:
\begin{equation}
  \label{active_set}
  \left\{
  \begin{array}{ll}
    \mathcal{I}_g(\mathbf{a},\tilde{\mathbf{x}})=\{j |j\in\mathcal{M}\backslash\{0\} , g_j(\mathbf{a},\tilde{\mathbf{x}})=0\},\\
    \mathcal{I}_h(\mathbf{a},\tilde{\mathbf{x}})=\{j |j\in\mathcal{M}, h_j(\mathbf{a},\tilde{\mathbf{x}})=0\},\\
    \mathcal{I}_G(\mathbf{a},\tilde{\mathbf{x}})=\{j |j\in\mathcal{M}, x_j=0 \},\\
    \mathcal{I}_H(\mathbf{a},\tilde{\mathbf{x}})=\{j |j\in\mathcal{M}, \lambda^u_j=0 \}.\\
  \end{array}\right.
\end{equation}
Let $G_j(\mathbf{a},\tilde{\mathbf{x}})=x_j$, $H_j(\mathbf{a},\tilde{\mathbf{x}})=\lambda_j$ and $\hat{h}(\mathbf{a},\tilde{\mathbf{x}})=\sum_{j\in\mathcal{M}}x_j-1$. Then, the Jacobian matrix of the active constraints can be derived based on (\ref{active_set}) with the following gradients: $\nabla_{(\mathbf{a},\tilde{\mathbf{x}})}g_i(\mathbf{a},\tilde{\mathbf{x}}): i\in\mathcal{I}_g(\mathbf{a},\tilde{\mathbf{x}})$,
$\nabla_{(\mathbf{a},\tilde{\mathbf{x}})}h_j(\mathbf{a},\tilde{\mathbf{x}}): j\in\mathcal{I}_h(\mathbf{a},\tilde{\mathbf{x}})$, $\nabla_{(\mathbf{a},\tilde{\mathbf{x}})}G_k(\mathbf{a},\tilde{\mathbf{x}}): k\in\mathcal{I}_G(\mathbf{a},\tilde{\mathbf{x}})$, $\nabla_{(\mathbf{a},\tilde{\mathbf{x}})}H_l(\mathbf{a},\tilde{\mathbf{x}}): l\in\mathcal{I}_g(\mathbf{a},\tilde{\mathbf{x}})$ and $\nabla_{(\mathbf{a},\tilde{\mathbf{x}})}\hat{h}(\mathbf{a},\tilde{\mathbf{x}})$. From (\ref{eq_proof_kkt}) in the proof of Theorem~\ref{thm_global_convergence},
we note that $\lambda^u_j\!=\!0$ only if $\pi^u_j(x_j,\mathbf{a})\!\ne\!0$, namely, $x_j\ne0$ at the parametric follower sub-game NE in the interior of $\mathcal{X}'$. Thus, we have $\{j:\lambda^u_j=0,x_j=0\}\!=\!\emptyset$, which implies that the NE strategies of the follower game satisfy the strict complementary conditions. Therefore, it is trivial to check that the aforementioned gradients are linearly independent when $\tilde{\mathbf{x}}$ is the follower sub-game NE. In other words, the Jacobian matrix of the active constraints has full row rank. Then, we know that the MPEC Linear Independence Constraint Qualification (MPEC-LICQ, see~\cite{doi:10.1137/S1052623499361233} for the definition) is satisfied by the optimization problem of each leader at any feasible point $(\mathbf{a},\tilde{\mathbf{x}})$, given that $\tilde{\mathbf{x}}$ corresponds to the follower sub-game NE in the interior of $\mathcal{X}'$. Then, we have the following threorem.
\begin{Theorem}
  \label{thm_existence_auxilary_NE}
  If $(\mathbf{a}^*,{\mathbf{x}}^*)$ is a (possibly local) SE of the hierarchical game described jointly by (\ref{eq_best_response_cp_KKT}) and (\ref{eq_best_response_mno_KKT}), then $(\mathbf{a}^*,{\mathbf{x}}^*)$ is an EPEC (Nash) strongly stationary point (see also~\cite{doi:10.1137/S1052623499361233}).
\end{Theorem}
\begin{proof}
  Since we have shown that MPEC-LICQ is satisfied at each leader $j$'s local optimization problem with equilibrium constraints ($j=0,1,\ldots,M$), Theorem~\ref{thm_existence_auxilary_NE} immediately follows theorem 4.2 in~\cite{su2004sequential}.
\end{proof}

By Theorem~\ref{thm_existence_auxilary_NE}, we are able to apply the classical KKT-based analysis to the hierarchical game again. From (\ref{eq_best_response_cp_KKT}) and (\ref{eq_best_response_mno_KKT}), we introduce the Lagrangian function for each CSP $j\in\mathcal{K}$ as follows:
\begin{equation}
   \label{eq_lagrange_dual_cp_full}
   L_j(\mathbf{a}, {\mathbf{x}}, \pmb\lambda,\pmb\mu)\!=\!\left\{\!\!
   \begin{array}{ll}
   \pi^c_j(x_j,a_j,a_0)\!-\!\lambda^{c,1}_jg_j(a_j, a_0, x_j)\!+\!\lambda^{c,2}_jx_j\!+\!\mu^{c,1}_jh_j(x_j,a_j, a_0)\!+\!\mu^{c,2}\displaystyle\sum_{j\in\mathcal{M}}x_j,
   &\textrm{if }j\!\ne\!0,\\
   \pi^o(\mathbf{x},\mathbf{a})\!-\!\displaystyle\sum_{j=1}^{M}\lambda^{o,1}_jg_j(a_j, a_0, x_j)\!+\!\sum_{j\in\mathcal{M}}\lambda_j^{o,2}x_j\!+\!
   \sum_{j=1}^{M}\mu^{o,1}_jh_j(x_j,a_j, a_0)\!+\!\mu^{o,2}\displaystyle\sum_{j\in\mathcal{M}}x_j,
   &\textrm{if }j\!=\!0,\\
   \end{array}\right.
 \end{equation}
where $\pmb\lambda=(\pmb\lambda^{c}, \pmb\lambda^{o})$ and $\pmb\mu=(\pmb\mu^{c}, \pmb\mu^{o})$ are the Lagrange multipliers associated with the inequality and equality constraints of the leaders, respectively. Note that in (\ref{eq_lagrange_dual_cp_full}), the complimentary condition-related constraints on $\lambda^u_j$ can be ignored, since forcing any CSP $j$ out of play (i.e., $x_j=0$ or $\pi^u_j=0$) is obviously a dominated strategy of the MNO. Then, based on (\ref{eq_lagrange_dual_cp_full}) we obtain the concatenated KKT system from all the CSPs and the MNO as follows:
\begin{equation}
\label{eq_concatenated_KKT_augmented}
\left\{\begin{array}{ll}
 \forall j\!\in\!\mathcal{K}:\; \nabla_{\mathbf{a}}L_j(\mathbf{a}, {\mathbf{x}}, \pmb\lambda,\pmb\mu)=0,\\
 \forall j\!\in\!\mathcal{K}:\; \nabla_{\mathbf{x}}L_j(\mathbf{a}, {\mathbf{x}}, \pmb\lambda,\pmb\mu)=0,\\
 \sum_{j\in\mathcal{M}}x_j=1,\\
 \forall j\!\in\!\mathcal{M}:\; h_j(x_j,a_j, a_0)=0, \\
 \forall j\!\in\!\mathcal{M}\backslash\{0\}:\; 0\ge g_j(x_j,a_j, a_0)\perp \lambda^{c,1}_j\ge0,\\
 \forall j\!\in\!\mathcal{M}\backslash\{0\}:\; 0\ge g_j(x_j,a_j, a_0)\perp \lambda^{o,1}_j\ge0,\\
 \forall j\!\in\!\mathcal{M}\backslash\{0\}:\; 0\le x_j\perp \lambda^{c,2}_j\ge0, \\
 \forall j\!\in\!\mathcal{M}\backslash\{0\}:\; 0\le x_j\perp \lambda^{o,2}_j\ge0,
\end{array}
\right.
\end{equation}
where the operator $\perp$ means component-wise orthogonality. Theorem~\ref{thm_existence_auxilary_NE}
ensures that the solution to the concatenated KKT system in (\ref{eq_concatenated_KKT_augmented}), $(\mathbf{a}^*, {\mathbf{x}}^*, \pmb\lambda^{*}, \pmb\mu^{*})$ provides a necessary optimal condition for the solution, i.e., the SE, to the EPEC problem defined by (\ref{eq_best_response_cp_KKT}) and (\ref{eq_best_response_mno_KKT}).

\begin{algorithm}[t]
  \caption{Sequential strategy updating}
 \begin{algorithmic}[1]
  \REQUIRE
  Randomly initialize the user population states as $\mathbf{x}(0)\!\in\!\mathcal{X}$ and the leader strategies as $\mathbf{a}(0)$.
  \ENSURE $\mathbf{x}(t)$, $\mathbf{a}(t)$
  \WHILE {the leader strategies $\mathbf{a}(t)$ do not converge, namely $\Vert \mathbf{a}(t+1)-\mathbf{a}(t) \Vert\ge\epsilon$}
  \FOR {$j=0,\ldots,M$}
  \STATE Given the fixed adversary leader strategy $a_{-j}(t)$, solve the local MPEC problem defined by (\ref{eq_best_response_cp_KKT}) or
  (\ref{eq_best_response_mno_KKT}) to obtain the solution $({a}_j(t+1),{a}_{-j}(t),\mathbf{x}_j(t))$, where $\mathbf{x}_j(t)$ is the intermediate solution for $\mathbf{x}$ given by provider $j$.
  \ENDFOR
  \STATE $\mathbf{x}(t+1)\leftarrow\mathbf{x}^M(t)$.
  \ENDWHILE
 \end{algorithmic}
  \label{alg0}
\end{algorithm}
The formulation of (\ref{eq_concatenated_KKT_augmented}) inspires the diagonalization methods~\cite{su2004sequential} for numerically approaching the SE of the multi-leader-single-follower game. More specifically, we can separate the local nonlinear complimentary problems for each leader in the form of KKT systems from (\ref{eq_concatenated_KKT_augmented}). A sequential (i.e., cyclicly repeated) procedure is performed in the manner of nonlinear Jacobi updating as in Algorithm~\ref{alg0}. Especially, the solution to the local MPEC problem in the inner loop of Algorithm~\ref{alg0} can be effectively obtained by relaxing the complementary constraints in each MPEC, e.g., through the standard augmented Lagrangian method~\cite{ruszczynski2006nonlinear}. However, by Theorem 4.3 of~\cite{su2004sequential}, if the sequence of solutions $\{(\mathbf{a}(t),\mathbf{x}(t))\}_t$ converges to a point $(\mathbf{a}^*,\mathbf{x}^*)$ for Algorithm~\ref{alg0}, with the MPEC-LICQ condition, $(\mathbf{a}^*,\mathbf{x}^*)$ is only guaranteed to be weakly stationary but may not be a local SE\footnote{More precisely, $(\mathbf{a}^*,\mathbf{x}^*)$ is a Bouligand stationary (B-stationary) point. See Definition 5.4 in~\cite{dempe2002foundations} or Section 2.1 in~\cite{doi:10.1287/moor.25.1.1.15213} for more details.}.
To further inspect whether $(\mathbf{a}^*,\mathbf{x}^*)$ is a (local) SE, we need to check if $(\mathbf{a}^*,\mathbf{x}^*)$ satisfies the Strong Sufficient Optimality Condition of second order (SSOC) based on (\ref{eq_lagrange_dual_cp_full}):

\begin{Definition}[SSOC~\cite{dempe2002foundations,doi:10.1137/S1052623499361233}]
 \label{def_SOSC}
 For each local problem given by (\ref{eq_best_response_cp_KKT}) or (\ref{eq_best_response_mno_KKT}), the SSOC is satisfied at a point $(\mathbf{a},\mathbf{x})$, if there exist multipliers $(\pmb\lambda, \pmb\mu)$ for (\ref{eq_concatenated_KKT_augmented}), such that for $j\in\mathcal{K}$ and every vector $d\ne0$ with
 \begin{equation}
   \left.
   \begin{array}{ll}
   \nabla_{(\mathbf{a},{\mathbf{x}})}g_j(\mathbf{a},{\mathbf{x}})d=0, & \lambda^{c,1}_j\ne0,\\
   \nabla_{(\mathbf{a},{\mathbf{x}})}x_jd=0, & \lambda^{c,2}_j\ne0,\\
   \nabla_{(\mathbf{a},{\mathbf{x}})}h_j(\mathbf{a}, {\mathbf{x}})d=0, &\\
   \nabla_{(\mathbf{a},{\mathbf{x}})}\sum_{i\in\mathcal{M}}x_id=0,&
 \end{array}\right\} j\ne0,\;
  \left.
  \begin{array}{ll}
  \nabla_{(\mathbf{a},{\mathbf{x}})}g_i(\mathbf{a},{\mathbf{x}})d=0, & \forall i\in\{\lambda^{o,1}_i\ne0\},\\
  \nabla_{(\mathbf{a},{\mathbf{x}})}x_id=0, & \forall i\in\{\lambda^{o,2}_i\ne0\},\\
  \nabla_{(\mathbf{a},{\mathbf{x}})}h_i(\mathbf{a}, {\mathbf{x}})d=0, & \forall i\in\mathcal{K},\\
  \nabla_{(\mathbf{a},{\mathbf{x}})}\sum_{i\in\mathcal{M}}x_id=0,&
\end{array}\right\} j=0\nonumber
\end{equation}
 the condition $d^{\top}\nabla^2_{(\mathbf{a},{\mathbf{x}})}L_j(\mathbf{a},{\mathbf{x}})d>0$ holds.
\end{Definition}

By Theorem 4.1 in~\cite{doi:10.1137/S1052623499361233}, if the SSOC is satisfied by $(\mathbf{a}^*,\mathbf{x}^*)$ for each local MPEC problem defined by (\ref{eq_best_response_cp_KKT}) or (\ref{eq_best_response_mno_KKT}), then Algorithm~\ref{alg0} converges to a local SE. With the logarithmic utility function for the MUs as given in (\ref{eq_negative_ne}) and (\ref{eq_positive_ne}), we note that the value of $\nabla^2_{(\mathbf{a},{\mathbf{x}})}L_j(\mathbf{a},{\mathbf{x}})$ are only dependent of the multipliers $\mu^{c,1}_j$ or $\mu^{o,1}_j$. Therefore, it is possible for us to check the SSOC in a simple manner by exhaustively enumerating the conditions of nonzero vectors $d$ for a given $(\mathbf{a}^*,\mathbf{x}^*)$ in each local MPEC.

\section{Distributed Equilibrium Searching in the Hierarchical Game}
\label{sec_algorithm}
\subsection{Protocol Design for Distributed Equilibrium Searching}
\label{sub_sec_algorithm}
Although Algorithm~\ref{alg0} provides a sequential approach at the leader level for numerically calculating the local SE, it is required that the ESS of the follower sub-game to be numerically solved in a centralized manner at each inner iteration in the algorithm. As a result, it is still impractical to deploy such a protocol in the real world. For this reason, in this subsection we aim to design a purely distributed equilibrium/stable point searching protocol that is able to emulate the interaction among the MU with bounded rationality in the real-world scenarios.

According to our analysis of the follower sub-game, the population evolution starting from the interior of $\mathcal{X}$ will always reach an ESS. Without assuming any centralized coordinator for providing the information about the payoff of subscribing to the other CSPs, we can describe in Algorithm~\ref{alg1} the strategy evolution of the MUs following the pairwise proportional imitation protocol. It is worth noting that the protocol will asymptotically lead to the replicator dynamics described by (\ref{eq_replicator_dynamics}) as the population increases. Furthermore, when the prices asked by the MNO and the states of the MU population are fixed, we note from the constraints in (\ref{eq_best_response_cp}) that the strategy space of the CSPs has a Cartesian product structure. Then, by Theorem~\ref{thm_existence_NE}, we introduce the distributed projection-based method (cf.,~\cite{Facchinei2003}) for the MNO's and CSPs' strategy searching in Algorithm~\ref{alg2}. The proposed Algorithms~\ref{alg1} and~\ref{alg2} jointly form an incentive-compatible protocol for decentralized strategy updating in the considered content/service market. The convergence condition for the strategy updating process in Algorithms~\ref{alg2} is provided in Theorem~\ref{thm_convergence}. Therefore, when imposing the protocol onto the considered market, no party has incentive to deviate from following the proposed strategy updating schemes without suffering from a revenue loss.

\begin{algorithm}[t]
  \caption{Pairwise proportional imitation protocol for CSP selection}
 \begin{algorithmic}[1]
  \REQUIRE
  Given the providers' strategies $\mathbf{a}$, each MU $i$ ($i\!\in\!\mathcal{N}$) randomly subscribes to a CSP $j_i(0)\!\in\!\mathcal{M}$
  \ENSURE $\mathbf{x}(t)$
  \WHILE {the evolutionary stable strategies are not reached}
  \FOR{each MU $n\in\mathcal{N}$}
    \STATE Randomly select a new CSP $j_n'\in\mathcal{M}$ and determine the switching probability $\rho({j_n(t),j'_n})$:
    \begin{equation}
      \label{eq_protocol_prob}
      \rho({j_n(t),j'_n})=x_j\max\left(\pi^u_{j_n'}(x_{j_n'}, \mathbf{a})-\pi^u_{j_n(t)}(x_{j_n(t)}, \mathbf{a}), 0\right)
    \end{equation}
    \IF {$\rho({j_n(t),j'_n})>0$}
      \STATE $j_n(t+1)\leftarrow j_n'$
    \ENDIF
  \ENDFOR
  \STATE $t\leftarrow t+1$ and update $\mathbf{x}(t+1)$
  \ENDWHILE
 \end{algorithmic}
  \label{alg1}
\end{algorithm}

\begin{algorithm}[t]
  \caption{Distributed provider strategy updating}
 \begin{algorithmic}[1]
  \REQUIRE
  Randomly initialize the user population states as $\mathbf{x}(0)\!\in\!\mathcal{X}$ and the leader strategies as $\mathbf{a}(0)$.
  \ENSURE $\mathbf{x}(t)$, $\mathbf{a}(t)$
  \WHILE {the provider strategies $\mathbf{a}(t)$ do not converge}
  \STATE Update the population state $\mathbf{x}(t\!+\!1)$ by Algorithm \ref{alg1}.
  \STATE Update the CSPs and the MNO's strategies as follows:
  \begin{eqnarray}
     \label{eq_extragradient_1}
     \left\{\!\begin{array}{ll}
     a_j(t\!+\!\frac{1}{2})\! =\! a_j(t)\!+\!\alpha \nabla_{a_j}\pi^o(\mathbf{x}(t\!+\!1),\mathbf{a}(t)), j\!=\!0,\\
     a_j(t\!+\!\frac{1}{2})\! =\! a_j(t)\!+\!\alpha \nabla_{a_j}\pi^c_j(\mathbf{x}(t\!+\!1),\mathbf{a}(t)), j\!\ne\!0,\\
     \end{array}\right.
   \end{eqnarray}
  where $\alpha>0$ is the updating step size and $j\in\mathcal{K}$.
  \STATE Denote $\mathbf{y}(t+\frac{1}{2})=(\mathbf{x}(t\!+\!1), \mathbf{a}(t+\frac{1}{2}))$. Project the averaged updating result of the strategies onto the feasible set
  ${\mathcal{C}}(\mathbf{x}(t+1))$:
    \begin{eqnarray}
     \label{eq_extragradient_2}
     \begin{array}{ll}
      \mathbf{y}(t\!+\!1)\! = \!\Pi_{\mathcal{C}(\mathbf{x}(t+1))}\left[\mathbf{y}(t)\!+\!\xi(\mathbf{y}(t\!+\!\frac{1}{2})\!-\!\mathbf{y}(t))\right],
     \end{array}
   \end{eqnarray}
    where $\Pi_{\mathcal{C}}[\cdot]$ is the projection operator onto the convex set $\mathcal{C}$, and $0<\xi<1$ is an averaging factor.
  \ENDWHILE
 \end{algorithmic}
  \label{alg2}
\end{algorithm}

\begin{Lemma}[2.9.25 in \cite{Facchinei2003}]
  \label{le_proximal_response}
  Suppose that $F(\mathbf{y})\!:\!\mathcal{D}\!\subseteq\!\mathbb{R}^n\!\rightarrow\!\mathbb{R}^n$ is continuously differentiable on the convex set
  $\mathcal{D}$. The following property (co-coercivity) holds if $F$ is the gradient mapping of a convex function:
  \begin{equation}
   \label{eq_proximal_mapping}
   \exists \xi>0: (\mathbf{y}-\mathbf{y}')^{\top}\left(F(\mathbf{y})-F(\mathbf{y}')\right)\ge\xi\Vert F(\mathbf{y})-F(\mathbf{y}')\Vert^2.
  \end{equation}
\end{Lemma}
\begin{Theorem}
 \label{thm_convergence}
 If the follower sub-game has a unique NE (i.e., the strict inequality condition given in (\ref{eq_strict_concavity_condition}) holds), Algorithm \ref{alg2} converges to an  NE as long as the initial price strategies and population states satisfy the feasible conditions given in (\ref{eq_feasible_space}), and the step size $\alpha$ is small enough.
\end{Theorem}
\begin{proof}
 The proof of Theorem~\ref{thm_convergence} employs the property of a GQVI solution as well as Lemma~\ref{le_proximal_response} to show that Algorithm~\ref{alg2} provides a contractive mapping of the strategies. See Appendix~\ref{Proof_thm_convergence} for the details.
\end{proof}

\subsection{Impact of Social Delays in the Evolutionary Sub-game}\label{Sec:Delay}
Note that in Algorithm~\ref{alg1}, it is implicitly assumed that an MU is able to immediately perceive the payoff of a CSP subscription without any time delay. However, in a practical scenario, when switching to a new CSP, the payoff information about the new subscription may present a certain fixed time delay $\tau_i$. Such time delay is unavoidable since it takes time for the MUs to evaluate their experience over the new social network services. For the replicator dynamics approximated by Algorithm~\ref{alg1}, it is also possible that the MUs in population fraction $j$ has to decide on switching to a new CSP $i$ at time $t$ according to the learned ``reputation'' about service $i$, while the reputation propagation may experience a time delay of $\tau_i$. This leads to the formation of the following delayed replicator dynamics from (\ref{eq_replicator_dynamics}) based on the pairwise proportional imitation protocol (cf., Example VII.1 in~\cite{Sandholm2009}):
\begin{equation}
 \label{eq_delayed_replicator_dynamics}
 \begin{array}{ll}
   \displaystyle\dv{x_j(t)}{t}\!=\!v_j(x_j(t),(x_i(t\!-\!\tau_i))_{i\in\mathcal{M}\backslash\{0\},i\ne j},\mathbf{a})\!=\!
   \!x_j(t)\!\displaystyle\sum_{i\in\mathcal{M}\backslash\{0\},i\ne j}\!x_i(t\!-\!\tau_i)\left(\pi_j^u(x_j(t), \mathbf{a})\!-\!\pi_i^u(x_i(t\!-\!\tau_i),\mathbf{a})\right),
 \end{array}
\end{equation}
where we assume that the upper-level sub-game strategies are fixed as $\mathbf{a}$. Note that the delay for the population states may potentially be asymmetric due to the variation of (side) information propagation for the service associated with different CSPs.

Since the stable rest points of the original ODE system in (\ref{eq_replicator_dynamics}) may not be stable in the delayed replicator dynamics given by
(\ref{eq_delayed_replicator_dynamics}), it is necessary to analyze the stability of the new Delayed Differential Equation (DDE) system. We consider the perturbations $x_j(t)=x_j^*+\tilde{x}_j(t)$, where $x^*_j$ is a stable rest point of (\ref{eq_replicator_dynamics}), and obtain the linear variational system of (\ref{eq_delayed_replicator_dynamics}) around $\mathbf{x}^*$ as follows $\forall j\in\mathcal{M}\backslash\{0\}$:
\begin{equation}
 \label{eq_variational_system}
 \displaystyle\dv{\tilde{x}_j(t)}{t}=\sum_{i\in\mathcal{M}\backslash\{0\}}\pdv{v_{j}(\mathbf{x}^*,\mathbf{a})}{x_i}\tilde{x}_i(t-\tau_i) +
 \pdv{v_{j}(\mathbf{x}^*,\mathbf{a})}{x_j}\tilde{x}_j(t).
\end{equation}
We first consider that the information delay is uniform for all the services. From (\ref{eq_variational_system}), we have:
\begin{equation}
 \label{eq_variational_system_2}
 \begin{array}{ll}
 \displaystyle\dv{\tilde{\mathbf{x}}_j(t)}{t}=A_0\tilde{\mathbf{x}}(t)+A_1\tilde{\mathbf{x}}(t-\tau)\\
 =\begin{bmatrix}
   \pdv{v_{1}(\mathbf{x}^*,\mathbf{a})}{x_1} & 0 & \ldots & 0\\
   0 & \pdv{v_{2}(\mathbf{x}^*,\mathbf{a})}{x_2} & \ldots & 0\\
   \vdots & \vdots & \ddots & \vdots\\
   0 & 0 & \ldots & \pdv{v_{M}(\mathbf{x}^*,\mathbf{a})}{x_M}
  \end{bmatrix}\tilde{\mathbf{x}}(t)+
  \begin{bmatrix}
   0 & \pdv{v_{1}(\mathbf{x}^*,\mathbf{a})}{x_2} & \ldots & \pdv{v_{1}(\mathbf{x}^*,\mathbf{a})}{x_M}\\
   \pdv{v_{2}(\mathbf{x}^*,\mathbf{a})}{x_1} & 0 & \ldots & \pdv{v_{2}(\mathbf{x}^*,\mathbf{a})}{x_M}\\
   \vdots & \vdots & \ddots & \vdots\\
   \pdv{v_{M}(\mathbf{x}^*,\mathbf{a})}{x_1} & \pdv{v_{M}(\mathbf{x}^*,\mathbf{a})}{x_2} & \ldots & 0
  \end{bmatrix}
\tilde{\mathbf{x}}(t-\tau).
 \end{array}
\end{equation}
It is known from the Hartman-Grobman Theorem that if the trivial solution of the linear system in (\ref{eq_variational_system}) is (locally) asymptotically stable, then the asymptotically stable rest point of the original replicator dynamics $\mathbf{x}^*$ in (\ref{eq_replicator_dynamics}) is also asymptotically stable for the delayed replicator dynamics in (\ref{eq_delayed_replicator_dynamics}) and therefore delay-independent. since $A_0+A_1$ is equivalent to the Jacobian of the delay-free replicator dynamics in (\ref{eq_replicator_dynamics}) at $\mathbf{x}^*$, we have the following proposition from \cite{Gu2003}:
\begin{Proposition}
 \label{prop1}
 For all sufficiently small uniform information delay $\tau$, the asymptotically stable states in the original replicator dynamics in (\ref{eq_replicator_dynamics}) is still asymptotically stable in the delayed system in (\ref{eq_delayed_replicator_dynamics}).
\end{Proposition}

From (\ref{eq_variational_system_2}), we can extend to the case of asymmetric delays and obtain the following delayed dynamics:
\begin{equation}
 \label{eq_variational_system_asymetric}
 \begin{array}{ll}
 \displaystyle\dv{\tilde{\mathbf{x}}_j(t)}{t}=A^0\tilde{\mathbf{x}}(t)+\sum_{j=1}^{M}A^j\tilde{\mathbf{x}}(t-\tau_j),
 \end{array}
\end{equation}
where $A^0$ is given by (\ref{eq_variational_system_2}) and $\forall j\ne0, \forall i\ne j$, $A_{j}(i,j)=\pdv{v_{i}(\mathbf{x}^*,\mathbf{a})}{x_j}$, otherwise $A_j(i,k)=0, \forall i, k\ne j$. We note that $\sum_{j=1}^{M}A_{j}$ in (\ref{eq_variational_system_asymetric}) is equal to $A_1$ in (\ref{eq_variational_system_2}). Therefore, it is straightforward that Proposition~\ref{prop1} also holds for (\ref{eq_variational_system_asymetric}). Further, based on (\ref{eq_variational_system_asymetric}) we now are able to investigate the stability of the DDEs with asymmetric non-small delays. By Proposition 7.1 in \cite{Gu2003}, we know that if the linearized system with fixed asymmetric non-small delays is stable, a Lyapunov-Krasovskii functional exists in the following form:
\begin{equation}
 \label{eq_Krasovskii}
 L(t) = \phi(0)^{\top}P \phi(0)
 +\sum_{j=1}^{M}\int_{h_j}^0{\phi(\epsilon)^{\top}R_j\phi(\epsilon) d\epsilon},
\end{equation}
where $\phi(t)^{\top}=\begin{bmatrix} \tilde{x}^{\top}(t) & \tilde{x}^{\top}(t-\tau_1) & \ldots & \tilde{x}^{\top}(t-\tau_M)\end{bmatrix}$, and $P$ and $R_j$ are symmetric and positive semi-definite. Since (\ref{eq_variational_system_asymetric}) is in the form of a typical retarded system, by Proposition 7.1 in~\cite{Gu2003}, it would be asymptotically stable if the following matrix inequality (i.e., the matrix being negative definite) holds:
\begin{eqnarray}
 \label{eq_LMI}
 \begin{array}{ll}
 \Phi=\begin{bmatrix}
  A_0^{{\top}}P\!+\!PA^{0}+\sum_{j=1}^{M}R_j & PA_1 &  PA_2 & \ldots & PA_M\\
  A_1^{\top}P & -R_1 & 0 & \ldots & 0 \\
  A_2^{\top}P & 0 & -R_2 & \ldots & 0 \\
  \vdots & \vdots & \vdots & \ddots & \vdots\\
  A_M^{\top}P & 0 & 0 & \ldots & -R_M \\
 \end{bmatrix}
 \end{array}\!<\!0.
\end{eqnarray}
Again, if the population fractions all experience uniform information delay as in (\ref{eq_variational_system_2}), the inequality condition in (\ref{eq_LMI}) is reduced to the following:
\begin{eqnarray}
 \label{eq_LMI_symmetric}
 \Phi=\begin{array}{ll}
 \begin{bmatrix}
  A_0^{{\top}}P\!+\!PA_{0}+R_1 & PA_1\\
  A_1^{\top}P & -R_1 \\
 \end{bmatrix}
 \end{array}\!<\!0.
\end{eqnarray}
Let $Q=P^{-1}$. By symmetric block matrix determinant calculation, (\ref{eq_LMI_symmetric}) is equivalent to
\begin{equation}
 \label{eq_converted_LMI}
 QA_0^{{\top}}+A_0Q+QR_1^{-1}Q+A_1R_1^{-1}A_1^{{\top}}<0.
\end{equation}
Due to the requirement that $R_1$ and $P$ are symmetric positive definite, we known that in (\ref{eq_converted_LMI}), $QR_1^{-1}Q+A_1R_1^{-1}A_1^{{\top}}$ is also positive definite. Therefore, it requires that $A_0$ is strictly negative definite. In general, the delayed replicator dynamic system is not always delay-invariant. However, we can simplify the situation by letting $P=I$ and $R=\omega I$, where $I$ is the identity matrix, and obtain the following sufficient condition for the follower evolutionary game to be delay-independent:

\begin{Proposition}
 \label{prop2}
 Denote the eigenvalues of $A_0+A_1$ by $\lambda_i(A_0+A_1)$. For a uniform deterministic delay, the asymptotic stability of the equilibria of the follower sub-game in (\ref{eq_replicator_dynamics}) is preserved in the delayed dynamics in  (\ref{eq_delayed_replicator_dynamics}), if we can find a scalar $\omega$ satisfying the following condition for each $\lambda_i(A_0+A_1)$:
 \begin{equation}
  \label{eq_delayed_sufficient_condition}
  \pdv{v_{i}(\mathbf{x}^*,\mathbf{a})}{x_i}+\omega+\frac{1}{\omega}\left(\lambda_i(A_0+A_1)-\pdv{v_{i}(\mathbf{x}^*,\mathbf{a})}{x_i}\right)^2<0.
 \end{equation}
\end{Proposition}
\begin{proof}
 With the assumption of $P=I$ and $R=\omega I$, (\ref{eq_converted_LMI}) is reduced to $2A_0+\omega I + \omega^{-1}(A_1A_1^{\top})<0$. Let $v_i$ denote the eigenvector associated with $\lambda_i(A_0+A_1)$. Then, we have $(A_0+A_1)v_i-A_0(i,i)Iv_i=\lambda_i(A_0+A_1)v_i-A_0(i,i)v_i$, which leads to (\ref{eq_delayed_sufficient_condition}).
\end{proof}

\section{Simulation Results}\label{Sec:Simulation}
For ease of exposition, in this section we adopt the logarithmic realization of the utility functions for the MUs and the CSPs given in (\ref{eq_negative_ne}), (\ref{eq_positive_ne}) and (\ref{eq_CP_utility}) for performing the numerical simulations.

\subsection{Convergence of the Proposed Algorithm and Stability of the Evolutionary Sub-game}\label{Subsec:Stability}
\begin{figure*}[t]
\centering     
\subfigure[]{\label{fig_converge_a}\includegraphics[width=.24\linewidth]{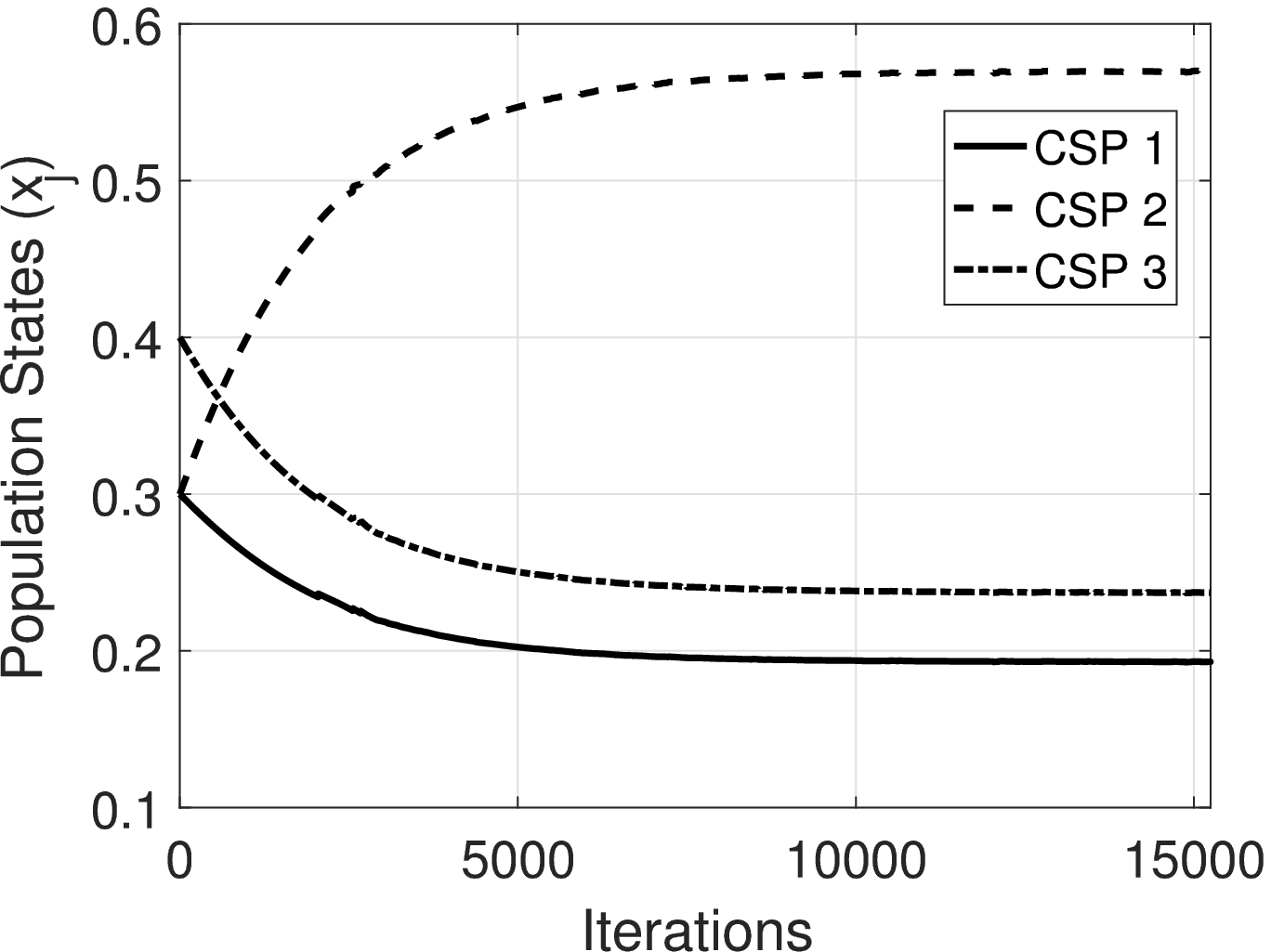}}
\subfigure[]{\label{fig_converge_b}\includegraphics[width=.24\linewidth]{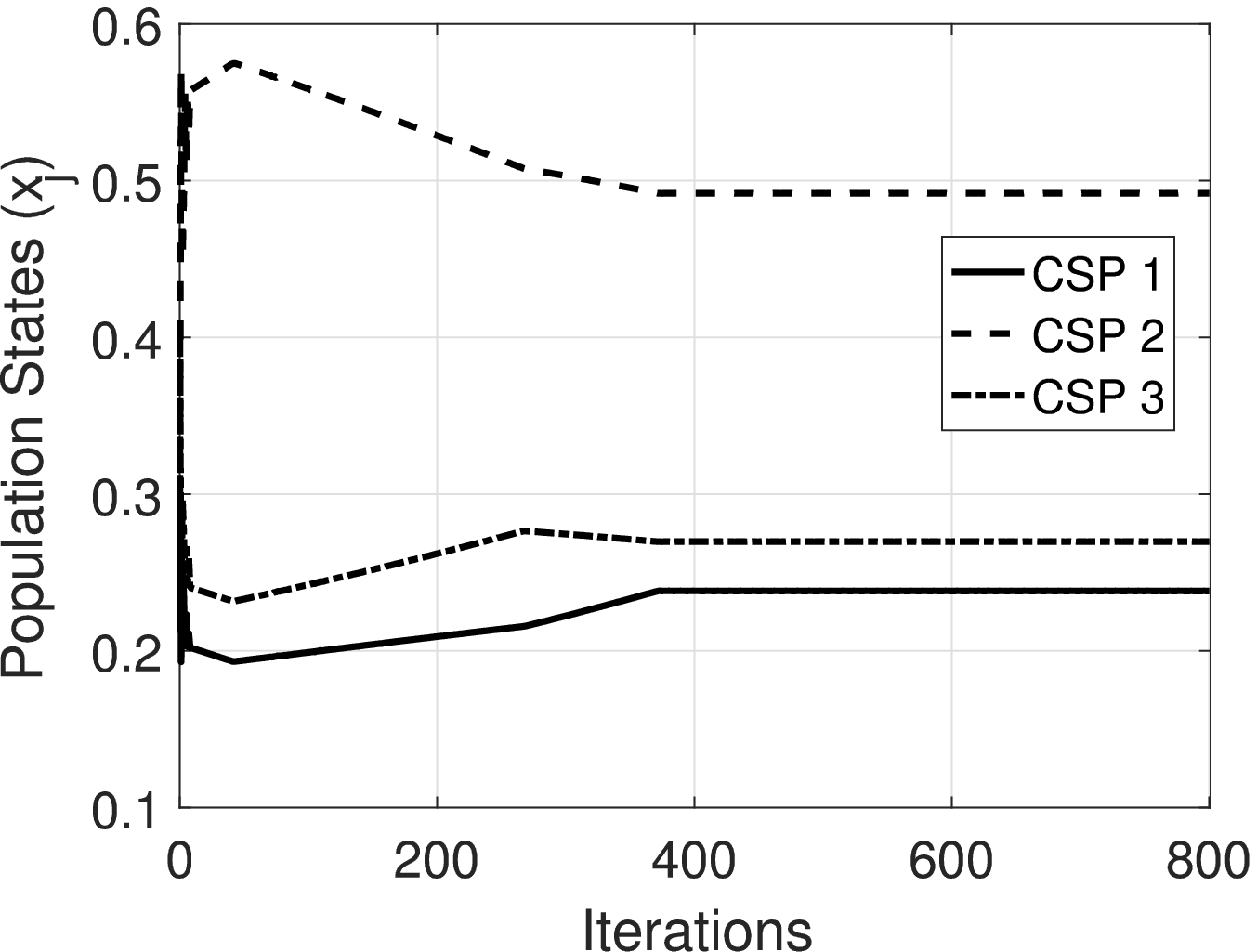}}
\subfigure[]{\label{fig_converge_c}\includegraphics[width=.24\linewidth]{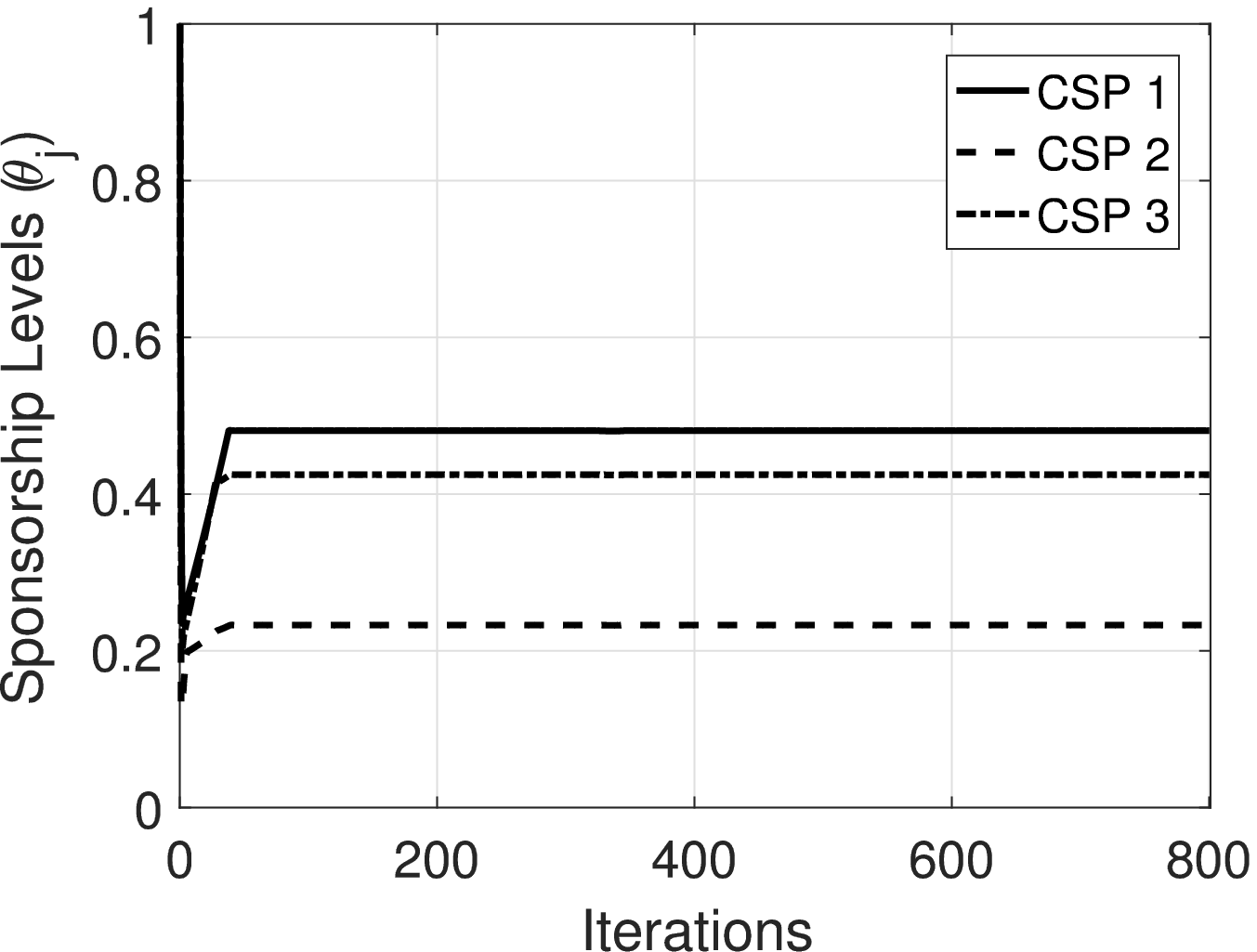}}
\subfigure[]{\label{fig_converge_d}\includegraphics[width=.24\linewidth]{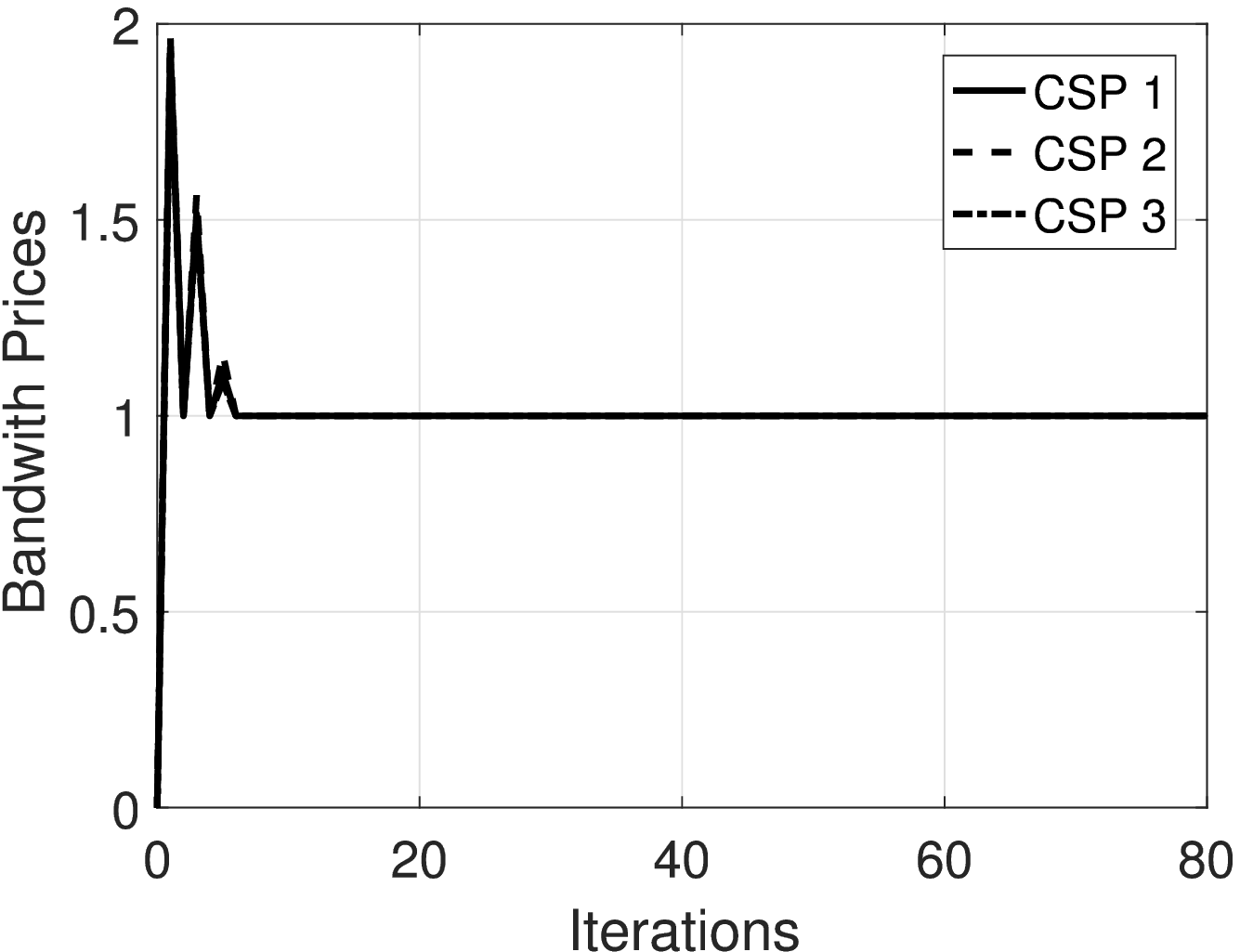}}
\caption{(a) A snapshot of the user state evolution at the inner loop of Algorithm~\ref{alg2} (i.e., Algorithm~\ref{alg1}). (b) The evolution of user states at the outer loop of Algorithm~\ref{alg2}. (c) The evolution of sponsorship levels ($\theta_j$) offered by the CSPs. (d) The evolution of subscription price $p_u$.}
\label{fig_convergence}
\end{figure*}

We first demonstrate the evolution of the MU population states and the providers' strategies with Algorithm~\ref{alg2} in Figure~\ref{fig_convergence}. For the purpose of visualization, we use a case of 3 CSPs to show the convergence property of the proposed strategy searching scheme. We choose the simulation parameters as $N\!=\!8000$, $o_1\!=\!2$, $o_2\!=\!5$, $o_3\!=\!4$, $\overline{c}_j\!=\!10^4$, $\mathbf{\gamma}_1\!=\!1.7$ and $\mathbf{\gamma}_2\!=\!1.1$.
To ensure that the MUs experience approximately the same level of utility, $\pi_j^u$, we set the initial delivery bandwidth reserved by each CSP to be ${b}_1(0)\!=\!1.4$MHz, ${b}_2(0)=\!2$MHz, ${b}_3(0)\!=\! 1.5$MHz. The selected parameter set satisfies the condition in (\ref{eq_strict_concavity_condition}) and creates a market of heterogeneous CSPs with discernible and imperfectly substitutable services.

Figure~\ref{fig_converge_a} shows a single round evolution of the MU population states in the follower sub-game (Algorithm~\ref{alg1}). Figure~\ref{fig_converge_b} demonstrates the convergence tendency of the MU population states in the outer loop of Algorithm~\ref{alg2}.
From Figures~\ref{fig_converge_d} and \ref{fig_converge_c}, we can observe that it takes about 10 iterations for the MNO to reach its equilibrium prices and it takes longer time (roughly 20 iterations) for the CSP to determine their equilibrium strategies.
Figures~\ref{fig_converge_c} and~\ref{fig_converge_d} provide important insight into the projection-based searching scheme in Algorithm~\ref{alg2}. Theoretically, the joint strategy-projection mechanism in (\ref{eq_extragradient_2}) prevents the gradient-based mechanism in (\ref{eq_extragradient_1}) to linearly increase the bandwidth price asked by the MNO and the sponsorship level offered by the CSPs. Figure~\ref{fig_converge_d} indicates that the MNO may experience short-term oscillation in determining the subscription prices.
The strategy evolution in Figures~\ref{fig_converge_c} and~{\ref{fig_converge_d}} matches well with our theoretical finding. Moreover, Figure~\ref{fig_converge_c} indicates that the CSP with a lower initial delivery bandwidth ${b}_j(0)$ tends to take more advantage of the global network effect to improve its profit. In order to compensate for the disadvantage in QoE due to a smaller ${b}_j(0)$ and attract users, The CSP will attempt to offer a higher level of sponsorship. Figure~\ref{fig_converge_b} indicates that although the CSPs providing a lower basic bandwidth (i.e., CSP 1 and CSP 3) are not able to increase their user utility to the same level as CSP 2 does due to the sponsoring cost, they do succeed in attracting more users by increasing their sponsorship levels. In comparison, thanks to a higher internal network effect, CSP 2 mainly relies on its better QoE among the CSPs to attract subscribers and does not need to offer a high level of subsidy.

\begin{figure*}[t]
\centering     
\subfigure[]{\label{fig_delay_a}\includegraphics[width=.24\linewidth]{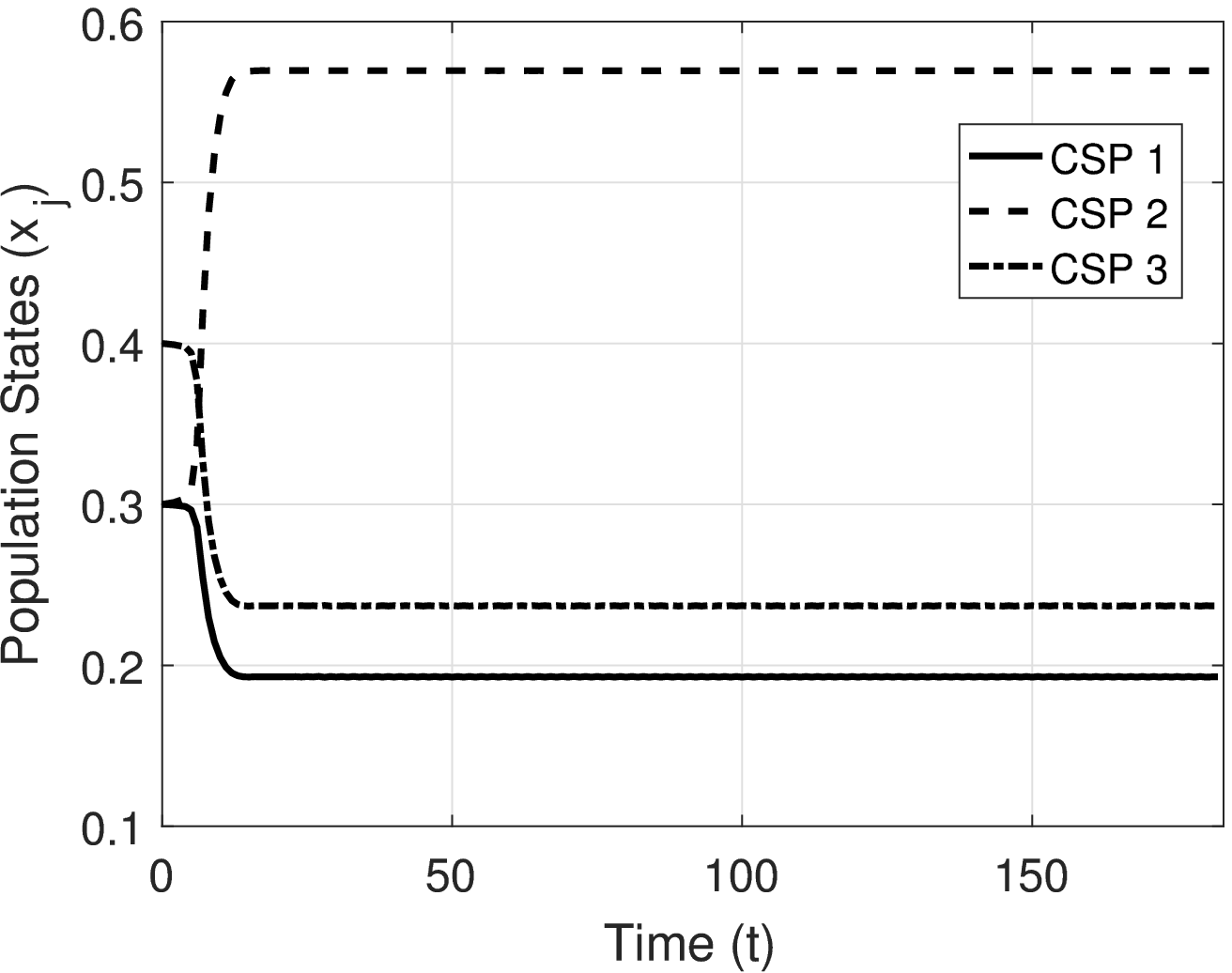}}
\subfigure[]{\label{fig_delay_b}\includegraphics[width=.24\linewidth]{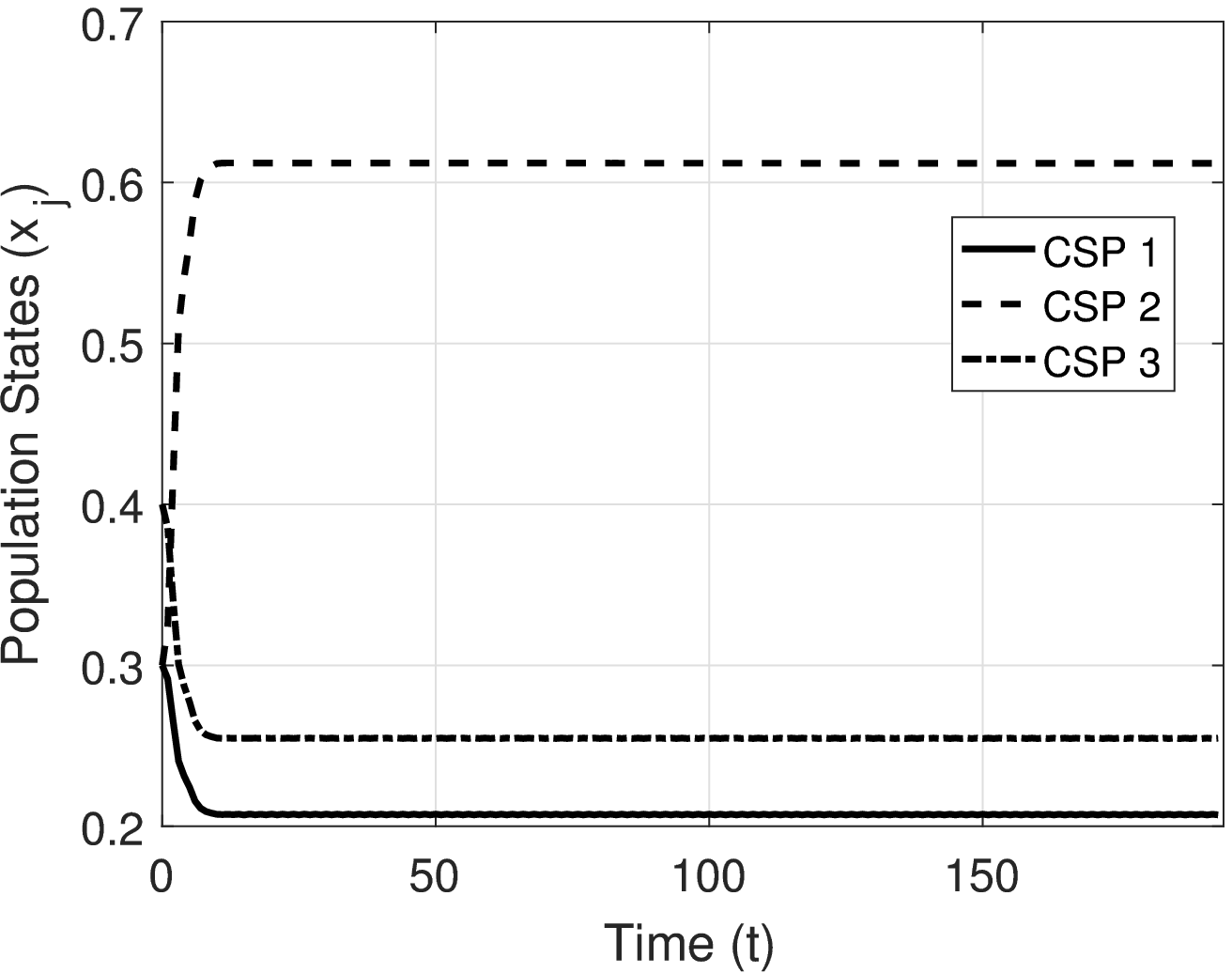}}
\subfigure[]{\label{fig_delay_c}\includegraphics[width=.24\linewidth]{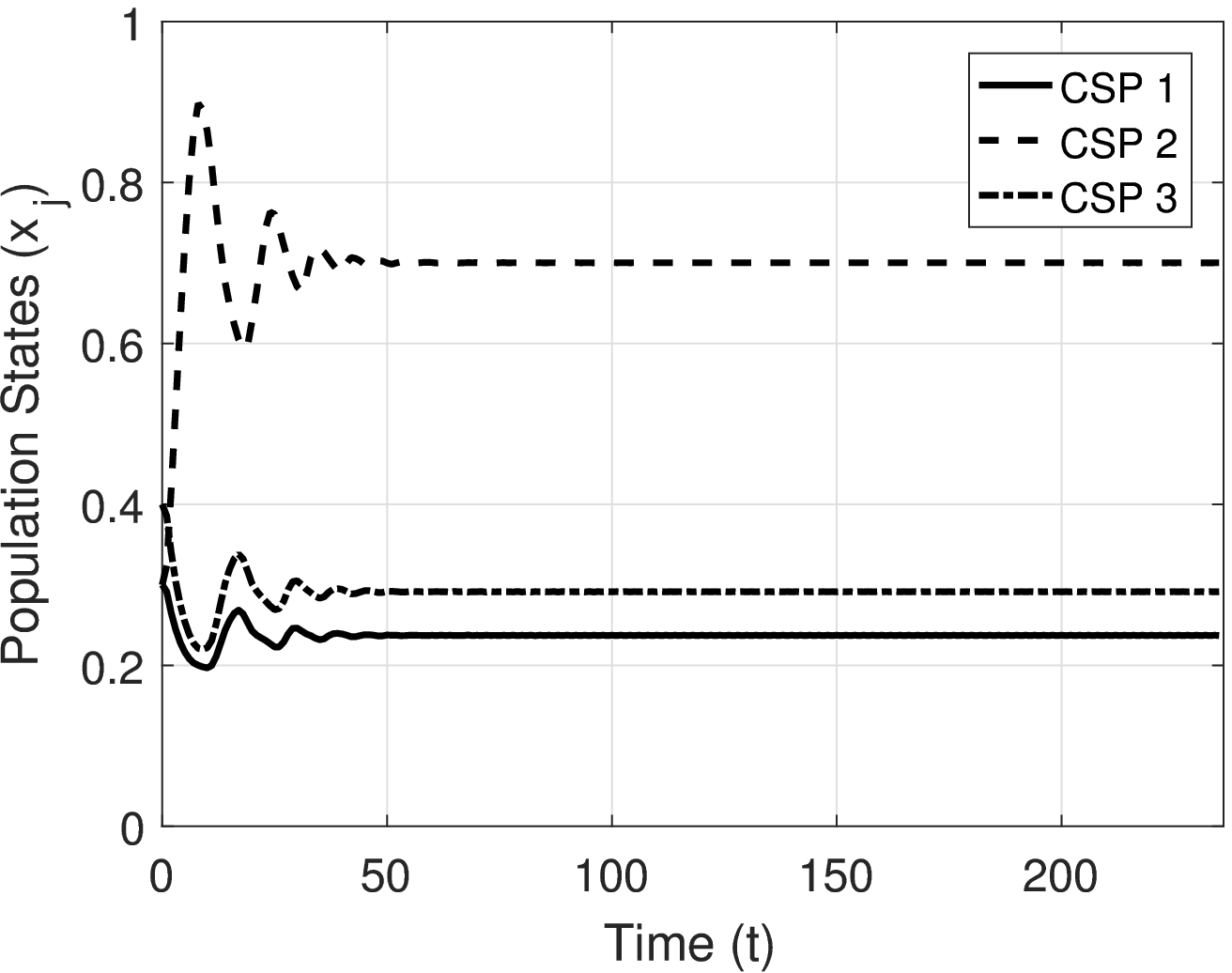}}
\subfigure[]{\label{fig_delay_d}\includegraphics[width=.24\linewidth]{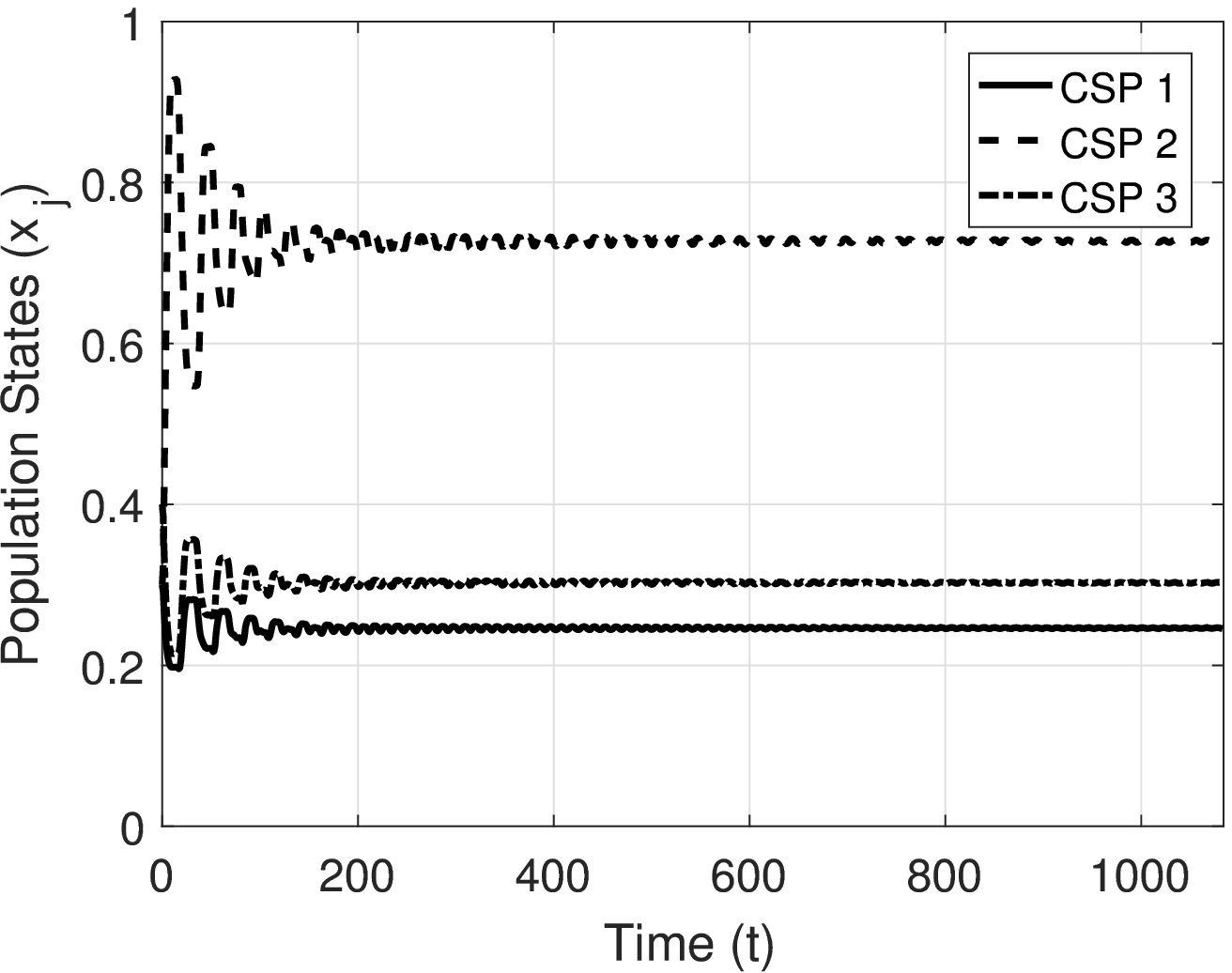}}
\caption{A snapshot of the MU state evolution at the inner loop of Algorithm~\ref{alg2} with different uniform delays $\tau$.
(a) $\tau=0.1$. (b) $\tau=1$. (c) $\tau=10$. (d) $\tau=50$.}
\label{fig_delay}
\end{figure*}
In Figure~\ref{fig_delay}, we present the population state evolution with fixed provider strategy and uniform information delays for the same system setup. As can be observed in Figure~\ref{fig_delay_a}, with small delays ($\tau=0.1$), the MU population will converge to the same evolutionary stable states as in Figure~\ref{fig_converge_a} via a different trajectory. Such a result provides the exact evidence of Proposition~\ref{prop1}. However, as we increase the value of delay, we can observe in Figures \ref{fig_delay_b}-\ref{fig_delay_d} that the evolutionary stable states will shift to a different position as we gradually increase the value of delay. Such a finding indicates that the evolutionary sub-game of the MUs is not delay-independent. However, in the studied cases there still exists a globally evolutionary stable state for the delayed replicator dynamics.

\subsection{Impact of Provider Strategies on the Utilities of Market Participants}\label{Subsec:Performance}
\begin{figure*}[t]
\centering     
\subfigure[]{\label{fig_sponsorship_a2}\includegraphics[width=.34\linewidth]{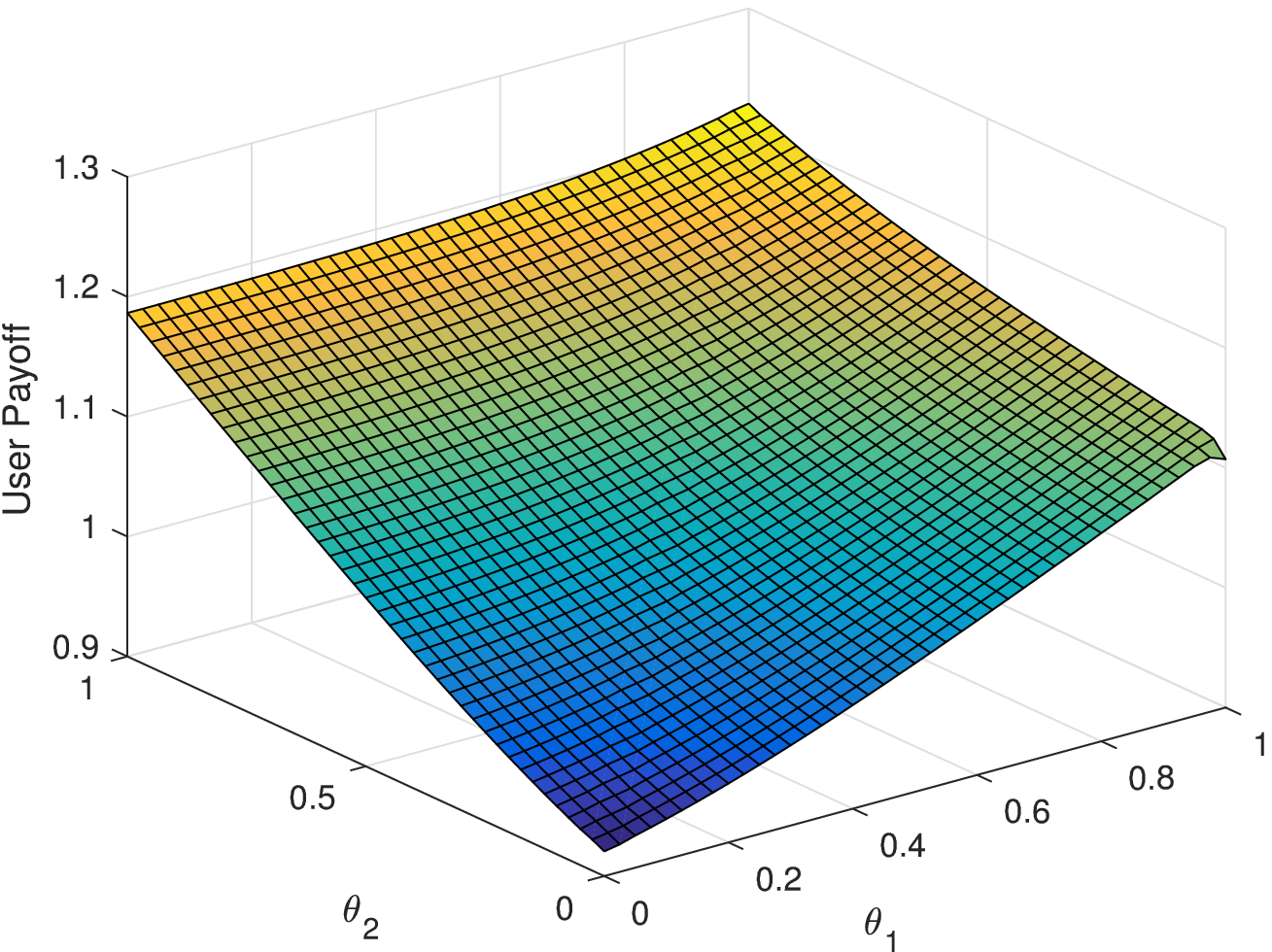}}
\subfigure[]{\label{fig_sponsorship_b2}\includegraphics[width=.34\linewidth]{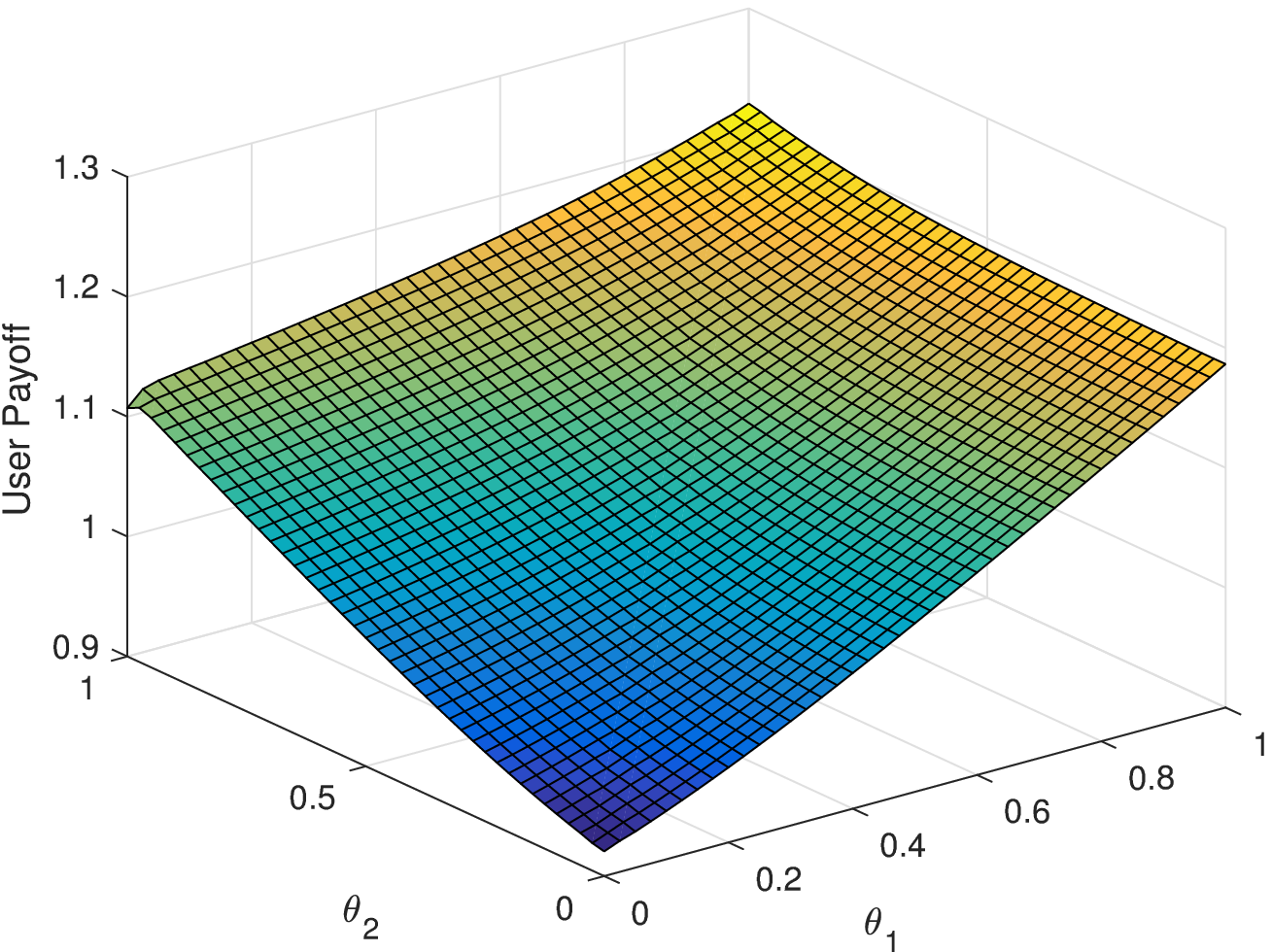}}
\subfigure[]{\label{fig_sponsorship_c2}\includegraphics[width=.31\linewidth]{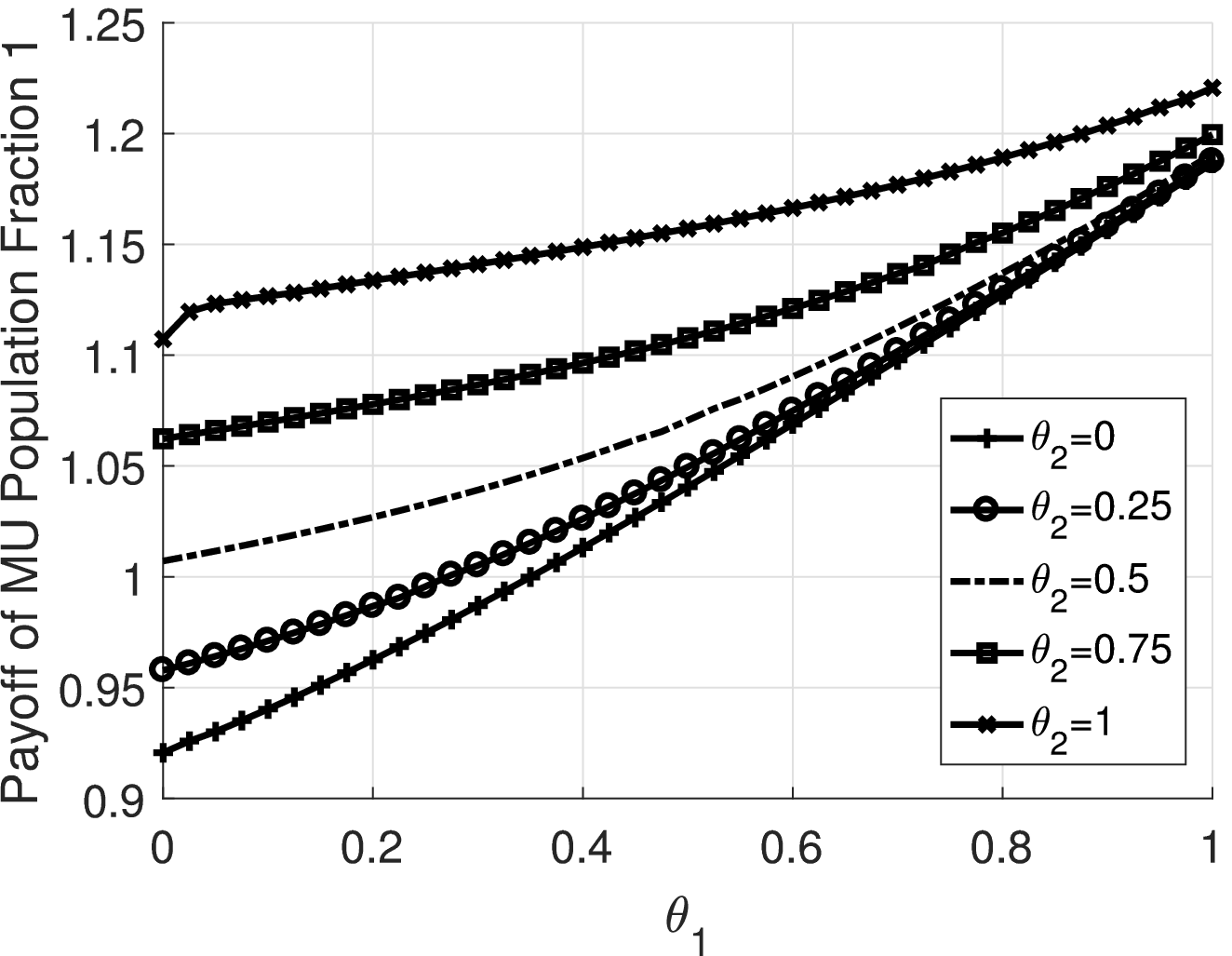}}
\subfigure[]{\label{fig_sponsorship_d2}\includegraphics[width=.31\linewidth]{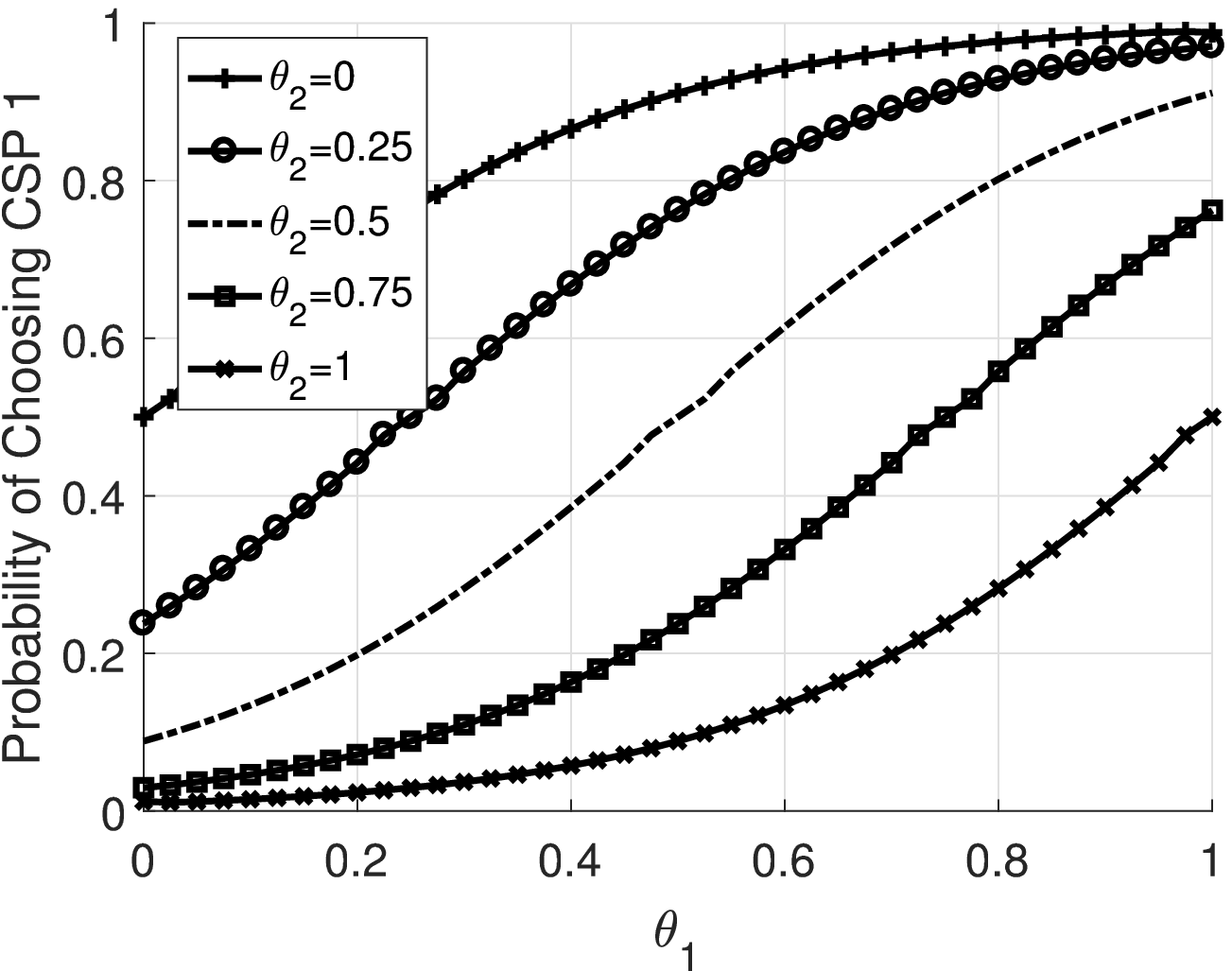}}
\subfigure[]{\label{fig_sponsorship_e2}\includegraphics[width=.31\linewidth]{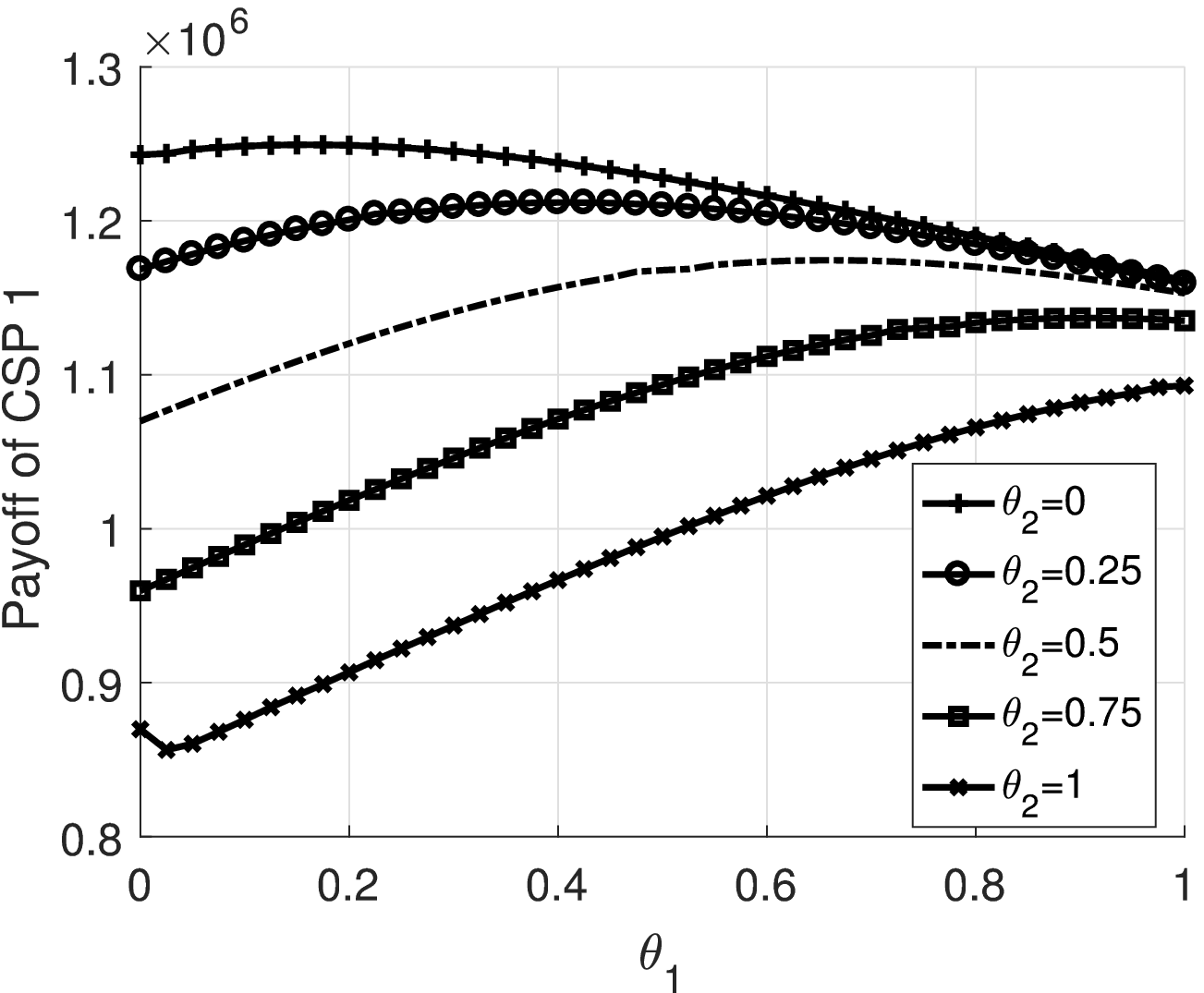}}
\caption{(a) MU's payoff for choosing CSP 1 at the NE with respect to $(\theta_1, \theta_2)$. (b) MU's payoff for choosing CSP 2 at the NE with respect to $(\theta_1, \theta_2)$. (c) MU's payoff for choosing CSP 1 at the NE with respect to flexible sponsorship levels ($\theta_1$) of CSP 1 and fixed sponsorship levels ($\theta_2$) of CSP 2. (d) Probability of the MUs subscribing to CSP 1 at the NE with respect to flexible sponsorship levels of CSP 1 and fixed sponsorship levels of CSP 2. (e) Payoff of CSP 1 with respect to its sponsorship levels at different levels of $\theta_2$.}
\label{fig_sponsorship_2}
\end{figure*}

We provide further analysis about the impact of the sponsorship levels and MNO's prices on the MUs' and the CSP's utilities. Again, for the purpose of visualization, we first consider a market of two symmetric CSPs with a fixed MNO pricing strategy. We assume that there are $50000$ MUs, and set the market parameters as $p_u\!=\!0.3$, $p_c\!=\!0$, $\sigma_j=10^5$, $\gamma_1\!=\!0.1$, $\gamma_2\!=\!0.05$ and a fixed ${b}_j\!=\!10$MHz. From Figures~\ref{fig_sponsorship_a2} and \ref{fig_sponsorship_b2}, we can observe that when the CSPs provide identical services, the performance of the two MUs are symmetric with respect to the sponsorship levels of the two CSPs. In Figure~\ref{fig_sponsorship_c2}, we note that when fixing one CSP's sponsorship level, increasing the other CSP's sponsorship level will always increase the equilibrium payoff of the corresponding MU population fraction. Consequently, this will lead to a higher probability for the MUs to subscribe to the CSP (see Figure~\ref{fig_sponsorship_d2}). However, as we can observe in Figure~\ref{fig_sponsorship_e2}, with the other CSP's sponsorship level is fixed, the CSP's payoff will first increase and then decrease as its sponsorship level is increased. This indicates that, when the subscription price asked by the MNO, $p_u$, is relatively small, the revenue earned through attracting more subscribers is able to compensate the payment made to the MNO. Therefore, we can predict that in the condition of a relatively large $\sigma_j$ and small $p_u$ for the CSPs, the provider-level sub-game equilibrium is reached around $\theta_j=1$. As the value of $p_u$ increases, the CSPs will gradually stop sponsoring the MUs. When the value of $p_u$ is very large, sub-game equilibrium is reached around $\theta_j=0$.

\begin{figure*}[t]
\centering     
\subfigure[]{\label{fig_sponsorship_g1}\includegraphics[width=.32\linewidth]{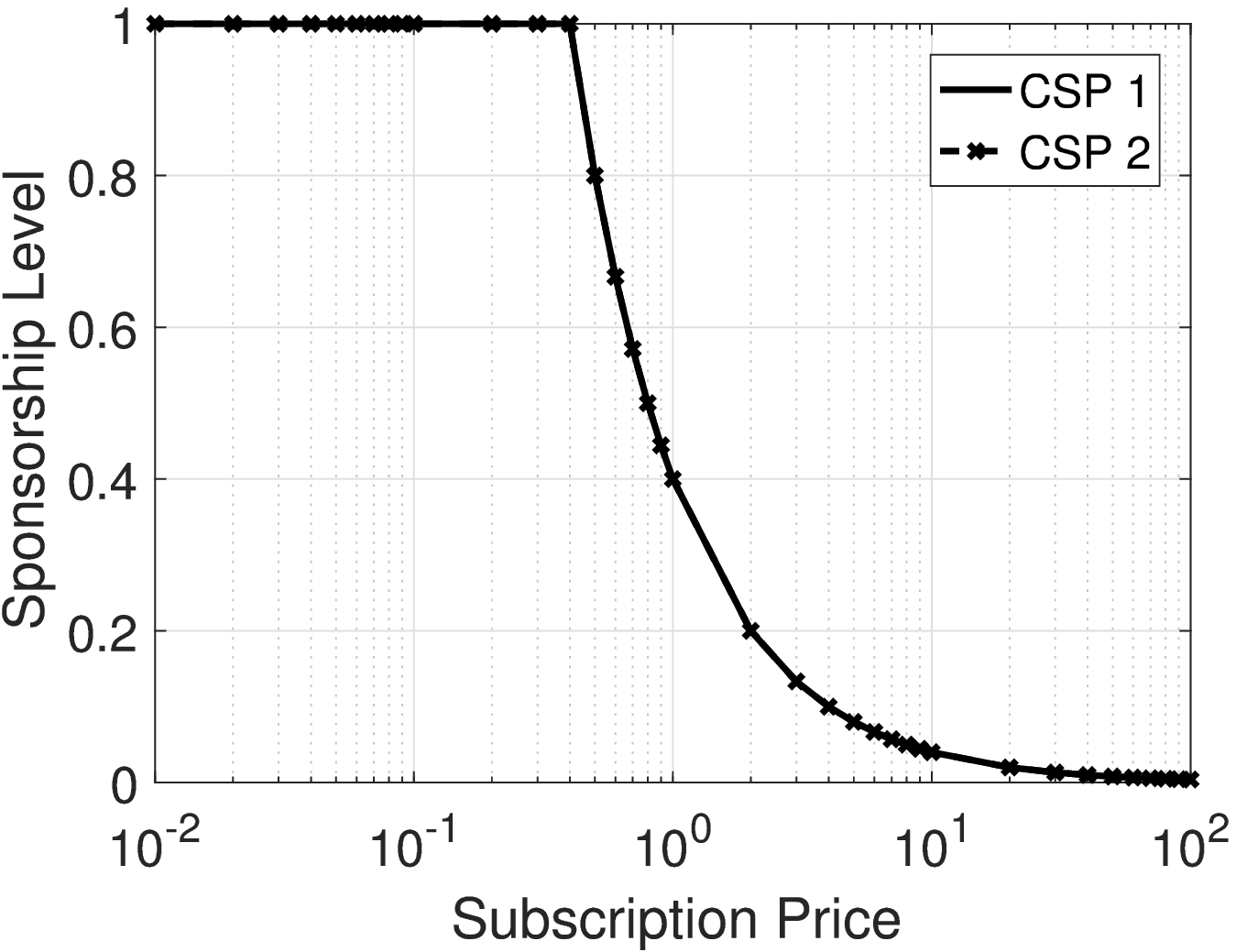}}
\subfigure[]{\label{fig_sponsorship_g2}\includegraphics[width=.32\linewidth]{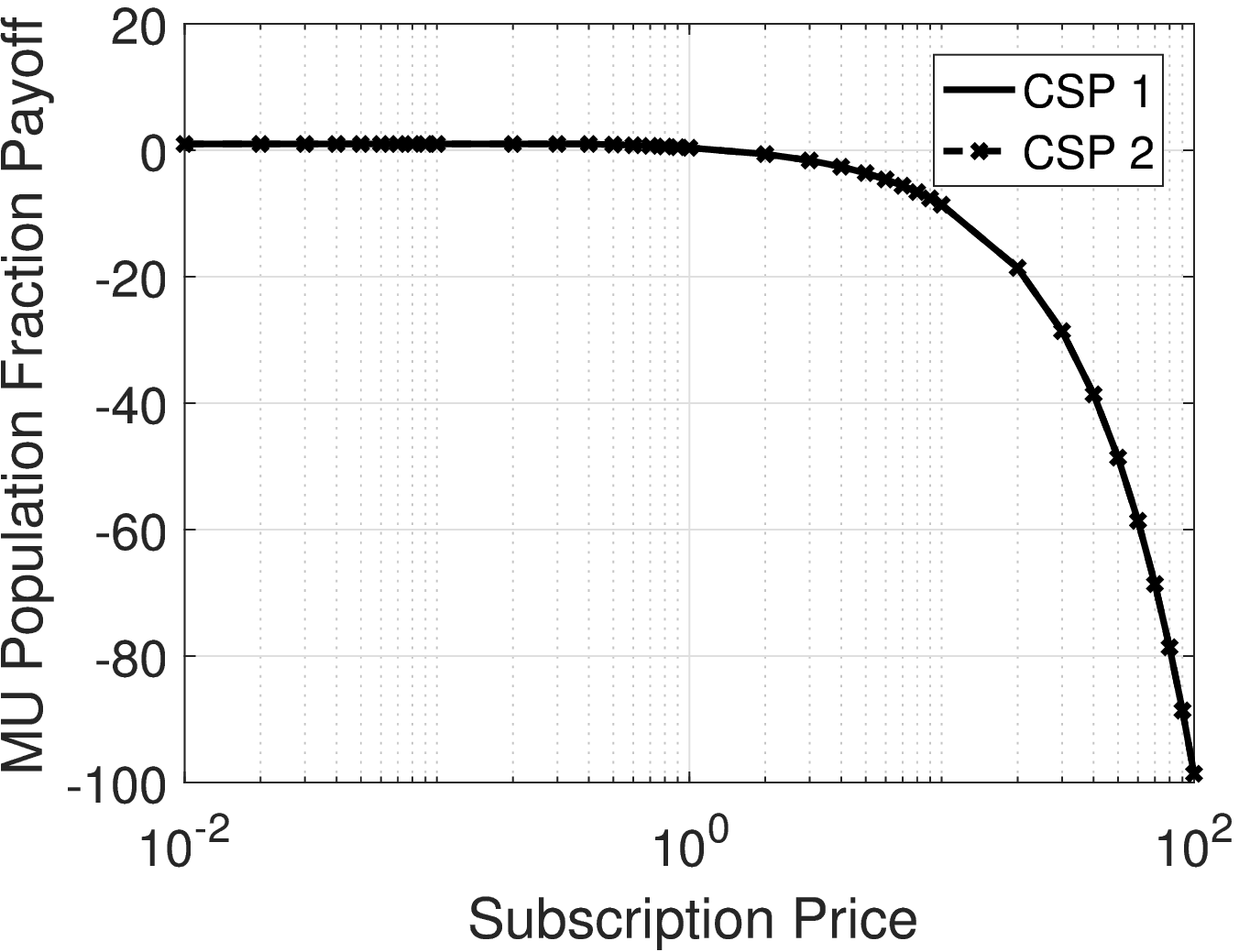}}
\subfigure[]{\label{fig_sponsorship_g3}\includegraphics[width=.32\linewidth]{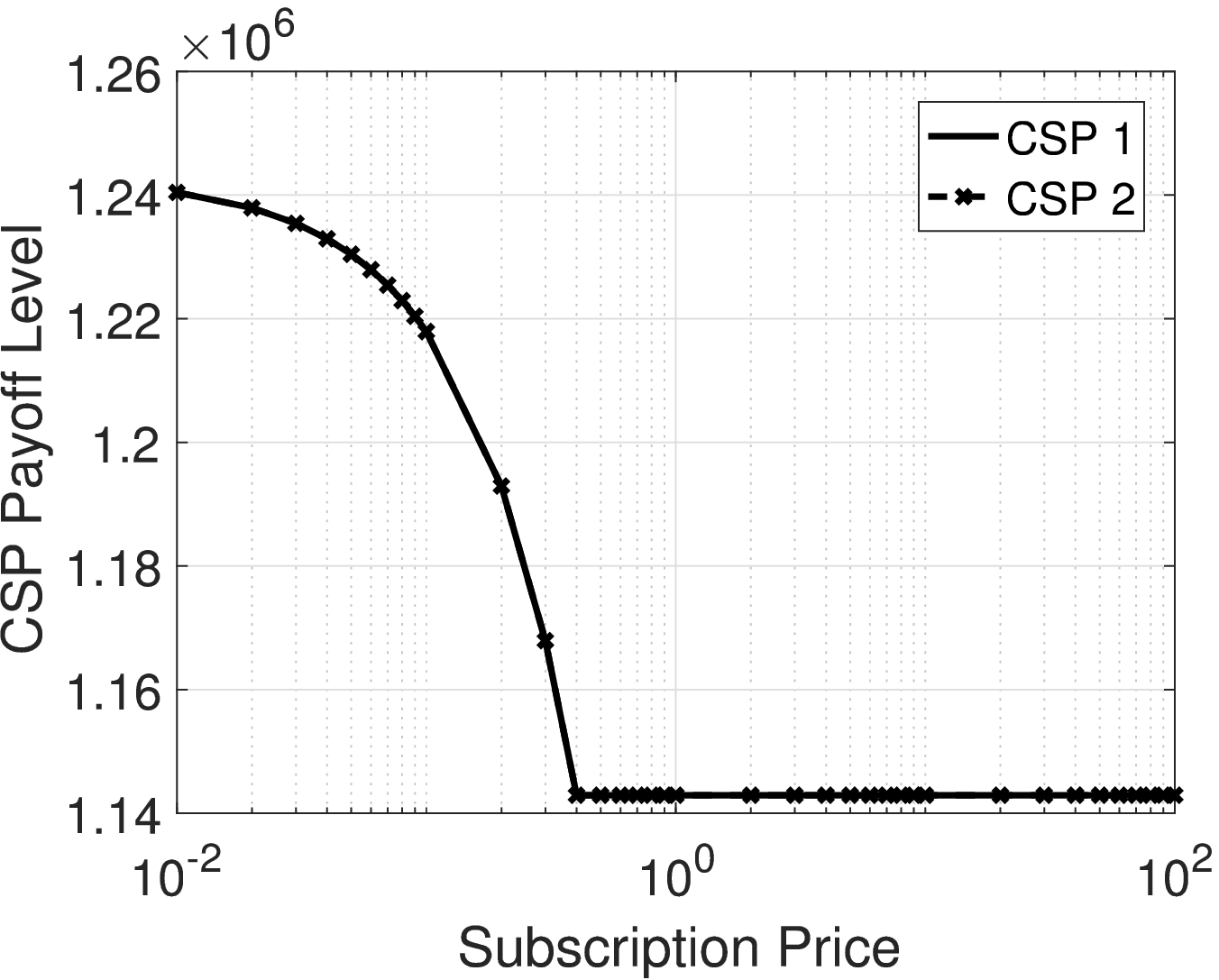}}
\caption{(a) Sponsorship levels at the equilibrium versus different fixed subscription prices $p_u$.
(b) MU's payoff at the equilibrium versus different fixed subscription prices $p_u$.
(c) CSP's payoff at the equilibrium versus different fixed subscription prices $p_u$.}
\label{fig_price_MNO}
\end{figure*}

To verify our prediction, we examine the impact of the MNO's pricing strategies on both the payoffs of the MUs and CSPs. To better illustrate the impact of the MNO's subscription price $p_u$ on the strategies and payoff levels of the MUs and the CSPs, we assume that the MUs are unable to refrain from playing the game and remove the strategy $m\!=\!0$ from $\mathcal{M}$ in the follower sub-game. The total demand of the services is also fixed. As shown in Figure~\ref{fig_sponsorship_g1}, when the subscription price $p_u$ asked by the MNO is low, the CSPs tend to offer full subsidies to the MUs since they will gain more from the increasing number of subscribers. As a result, the competition between the (homogeneous) CSPs become destructive and result in a prisoner's dilemma-like situation (see the plateau part in Figure~\ref{fig_sponsorship_g1} and the corresponding part in Figure~\ref{fig_sponsorship_g3}). On the other hand, the ``arm-racing'' between the CSPs will benefit the MUs since the increased sponsorship will compensate for the increasing subscription cost (see Figure~\ref{fig_sponsorship_g2}). However, as the subscribing price $p_u$ keeps increasing, the revenue received by the CSPs will finally be unable to cover the cost of attracting subscribers. Then, the CSPs will gradually reduce the sponsorship offered to the MUs. Since the MUs are forced to subscribe to one of the two CSPs, the MUs will eventually suffer from negative payoff. The CSPs will still enjoy a positive revenue at the equilibrium by finally stop offering any subsidies (see Figure~\ref{fig_sponsorship_g3})).

\subsection{Impact of Network Effects}
\begin{figure*}[t]
\centering     
\subfigure[]{\label{fig_compare_1}\includegraphics[width=.34\linewidth]{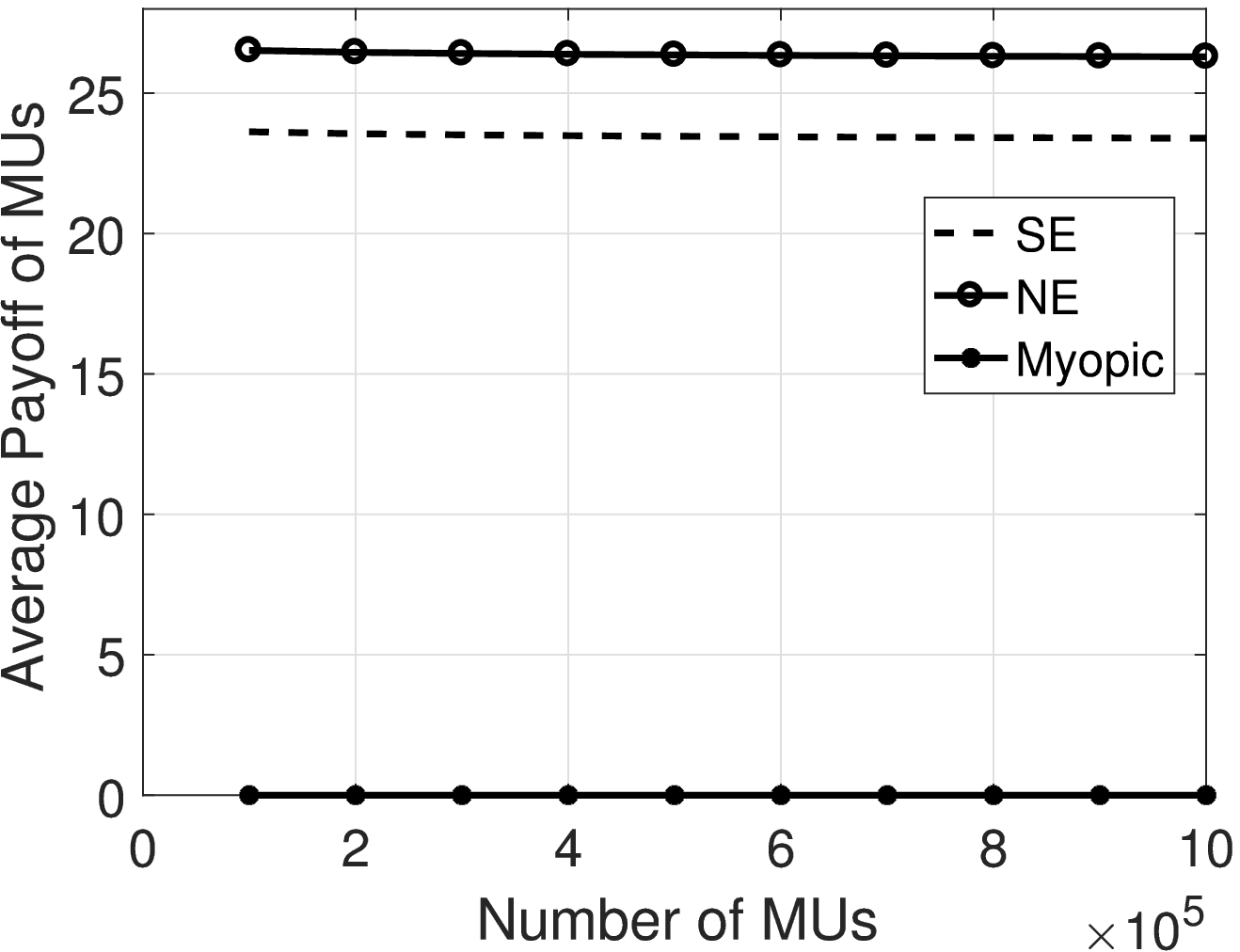}}
\subfigure[]{\label{fig_compare_2}\includegraphics[width=.34\linewidth]{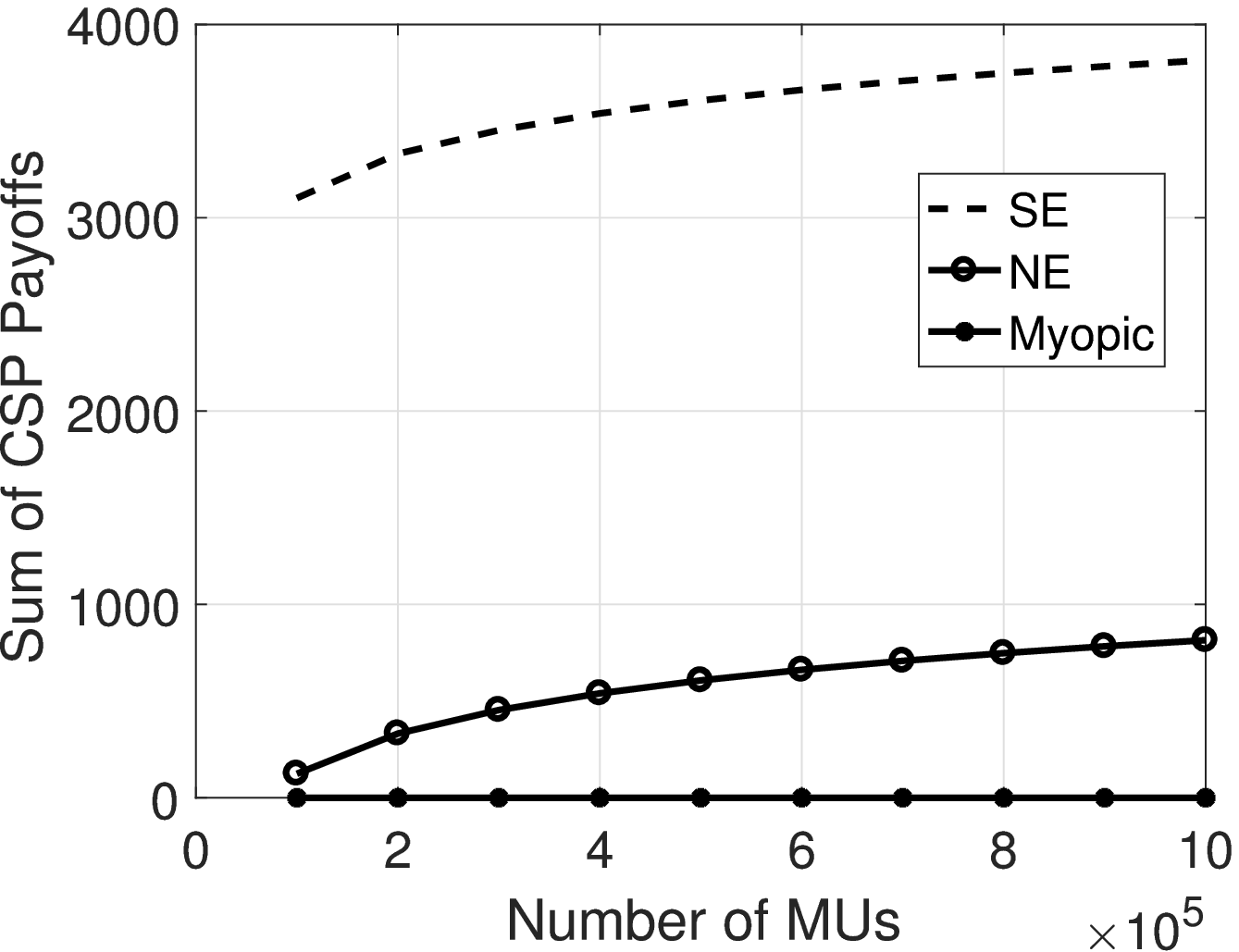}}
\caption{(a) Average MU payoff at the equilibrium versus number of MUs.
(b) Average CSP payoff at the equilibrium versus number of MUs.}
\label{fig_compare_CSP_MU}
\end{figure*}

In this section, we evaluate the impact of the global network effects on the performance of the MUs and the CSPs at the equilibrium. We assume a market of three homogeneous CSPs and set the simulation parameters as ${b}_j\!=\!2$MHz, $o_j\!=\!1$, $\overline{c}_j\!=\!10^4$, ${\gamma}_1\!=\!2.5$ and ${\sigma}_j\!=\!20$. We also set the range of MNO prices by $0\!\le\!p_u\!\le\!20$ and $0\!\le\!p_c\!\le\!200$. In addition to the equilibrium obtained through Algorithm~\ref{alg2}, we also examine the player performance in the following cases:
\begin{itemize}
 \item SE: the strategies are updated following the Gauss-Seidel updating scheme described in Algorithm~\ref{alg0}.
 \item NE: all the players update their strategy in the distributed manner given by Algorithm~\ref{alg2}.
 \item The myopic strategies: at each round, the MNO and the CSPs search for their best-response with respect to the MU population states.
\end{itemize}
As can be observed in Figure~\ref{fig_compare_CSP_MU}, myopically updating the strategies through best-response search will finally enforce all the MUs to subscribe to no service. Therefore, both the MUs and the CSPs receive zero payoff. At the SE, the leaders' performance are significantly better than that at the NE, and such an improvement is at the loss of MUs. In the mean time, due to the existence of the global network effect, when the number of MUs increases, the worsen congestion effect among the MUs is compensated by the increasing network effect. The performance of the MUs at both the SE and the NE remains almost unchanged.

\section{Conclusion}\label{Sec:Conclusion}
In this paper, we have studied the structure of a sponsored content/service market in mobile networks, which is featured by the global network effect in the perceived utility of a large population of users and the existence of multiple content/service providers. We have proposed a hierarchical game-based framework to model the interactions among the three parties in the market, namely, the mobile network operator, the content/service providers and the mobile users. By exploiting the structure of the proposed hierarchical game, we have modeled the user-level sub-game as an evolutionary game and the interaction among the content/service provider and the mobile network operator as a non-cooperative equilibrium searching problem. We have discovered a few important properties regarding the equilibria of the game and its existence condition. Based on our discoveries, we have designed a distributed, iterative scheme for equilibrium searching. Both the theoretical analysis and numerical simulation results have shown the effectiveness of the proposed algorithm. The simulation results have provided important insight into the formation of the market equilibrium as well as the impact of sponsorship levels on the payoffs of both the mobile users and the service providers.

\appendix
\subsection{Proof of Theorem \ref{thm_convergence}}
\label{Proof_thm_convergence}
Consider a fixed feasible provider strategies $\mathbf{a}$. If the concavity condition in Theorem~\ref{thm_existence_NE} is satisfied, the replicator dynamics described by Algorithm~\ref{alg1} ensures the convergence to a solution of the GQVI problem given in (\ref{eq_evolutionary_equilibrium}): $({\mathbf{x}}-\mathbf{x}')^{\top}F_f(\mathbf{x}',\mathbf{a})\ge0, \forall {\mathbf{x}}\in\mathcal{C}_{\mathbf{x}}(\mathbf{x}', \mathbf{a})$. The concavity condition in Theorem~\ref{thm_existence_NE} also ensures the existence of a non-empty set of solutions to the GQVI problem given by (\ref{eq_auxiliary_gqvi}). We denote the joint strategies across the two levels by $\mathbf{y}\!=\![\mathbf{x}^{\top}, \mathbf{a}^{\top}]^{\top}$ and the gradient-based mapping for the joint strategies by $H(\mathbf{x},\mathbf{a}) =[F^{\top}_l(\mathbf{x},\mathbf{a}), F^{\top}_f(\mathbf{x}, \mathbf{a})]^{\top}$. Now the lower-level potential function $f(\mathbf{x},\mathbf{a})$ defined in (\ref{eq_proof_global_maximization}) is concave with respect to $\mathbf{x}$ over the simplex $\mathcal{X}$, and $\pi_j^c(\mathbf{x}, \mathbf{a})$ and $\pi^o(\mathbf{x}, \mathbf{a})$ are linear with respect
to the local strategies $a_j$ on a compact convex set $\mathcal{C}_{a_j}(\mathbf{x})$. Then, the equivalent GQVI problem for the Stackelberg solution in (\ref{eq_auxiliary_gqvi}) can be rewritten in the following form:
 \begin{equation}
  \label{eq_QVI_equilibrium}
  (\mathbf{y}-\mathbf{y}^*)^{\top}H(\mathbf{y}^*)\ge0, \forall \mathbf{y}\!=\![\mathbf{x}^{\top}, \mathbf{a}^{\top}]^{\top}\in\mathcal{C}({\mathbf{x}^*,\mathbf{a}^*}).
 \end{equation}

In the inner loop of Algorithm \ref{alg2}, the result of Algorithm \ref{alg1}, $\mathbf{x}(t+1)$, is the solutions to (\ref{eq_evolutionary_equilibrium}) given the upper-level strategy $\mathbf{a}(t)$ at time $t$. Let $\{\mathbf{y}^1(t)=(\mathbf{x}(t),\mathbf{a}(t))\}$ and $\{\mathbf{y}^2(t)=(\mathbf{z}(t),\mathbf{b}(t))\}$ be two sequences generated by Algorithm \ref{alg2} from different initial strategies. Since $\mathbf{x}(t+1)$ and $\mathbf{z}(t+1)$ are the equilibrium states in the follower sub-game given $\mathbf{a}(t)$ and $\mathbf{b}(t)$, respectively, by (\ref{eq_evolutionary_equilibrium}), we have the following inequality:
  \begin{equation}
  \label{eq_proof_1}
  \begin{array}{ll}
  0{\le} (\mathbf{x}(t+1)-\mathbf{z}(t+1))^{\top}F_f(\mathbf{z}(t+1),\mathbf{b}(t))
  +(\mathbf{z}(t+1)-\mathbf{x}(t+1))^{\top}F_f(\mathbf{x}(t+1),\mathbf{a}(t))\\
  =\!(\mathbf{x}(t\!+\!1)\!-\!\mathbf{z}(t\!+\!1))^{\top}\big(F_f(\mathbf{z}(t\!+\!1),\mathbf{b}(t))\!-\!F_f(\mathbf{x}(t\!+\!1),\mathbf{a}(t))\big).
  \end{array}
 \end{equation}
 By the non-expansive property of the projection operation over a convex sub-space~\cite{Facchinei2003}, we can obtain the following inequality by applying Algorithm~\ref{alg2} when ignoring the strategy averaging process in (\ref{eq_extragradient_2}),
 \begin{eqnarray}
  \label{eq_non_expansive}
  \begin{array}{ll}
  \Vert \mathbf{y}^1(t\!+\!1)\!-\!\mathbf{y}^2(t\!+\!1)\Vert^2
  =\!\left\Vert\Pi_{\mathcal{C}}(\mathbf{y}^1(t\!+\!\frac{1}{2}))\!-\!\Pi_{\mathcal{C}}(\mathbf{y}^2(t\!+
  \!\frac{1}{2}))\right\Vert^2\\
 \!\le\!\left\Vert \mathbf{y}^1(t\!+\!\frac{1}{2})\!-\!\mathbf{y}^2(t\!+\!\frac{1}{2}) \right\Vert^2
  =\left\Vert\mathbf{x}(t+1)-\mathbf{z}(t+1)\right\Vert^2+\left\Vert\mathbf{a}(t+\frac{1}{2})-\mathbf{b}(t+\frac{1}{2})\right\Vert^2.
  \end{array}
 \end{eqnarray}
 Regarding the second term on the right-hand side in (\ref{eq_non_expansive}), the following inequality holds,
  \begin{equation}
  \label{eq_proof_2}
  \begin{array}{ll}
  \Vert \mathbf{a}(t\!+\!\frac{1}{2})\!-\!\mathbf{b}(t\!+\!\frac{1}{2})\Vert^2
  =\left\Vert\mathbf{a}(t)\!-\!\alpha F_l(\mathbf{x}(t\!+\!1),\mathbf{a}(t))
  \!-\!\left(\mathbf{b}(t)\!-\!\alpha F_l(\mathbf{z}(t\!+\!1),\mathbf{b}(t)\right)\right\Vert^2\\
  \le\!\left\Vert \mathbf{a}(t)\!-\!\mathbf{b}(t)\!-\!\alpha(F_l(\mathbf{x}(t\!+\!1),\mathbf{a}(t))\!-\!F_l(\mathbf{z}(t\!+\!1),\mathbf{b}(t)))\right\Vert^2\\
  =\!\left\Vert \mathbf{a}(t)\!-\!\mathbf{b}(t)\right\Vert^2\!+\!\alpha^2\left\Vert F_l(\mathbf{x}(t\!+\!1),\mathbf{a}(t))\!-\!F_l(\mathbf{z}(t\!+\!1),\mathbf{b}(t))\right
  \Vert^2\\
  -\!2\alpha\left(\mathbf{a}(t)\!-\!\mathbf{b}(t)\right)^{\top}\!\!\big(F_l(\mathbf{x}(t\!+\!1),\mathbf{a}(t))\!-\!F_l(\mathbf{z}(t\!+\!1),\mathbf{b}(t))\big).
  \end{array}
 \end{equation}
 After multiplying (\ref{eq_proof_1}) by $2\alpha$ and adding it up with (\ref{eq_proof_2}), we obtain
 \begin{equation}
  \label{eq_proof_4}
  \begin{array}{ll}
   \Vert \mathbf{a}(t\!+\!\frac{1}{2})\!-\!\mathbf{b}(t\!+\!\frac{1}{2})\Vert^2
   \le\!\Vert \mathbf{a}(t)\!-\!\mathbf{b}(t)\Vert^2\!+\!\alpha^2\Vert F_l(\mathbf{x}(t\!+\!1),\mathbf{a}(t))\!-\!F_l(\mathbf{z}(t\!+\!1),\mathbf{b}(t))\Vert^2\\
   -\!2\alpha\left([\mathbf{x}^{\top}(t\!+\!1),\mathbf{a}^{\top}(t)]^{\top}\!-\![\mathbf{z}^{\top}(t\!+\!1),\mathbf{b}^{\top}(t)]^{\top}\right)^{\top}\!
   \!\left(H(\mathbf{x}(t\!+\!1),\mathbf{a}(t))\!-\!H(\mathbf{z}(t\!+\!1),\mathbf{b}(t))\right).
  \end{array}
 \end{equation}
  Since the potential function of the lower-level sub-game $f(\mathbf{x},\mathbf{a})$ is concave, the gradient-based mapping $F_f(\mathbf{x},\mathbf{a})$ is monotone and continuously differentiable in $\mathbf{x}$. Similarly, the gradient mapping obtained from the linear objective function from the provider level, $F_l(\mathbf{x},\mathbf{a})$, is also monotone and continuously differentiable. By (\ref{eq_proximal_mapping}), there exists a positive parameter $\xi$ such that the following holds $\forall \mathbf{y}, \mathbf{y}'$:
    \begin{equation}
   \label{eq_proximal_mapping_2}
   (\mathbf{y}-\mathbf{y}')^{\top}\left(H(\mathbf{y})-H(\mathbf{y}')\right)\ge\xi\Vert H(\mathbf{y})-H(\mathbf{y}')\Vert^2.
  \end{equation}
  Then, from (\ref{eq_proof_4}) and (\ref{eq_proximal_mapping_2}) we obtain
   \begin{equation}
  \label{eq_proof_5}
  \begin{array}{ll}
   \Vert \mathbf{a}(t\!+\!\frac{1}{2})\!-\!\mathbf{b}(t\!+\!\frac{1}{2})\Vert^2\\
   \le\!\Vert \mathbf{a}(t)\!-\!\mathbf{b}(t)\Vert^2\!+\!\alpha^2\Vert F_l(\mathbf{x}(t\!+\!1),\mathbf{a}(t))\!-\!F_l(\mathbf{z}(t\!+\!1),\mathbf{b}(t))\Vert^2
   -\!2\alpha\xi\left\Vert H(\mathbf{x}(t\!+\!1),\mathbf{a}(t))\!-\!H(\mathbf{z}(t\!+\!1),\mathbf{b}(t))\right\Vert^2\\
   \le\Vert \mathbf{a}(t)\!-\!\mathbf{b}(t)\Vert^2\!
   -(2\alpha\xi-\alpha^2)\Vert F_l(\mathbf{x}(t\!+\!1),\mathbf{a}(t))\!-\!F_l(\mathbf{z}(t\!+\!1),\mathbf{b}(t))\Vert^2\\
    -\!2\alpha\xi\left\Vert F_f(\mathbf{x}(t\!+\!1),\mathbf{a}(t))\!-\!F_f(\mathbf{z}(t\!+\!1),\mathbf{b}(t))\right\Vert^2.
  \end{array}
 \end{equation}
 Therefore, as long as $\alpha$ is sufficiently small (i.e., $0\!<\!\alpha\!\le\!2\xi$), and the values of $\mathbf{x}(t+1)$ obtained from the inner-loop
 is the equilibrium of the lower-level sub-game given $\mathbf{a}(t)$, Algorithm~\ref{alg2} provides a non-expansive mapping for the sequence  $\{\mathbf{a}(t)\}$. Then, after applying the averaging scheme, Algorithm~\ref{alg2} guarantees the convergence of $\mathbf{a}(t)$: $\lim\limits_{t\rightarrow\infty}\Vert
 \mathbf{a}(t)\!-\!\mathbf{b}(t)\Vert\!=\!0$ (cf.,~\cite{facchinei201012}). With a converging sequence of $\mathbf{a}(t)$, the following holds when the user-level sub-game has a unique NE,
 \begin{equation}
  \label{eq_proof_6}
  \left\Vert\mathbf{x}(t\!+\!1)\!-\!\mathbf{z}(t\!+\!1)\right\Vert\le\left\Vert\mathbf{x}(t)\!-\!\mathbf{z}(t)\right\Vert.
 \end{equation}
 Plugging (\ref{eq_proof_5}) and (\ref{eq_proof_6}) into (\ref{eq_non_expansive}), we obtain
 \begin{eqnarray}
  \label{eq_proof_final}
  \begin{array}{ll}
   \Vert \mathbf{y}^1(t\!+\!1)\!-\!\mathbf{y}^2(t\!+\!1)\Vert^2\le\left\Vert\mathbf{y}^1(t)\!-\!(\mathbf{y}^2(t))\right\Vert^2\\
   -(2\alpha\xi-\alpha^2)\Vert F_l(\mathbf{x}(t\!+\!1),\mathbf{a}(t))\!-\!F_l(\mathbf{z}(t\!+\!1),\mathbf{b}(t))\Vert^2
    -\!2\alpha\xi\left\Vert F_h(\mathbf{x}(t\!+\!1),\mathbf{a}(t))\!-\!F_h(\mathbf{z}(t\!+\!1),\mathbf{b}(t))\right\Vert^2.
  \end{array}
 \end{eqnarray}
 Therefore, Algorithm~\ref{alg2} provides a non-expansive mapping of the joint strategies. With the perturbation introduced by an averaging process, it is known that a non-expansive mapping guarantees the convergence to the solution of the GQVI~\cite{facchinei201012}. Then, the proof of Theorem \ref{thm_convergence} is completed.

\linespread{1.27}
\bibliographystyle{IEEEtran}
\bibliography{bibfile}

\begin{thebibliography}{10}
\providecommand{\url}[1]{#1}
\csname url@samestyle\endcsname
\providecommand{\newblock}{\relax}
\providecommand{\bibinfo}[2]{#2}
\providecommand{\BIBentrySTDinterwordspacing}{\spaceskip=0pt\relax}
\providecommand{\BIBentryALTinterwordstretchfactor}{4}
\providecommand{\BIBentryALTinterwordspacing}{\spaceskip=\fontdimen2\font plus
\BIBentryALTinterwordstretchfactor\fontdimen3\font minus
  \fontdimen4\font\relax}
\providecommand{\BIBforeignlanguage}[2]{{%
\expandafter\ifx\csname l@#1\endcsname\relax
\typeout{** WARNING: IEEEtran.bst: No hyphenation pattern has been}%
\typeout{** loaded for the language `#1'. Using the pattern for}%
\typeout{** the default language instead.}%
\else
\language=\csname l@#1\endcsname
\fi
#2}}
\providecommand{\BIBdecl}{\relax}
\BIBdecl

\bibitem{Wang1712:Hierarchical}
W.~Wang, Z.~Xiong, D.~Niyato, and P.~Wang, ``A hierarchical game with strategy
  evolution for mobile sponsored {content/service} markets,'' in \emph{GLOBECOM
  2017 - 2017 IEEE Global Communications Conference}, Singapore, Dec. 2017, pp.
  1--6.

\bibitem{index2016global}
Cisco, ``Cisco visual networking index: Global mobile data traffic forecast
  update, 2016-2021,'' Tech. Rep., Mar. 2017.

\bibitem{musacchio2007network}
J.~Musacchio, J.~Walrand, and G.~Schwartz, ``Network neutrality and provider
  investment incentives,'' in \emph{41st Asilomar Conference on Signals,
  Systems and Computers}, Pacific Grove, CA, Nov. 2007, pp. 1437--1444.

\bibitem{1653003}
B.~Briscoe, A.~Odlyzko, and B.~Tilly, ``Metcalfe's law is wrong -
  communications networks increase in value as they add members-but by how
  much?'' \emph{IEEE Spectrum}, vol.~43, no.~7, pp. 34--39, Jul. 2006.

\bibitem{7218528}
C.~Joe-Wong, S.~Ha, and M.~Chiang, ``Sponsoring mobile data: An economic
  analysis of the impact on users and content providers,'' in \emph{2015 IEEE
  Conference on Computer Communications (INFOCOM)}, Apr. 2015, pp. 1499--1507.

\bibitem{zhang2014sponsoring}
L.~Zhang and D.~Wang, ``Sponsoring content: Motivation and pitfalls for content
  service providers,'' in \emph{IEEE Conference on Computer Communications
  Workshops (INFOCOM WKSHPS)}, Toronto, Canada, Apr. 2014, pp. 577--582.

\bibitem{zhang2015sponsored}
L.~Zhang, W.~Wu, and D.~Wang, ``Sponsored data plan: A two-class service model
  in wireless data networks,'' in \emph{ACM SIGMETRICS Performance Evaluation
  Review}, vol.~43, no.~1, 2015, pp. 85--96.

\bibitem{eldelgawy2015interaction}
R.~ElDelgawy and R.~J. La, ``Interaction between a content provider and a
  service provider and its efficiency,'' in \emph{2015 IEEE International
  Conference on Communications (ICC)}, London, UK, Jun. 2015, pp. 5890--5895.

\bibitem{zhang2016tds}
L.~Zhang, W.~Wu, and D.~Wang, ``Tds: Time-dependent sponsored data plan for
  wireless data traffic market,'' in \emph{IEEE International Conference on
  Computer Communications}, San Francisco, CA, Apr. 2016.

\bibitem{7524559}
M.~Andrews, Y.~Jin, and M.~I. Reiman, ``A truthful pricing mechanism for
  sponsored content in wireless networks,'' in \emph{IEEE INFOCOM 2016 - The
  35th Annual IEEE International Conference on Computer Communications}, San
  Francisco, CA, Apr. 2016, pp. 1--9.

\bibitem{katz1994systems}
M.~L. Katz and C.~Shapiro, ``Systems competition and network effects,''
  \emph{The journal of economic perspectives}, vol.~8, no.~2, pp. 93--115,
  1994.

\bibitem{brake2016}
D.~Brake, ``Mobile zero rating: The economics and innovation behind free
  data,'' in \emph{Net Neutrality Reloaded: Zero Rating, Specialised Service,
  Ad Blocking and Traffic Management}, L.~Belli, Ed.\hskip 1em plus 0.5em minus
  0.4em\relax Internet Governance Forum, 2016, ch.~7, p. 132.

\bibitem{7835123}
X.~Gong, L.~Duan, X.~Chen, and J.~Zhang, ``When social network effect meets
  congestion effect in wireless networks: Data usage equilibrium and optimal
  pricing,'' \emph{IEEE Journal on Selected Areas in Communications}, vol.~35,
  no.~2, pp. 449--462, Feb. 2017.

\bibitem{odlyzko2014will}
A.~Odlyzko, ``Will smart pricing finally take off?'' Indiana University at
  Bloomington, Tech. Rep., 2014.

\bibitem{chen2016incentivizing}
Y.~Chen, B.~Li, and Q.~Zhang, ``Incentivizing crowdsourcing systems with
  network effects,'' in \emph{IEEE International Conference on Computer
  Communications}, San Francisco, CA, Apr. 2016.

\bibitem{4575128}
Q.~T. Nguyen-Vuong, Y.~Ghamri-Doudane, and N.~Agoulmine, ``On utility models
  for access network selection in wireless heterogeneous networks,'' in
  \emph{IEEE Network Operations and Management Symposium}, Salvador, Bahia,
  Apr. 2008, pp. 144--151.

\bibitem{Courcoubetis:2016:NPP:2909066.2883610}
C.~Courcoubetis, L.~Gyarmati, N.~Laoutaris, P.~Rodriguez, and K.~Sdrolias,
  ``Negotiating premium peering prices: A quantitative model with
  applications,'' \emph{ACM Trans. Internet Technol.}, vol.~16, no.~2, pp.
  14:1--14:22, Apr. 2016.

\bibitem{Hofbauer20091665}
J.~Hofbauer and W.~H. Sandholm, ``Stable games and their dynamics,''
  \emph{Journal of Economic Theory}, vol. 144, no.~4, pp. 1665 -- 1693.e4, Feb.
  2009.

\bibitem{Sandholm2009}
W.~H. Sandholm, ``Evolutionary game theory,'' in \emph{Encyclopedia of
  Complexity and Systems Science}, R.~A. Meyers, Ed.\hskip 1em plus 0.5em minus
  0.4em\relax New York, NY: Springer New York, 2009, pp. 3176--3205.

\bibitem{dempe2002foundations}
S.~Dempe, \emph{Foundations of bilevel programming}.\hskip 1em plus 0.5em minus
  0.4em\relax Dordrecht, The Netherlands: Kluwer Academic Publishers, 2002.

\bibitem{Leyffer:2010:SMG:1744742.1744748}
S.~Leyffer and T.~Munson, ``Solving multi-leader-common-follower games,''
  \emph{Optimization Methods Software}, vol.~25, no.~4, pp. 601--623, Aug.
  2010.

\bibitem{SANDHOLM200181}
W.~H. Sandholm, ``Potential games with continuous player sets,'' \emph{Journal
  of Economic Theory}, vol.~97, no.~1, pp. 81 -- 108, Mar. 2001.

\bibitem{Weibull1997}
J.~W. Weibull, \emph{Evolutionary game theory}.\hskip 1em plus 0.5em minus
  0.4em\relax MIT press, 1997.

\bibitem{Sastry1999}
S.~Sastry, ``Lyapunov stability theory,'' in \emph{Nonlinear Systems: Analysis,
  Stability, and Control}.\hskip 1em plus 0.5em minus 0.4em\relax New York, NY:
  Springer New York, 1999, pp. 182--234.

\bibitem{sundaram1996first}
R.~K. Sundaram, \emph{A first course in optimization theory}.\hskip 1em plus
  0.5em minus 0.4em\relax Cambridge university press, 1996.

\bibitem{REGGIANI201616}
C.~Reggiani and T.~Valletti, ``Net neutrality and innovation at the core and at
  the edge,'' \emph{International Journal of Industrial Organization}, vol.~45,
  pp. 16 -- 27, 2016.

\bibitem{facchinei201012}
F.~Facchinei and J.-S. Pang, ``Nash equilibria: the variational approach,'' in
  \emph{Convex optimization in signal processing and communications}, D.~P.
  Palomar and Y.~C. Eldar, Eds.\hskip 1em plus 0.5em minus 0.4em\relax
  Cambridge University Press, 2010, ch.~12, p. 443.

\bibitem{doi:10.1137/S1052623499361233}
S.~Scholtes, ``Convergence properties of a regularization scheme for
  mathematical programs with complementarity constraints,'' \emph{SIAM Journal
  on Optimization}, vol.~11, no.~4, pp. 918--936, 2001.

\bibitem{su2004sequential}
C.-L. Su, ``A sequential ncp algorithm for solving equilibrium problems with
  equilibrium constraints,'' Department of Management Science and Engineering,
  Stanford University, Stanford, CA, Tech. Rep., 2004.

\bibitem{ruszczynski2006nonlinear}
A.~P. Ruszczy{\'n}ski, \emph{Nonlinear optimization}, Princeton, New Jersey,
  2006, vol.~13.

\bibitem{doi:10.1287/moor.25.1.1.15213}
H.~Scheel and S.~Scholtes, ``Mathematical programs with complementarity
  constraints: Stationarity, optimality, and sensitivity,'' \emph{Mathematics
  of Operations Research}, vol.~25, no.~1, pp. 1--22, 2000.

\bibitem{Facchinei2003}
F.~Facchinei and J.-S. Pang, Eds., \emph{Finite-Dimensional Variational
  Inequalities and Complementarity Problems}.\hskip 1em plus 0.5em minus
  0.4em\relax New York, NY: Springer New York, 2003.

\bibitem{Gu2003}
K.~Gu, V.~L. Kharitonov, and J.~Chen, \emph{Systems with Multiple and
  Distributed Delays}.\hskip 1em plus 0.5em minus 0.4em\relax Boston, MA:
  Birkh{\"a}user Boston, 2003, pp. 233--271.

\end{thebibliography}

\end{document}